\RequirePackage[T1]{fontenc}
\documentclass[12pt]{article}

\usepackage[height=8.85in,width=6.45in]{geometry}

\usepackage{times}

\usepackage[utf8]{inputenc}
\usepackage{amsmath}
\usepackage{amssymb}
\usepackage{mathtools}
\numberwithin{equation}{section}
\usepackage{slashed}
\usepackage{braket}
\usepackage[svgnames,psnames]{xcolor}
\usepackage[colorlinks,citecolor=DarkGreen,linkcolor=FireBrick,linktocpage]{hyperref}
\usepackage{cite}
\usepackage{graphicx}
\usepackage{tikz}
\usetikzlibrary{arrows.meta}
\usepackage{tikz-cd}
\usepackage{floatrow} 
\usepackage{subcaption}
\usepackage{courier}
\usepackage{bm}
\usepackage{colortbl}
\usepackage{dashbox}

\usepackage{mdframed}

\renewenvironment{figure}[1][]{
  \begin{originalfigure}[#1]
    \begin{mdframed}[linecolor=black!0,backgroundcolor=black!1]
}{
    \end{mdframed}
  \end{originalfigure}
}
\renewenvironment{table}[1][]{
  \begin{originaltable}[#1]
    \begin{mdframed}[linecolor=black!0,backgroundcolor=black!1]
}{
    \end{mdframed}
  \end{originaltable}
}

\tikzstyle{na} = [baseline=-.5ex]
\def\Llleftarrow{%
\lower2pt\hbox{\begingroup
\tikz
\draw[shorten >=0pt,shorten <=0pt] (0,3pt) -- ++(-1em,0) (0,1pt) -- ++(-1em-1pt,0) (0,-1pt) -- ++(-1em-1pt,0) (0,-3pt) -- ++(-1em,0) (-1em+1pt,5pt) to[out=-105,in=45] (-1em-2pt,0) to[out=-45,in=105] (-1em+1pt,-5pt);
\endgroup}
}
\def\node#1#2{\overset{#1}{\underset{#2}{\circ}}}
\def\ver#1#2{\overset{{\llap{$\scriptstyle#1$}\displaystyle\circ{\rlap{$\scriptstyle#2$}}}}{\scriptstyle\vert}}

\newcommand*{\bZ}{\mathbb{Z}}
\newcommand*{\bR}{\mathbb{R}}

\newcommand*{\cH}{\mathcal{H}}

\def\Nequals#1{$\mathcal{N}{=}#1$}

\def\bC{\mathbb{C}}
\def\bH{\mathbb{H}}

\def\cN{\mathcal{N}}
\def\cM{\mathcal{M}}

\def\D{\Delta}
\usepackage{arydshln}
\definecolor{amaranth}{rgb}{0.9, 0.17, 0.31}
\definecolor{coolblack}{rgb}{0.0, 0.18, 0.39}
\definecolor{gold(web)(golden)}{rgb}{1.0, 0.84, 0.0}
\definecolor{deepcarmine}{rgb}{0.66, 0.13, 0.24}

\def\green#1{{\color{green!70!black}{#1}}}
\def\blue#1{{\color{blue}{#1}}}
\def\cblack#1{{\color{coolblack}{#1}}}
\def\XSb{{\bar\cS}}
\def\bu{{\boldsymbol u}}
\def\XS{\mathfrak{S}}
\def\XSb{\bar{\XS}}
\def\H{\mathbb{H}}
\def\hTE{\hat{\CT}_{E_6,2}}
\def\tTE{\widetilde{\CT}_{E_6,2}}
\def\hTG{\hat{\CT}_{D_4,3}}
\def\hTA{\hat{\CT}_{A_2,4}}
\def\ccr{\cellcolor{red!15}} 
\def\ccg{\cellcolor{green!15}} 

\newcommand{\be}{\begin{equation}}
\newcommand{\ee}{\end{equation}}
\def\Tr{\mathop{\mathrm{Tr}}\nolimits}
\def\CN{{\cal N}}
\def\CS{{\cal S}}
\def\CT{{\cal T}}

\usepackage{amsthm}
\usepackage{tikz}
\usetikzlibrary{arrows,snakes,shapes.arrows,decorations.markings}
     \tikzset{>=triangle 90}
     \tikzstyle{bbc}=[draw,circle,fill=black,scale=.75]
     \tikzstyle{rc}=[circle,fill=red,scale=.6]
     \tikzstyle{wc}=[draw,circle,scale=.75]

\def\blue#1{{\color{blue}{#1}}}

\def\green#1{{\color{black!25!green}{#1}}}

\def\bar{\overline}

\def\hat{\widehat}

\def\^{\wedge}

\def\dim{{\rm dim}}

\def\Tr{{\rm Tr}}

\def\U{{\rm U}}
\def\SU{{\rm SU}}

\def\D{{\Delta}}

\def\L{{\Lambda}}

\def\s{{\sigma}}

\def\bh{{\boldsymbol h}}

\def\bu{{\boldsymbol u}}

\def\ff{\mathfrak{f}}

\def\cC{{\mathcal C}}

\def\cCrg{\cC_{\rm reg}}

\def\cH{{\mathcal H}}

\def\cM{{\mathcal M}}

\def\cN{{\mathcal N}}

\def\cS{{\mathcal S}}

\def\cT{{\mathcal T}}

\def\C{\mathbb{C}} 
 
\def\H{\mathbb{H}}

\def\R{\mathbb{R}}

\def\beq{\begin{equation}}
\def\eeq{\end{equation}}

\newcommand{\bpmat}{\begin{pmatrix}}
\newcommand{\epmat}{\end{pmatrix}}
\newcommand{\bsmat}{\begin{smallmatrix}}
\newcommand{\esmat}{\end{smallmatrix}}
\newfloatcommand{capbtabbox}{table}[][\FBwidth]

\def\green#1{{\color{green!60!black}{#1}}}

\def\XT{\mathfrak{T}}

\begin{document}

\begin{titlepage}

\begin{flushright}
\end{flushright}

\vskip 3cm

\begin{center}

{\Large\bfseries More on $\mathcal{N}{=}2$ S-folds}

\vskip 1cm
Simone Giacomelli$^1$,
Mario Martone$^{2}$,
Yuji Tachikawa$^3$,
and
Gabi Zafrir$^4$
\vskip 1cm

\begin{tabular}{ll}
$^1$ & Mathematical Institute, University of Oxford, \\
&Woodstock Road, Oxford, OX2 6GG, United Kingdom\\
$^2$ & C.~N.~Yang Institute for Theoretical Physics and \\
 & Simons Center for Geometry and Physics, Stony Brook University,\\
&Stony Brook, NY 11794-3840, USA\\
$^3$ & Kavli Institute for the Physics and Mathematics of the Universe (WPI), \\
& University of Tokyo,  Kashiwa, Chiba 277-8583, Japan\\
$^4$ & Dipartimento di Fisica, Universit\`a di Milano-Bicocca \& INFN, \\
&Sezione di Milano-Bicocca, I-20126 Milano, Italy
\end{tabular}

\vskip 1cm

\end{center}

\noindent 

We carry out a systematic study of 4d $\mathcal{N}=2$ preserving S-folds of F-theory 7-branes
and the worldvolume theories on D3-branes probing them.
They consist of two infinite series of theories, which we denote  
following \cite{Apruzzi:2020pmv,Giacomelli:2020jel} by
 $\mathcal{S}^{(r)}_{G,\ell}$ for $\ell=2,3,4$
and
 $\mathcal{T}^{(r)}_{G,\ell}$  for $\ell=2,3,4,5,6$.
Their distinction lies in the discrete torsion carried by the S-fold
and in the difference in the asymptotic holonomy of the gauge bundle on the 7-brane.
We study various properties of these theories, using diverse field theoretical and string theoretical methods.

\end{titlepage}

\setcounter{tocdepth}{2}
\tableofcontents

\bigskip

\section{Introduction and summary}
Four-dimensional quantum field theories with \Nequals3 supersymmetry were first constructed in \cite{Garcia-Etxebarria:2015wns} within string theory, 
by considering D3-branes probing certain F-theoretic singularities $\bC^3/\bZ_\ell$ which generalize ordinary orientifold planes. 
These singularities, now commonly called S-folds, were studied in more detail in \cite{Aharony:2016kai}, which found the following:
The order of the quotient is restricted to $\ell=2,3,4,6$, where $\ell=2$ corresponds to the known \Nequals4 orientifolds.
Furthermore, for $\ell=2,3,4$, there is a possibility of turning on a discrete flux, producing another variant for each $\ell$. 
Then rank-$r$ \Nequals3 theories are obtained by probing these S-folds by $r$ D3-branes.
Their properties are summarized in Table~\ref{table:N3charges}.
\begin{table}[h]
\[
\begin{array}{|c|c||c|l|}
\hline
\ell & \text{discrete flux} & \text{D3 charge} & \text{CB spectrum} \\
\hline
\hline
2 & \ccr \text{no} & -1/4 & 2,4,\ldots, 2(r-1), r \\
2 & \ccg \text{yes} & +1/4 & 2,4,\ldots, 2(r-1),2r \\
\hline
3 & \ccr\text{no} & -1/3 & 3,6,\ldots, 3(r-1), r \\
3 &\ccg \text{yes} & +1/3 & 3,6,\ldots, 3(r-1),3r \\
\hline
4 & \ccr\text{no} & -3/8 & 4,8,\ldots, 4(r-1), r \\
4 & \ccg \text{yes} & +3/8 & 4,8,\ldots, 4(r-1),4r \\
\hline
6 & \ccr\text{no} & -5/12 & 6,12,\ldots, 6(r-1), r \\
\hline
\end{array}
\]
\caption{%
Basic properties of \Nequals3 S-folds.
Note that the D3-charge $\epsilon$ is given by the uniform formulas 
$\epsilon_\text{fluxless}=-\frac{\ell-1}{2\ell}$,
$\epsilon_\text{fluxful}=+\frac{\ell-1}{2\ell}$.
\label{table:N3charges}}
\end{table}

\begin{table}[h]
\[
\begin{array}{|c||c|c|c|c|c|c|c|c|c|}
\hline
\text{Kodaira type} & I_0 & II & III & IV & I_0^* & IV^* & III^*& II^* \\
\hline
G & \varnothing & \varnothing & A_1 & A_2 & D_4 & E_6 & E_7 & E_8 \\
\hline
\Delta_7 & 1 & 6/5 & 4/3 & 3/2 & 2 & 3 & 4 &  6 \\
\hline
\text{axiodilaton} & 
\tau & e^{\pi i/3} & e^{\pi i/2} & e^{\pi i/3 }&
\tau &
e^{\pi i/3} & e^{\pi i/2} & e^{\pi i/3 }\\
\hline 
\end{array}
\]
\caption{Basic properties of F-theory 7-branes. $I_0$ corresponds to the absence of the 7-brane.
Only in the $I_0$ and $I_0^*$ cases  the axiodilaton (i.e.~the 4d gauge coupling) is not frozen to a specific value.
\label{table:7brane}}
\end{table}

In \cite{Apruzzi:2020pmv,Giacomelli:2020jel}, one of the authors (Giacomelli) and his collaborators, generalized these constructions by considering \emph{fluxful} S-folds of F-theory 7-branes. 
As is well known, F-theory 7-branes can be characterized by their deficit angle $\Delta_7$, where angular coordinate around the 7-brane has periodicity $2\pi/\Delta_7$.
The value of the type IIB axiodilaton $\tau$ and the gauge algebra $G$ on the 7-brane are 
then determined as given in Table~\ref{table:7brane}.

With S-folds these configurations only preserve 8 supercharges and they were therefore named $\cN=2$ S-folds.
In \cite{Apruzzi:2020pmv}, only the \emph{fluxful} S-folds were considered,
and the worldvolume theories on $r$ D3-branes probing them were analyzed
and labeled as $\mathcal{S}^{(r)}_{G,\ell}$.
Slightly later, in  \cite{Giacomelli:2020jel}, 
a close cousin of these theories, labeled as $\mathcal{T}^{(r)}_{G,\ell}$,
was obtained by Higgsing from $\mathcal{S}^{(r)}_{G,\ell}$.

The aim of this paper is first to show that these $\mathcal{T}^{(r)}_{G,\ell}$ theories are obtained by probing \emph{flux-less} S-folds by D3-branes\footnote{%
While this work is nearing completion, the same statement appeared in \cite{Heckman:2020svr}.
The supporting pieces of evidence provided in \cite{Heckman:2020svr} and 
those given in this paper are largely complementary.
}.
From this perspective, the original \Nequals3 theories of \cite{Garcia-Etxebarria:2015wns,Aharony:2016kai} can be denoted as 
$\mathcal{S}^{(r)}_{\varnothing,\ell}$ 
and
$\mathcal{T}^{(r)}_{\varnothing,\ell}$.
We also point out that there is also another infinite series of theories $\mathcal{T}^{(r)}_{\varnothing,5}$,
for which there is no counterpart with \Nequals3 supersymmetry.\footnote{%
Here we use $G=\varnothing$ to denote the two cases $\Delta_7=1$ and $\Delta_7=\frac65$. 
This leads to no ambiguity, since we can only have $\ell=2,3,4,6$ for the former and $\ell=5$ for the latter.
}
We summarize the properties of these theories in Table~\ref{table:table}. 
Generically, they have flavor symmetry of the form $H\times \blue{SU(2)}$ when $\ell=2$
or $H\times \blue{U(1)}$ when $\ell\neq 2$,
where $H$ is a subgroup of $G$ fixed by a certain order-$\ell$ automorphism,
and $\blue{SU(2)}$ or $\blue{U(1)}$ come from the hyperk\"ahler isometry of $\bC^2/\bZ_\ell$.
We also listed the conformal central charges $a$ and $c$,
the central charges for the nonabelian part of the flavor symmetry,
the dimension of the Higgs branch,
and the spectrum of the Coulomb branch operators.

\begin{table}
\[
\begin{array}{|c|c|c||l@{}l@{}l|c|c|c|}
\hline
\multicolumn{3}{|c||}{\cS^{(r)}_{G,\ell}}&
\multicolumn{6}{c|}{\text{CB spectrum:}\quad \ell\Delta_7,
2\ell\Delta_7, \cdots, (r-1)\ell\Delta_7, r\ell\Delta_7}\\
\hline
\hline 
\ell & G &\Delta_7& \multicolumn{3}{c|}{\text{Flavor Symmetry}} & a & c & \text{Dim. Higgs}\\
\hline
2& E_6& 3&Sp(4)_{6r+1}&\times& \blue{SU(2)}_{6r^2+r}&\frac{36 r^2+42 r+4}{24}&\frac{36r^2+54r+8}{24}& 12r+4\\
2& D_4& 2& Sp(2)_{4r+1}\times SU(2)_{8r}&\times& \blue{SU(2)}_{4r^2+r}& \frac{24r^2+24r+2}{24}&\frac{24r^2+30r+4}{24}& 6r+2\\
2& A_2& \frac32&Sp(1)_{3r+1}\times U(1)&\times& \blue{SU(2)}_{3r^2+r}&\frac{18r^2+15r+1}{24}&\frac{18r^2+18r+2}{24}& 3r+ 1\\
2 & \varnothing & 1 & && \blue{SU(2)}_{2r^2+r} & \frac{12r^2+6r}{24}& \frac{12r^2+6r}{24}& r \\
\hline
3& D_4&2 &SU(3)_{12r+2}&\times &\blue{U(1)} &\frac{ 36r^2+36r+3}{24}&\frac{36r^2+42r+6}{24}&6r+3\\
3& A_1& \frac43& U(1)&\times &\blue{U(1)}&\frac{24r^2+20r+1}{24}&\frac{24r^2+22r+2}{24}& 2r+1\\
3 & \varnothing & 1 & & &\blue{U(1)}_{} &\frac{18r^2+12r}{24} & \frac{18r^2+12r}{24}& r \\
\hline
4& A_2& \frac32& SU(2)_{12r+2}&\times &\blue{U(1)} &\frac{36r^2+33r+2}{24}&\frac{36r^2+36r+4}{24}& 3r+2\\
4 & \varnothing & 1 & && \blue{U(1)}_{} &\frac{24r^2+18r}{24} &\frac{24r^2+18r}{24} & r \\
\hline
\end{array}\]
\[
\begin{array}{|c|c|c||r@{}l|c|c|c|}
\hline
\multicolumn{3}{|c||}{\cT^{(r)}_{G,\ell}}&
\multicolumn{5}{c|}{\text{CB spectrum:}\quad \ell\Delta_7,
2\ell\Delta_7, \cdots, (r-1)\ell\Delta_7, r\Delta_7}\\
\hline
\hline
\ell & G &\Delta_7&  \multicolumn{2}{c|}{\text{Flavor Symmetry}} & a & c & \text{Dim. Higgs}\\
\hline 
2& E_6&3 &(F_4)_{6r}\times& \blue{SU(2)}_{6r^2-5r}&\frac{6 r^2+r}{4}&\frac{6r^2+3r}{4}& 12r\\
2& D_4&2& SO(7)_{4r}\times& \blue{SU(2)}_{4r^2-3r}& r^2&\frac{4r^2+r}{4}& 6r\\
2& A_2&\frac32&SU(3)_{3r}\times& \blue{SU(2)}_{3r^2-2r}&\frac{6r^2-r}{8}&\frac{3r^2}{4}& 3r\\
2 & \varnothing & 1 & & \blue{SU(2)}_{2r^2-r} & \frac{2r^2-r}{4}&\frac{2r^2-r}{4} & r \\
\hline
3& D_4&2&(G_2)_{4r}\times& \blue{U(1)} &\frac{ 3r^2-r}{2}&\frac{6r^2-r}{4}& 6r\\
3& A_1&\frac43 & SU(2)_{\frac{8r}{3}}\times& \blue{U(1)}&\frac{2r^2-r}{2}&\frac{12r^2-5r}{12}& 2r\\
3 & \varnothing & 1 & & \blue{U(1)}_{} &\frac{3r^2-2r}{4} &\frac{3r^2-2r}{4} & r \\
\hline
4& A_2&\frac32&  SU(2)_{3r}\times& \blue{U(1)} &\frac{12r^2-7r}{8}&\frac{6r^2-3r}{4}& 3r\\
4 & \varnothing & 1 & & \blue{U(1)}_{} & \frac{4r^2-3r}{4}&\frac{4r^2-3r}{4} & r \\
\hline
5& \varnothing &\frac65&   &\blue{U(1)} &\frac{15r^2-11r}{10}&\frac{30r^2-21r}{20}& r\\
\hline 
6 & \varnothing & 1 & & \blue{U(1)}_{} & \frac{6r^2-5r}{4}&  \frac{6r^2-5r}{4}& r \\
\hline
\end{array}\]
\caption{%
Basic properties of theories $\cS^{(r)}_{G,\ell}$ and $\cT^{(r)}_{G,\ell}$.
In the flavor symmetry, the last factor $\blue{SU(2)}$ or $\blue{U(1)}$, shown in \blue{blue}, comes from the isometry of $\bC^2/\bZ_\ell$ and the rest comes from the part of $G$ commuting with the asymptotic holonomy.
We use subscripts to denote the levels of the flavor symmetries.
The cases with $\Delta_7=1$ have \Nequals4 when $\ell=2$ and \Nequals3 otherwise.
The flavor symmetries listed here are for generic rank.
For the enhancements in low rank, see Table~\ref{table:enhancement}.
\label{table:table}}
\end{table}

In the rest of the paper, we study various detailed properties of these theories using diverse methods.
Before proceeding, we would like to make two remarks.

The first is on the possible discrete gauging on these theories.
Our analysis strongly suggests that both $\cS^{(r)}_{G,\ell}$ and $\cT^{(r)}_{G,\ell}$ admits a discrete gauging by a $\bZ_\ell$ symmetry,
which always acts non-trivially on the Higgs branch, 
while for $\cT^{(r)}_{G,\ell}$ also acts non-trivially on the Coulomb branch such that the operator of dimension $r\Delta_7$ becomes of dimension $r\ell\Delta_7$.
Just to be consistent, we reserve our letters $\cS^{(r)}_{G,\ell}$ and $\cT^{(r)}_{G,\ell}$ for the ungauged versions of these theories.

The second is on the special behaviors of these theories when the rank is low enough.
In a series of papers \cite{Argyres:2015ffa,Argyres:2015gha,Argyres:2016xmc,Argyres:2016xua} by one of the authors (Martone) and his collaborators,
a purely field-theoretical classification of rank-1 4d \Nequals2 superconformal theories was performed,
where a number of theories unknown at that time were found. 
The $\mathcal{S}^{(1)}_{G,\ell}$ theories neatly reproduce all of them, 
via an interesting enhancement of symmetries, summarized in Table~\ref{table:enhancement}.
\begin{table}
\[
\begin{array}{|c|c|c||l@{}l|c|c|c|}
\hline
\multicolumn{3}{|c||}{\cS^{(r=1)}_{G,\ell}}&
\multicolumn{4}{c|}{\text{CB spectrum:}\quad \ell\Delta_7}\\
\hline 
\ell & G &\Delta_7 & \multicolumn{2}{c|}{\text{Generic Flavor Symmetry}} & \text{Enhanced Flavor Symmetry}\\
\hline
\hline
2& E_6&3&Sp(4)_{7}&\times \blue{SU(2)}_{7} & Sp(5)_7 \\
2& D_4&2& Sp(2)_{5}\times SU(2)_{8}&\times \blue{SU(2)}_{5}& Sp(3)_5\times SU(2)_8\\
2& A_2&\frac32 &Sp(1)_{4}\times U(1)&\times \blue{SU(2)}_{4}&Sp(2)_4\times U(1)\\
\hline
3& D_4&2 &SU(3)_{14}&\times \blue{U(1)} & SU(4)_{14} \\
3& A_1&\frac43& U(1)&\times \blue{U(1)}&  SU(2)_{10}\times U(1) \\
\hline
4& A_2& \frac32& SU(2)_{14}&\times \blue{U(1)} & SU(3)_{14} \\
\hline
\end{array}
\]
\caption{
The pattern of symmetry enhancement of $\cS^{(r=1)}_{G,\ell}$.
Note that often a subgroup of $G$ and the isometry of $\bC^2/\bZ_\ell$ combine to form a larger simple component of the enhanced flavor symmetry.
We also note that the symmetry of low-rank $\cT$ theories enhance:
the $\cT^{(r=1)}_{G,\ell}$ theories have the full symmetry $G$,
since they are equal to the old rank-1 theory with $G$ symmetry plus a free hypermultiplet,
whereas for the $\cT^{(r=2)}_{G,\ell}$ theories the isometry $\blue{SU(2)}$ symmetry for $\ell=2$ enhances to $SU(2)^2$ while the isometry $\blue{U(1)}$ symmetry for $\ell\neq 2$ enhances to $SU(2)$.
The subscripts are for the flavor central charges.
\label{table:enhancement}}
\end{table}

We can now Higgs $\mathcal{S}^{(1)}_{G,\ell}$ theories to $\mathcal{T}^{(1)}_{G,\ell}$,
which are in fact equivalent to the old rank-1 theory of type $G$ plus a single free hypermultiplet.
We note, for example, the level of the $SU(2)$ symmetry for all the $\ell=2$ cases is $1$, which acts only on the free hypermultiplet.
Also, the generic flavor symmetry enhances to the entirety of $G$.
Finally, we can further Higgs it to the $\mathcal{S}^{(0)}_{G,\ell}$ theories,
which are simply free hypermultiplets.
These and other info are also summarized in the flow diagram in Fig.~\ref{ThrRl2}.

\begin{figure}[tbp]
\centering
\begin{tikzpicture}
[
auto,
good/.style={rectangle,rounded corners,fill=green!50,inner sep=2pt},
bad/.style={rectangle,rounded corners,fill=red!15,inner sep=2pt},
ugly/.style={rectangle,rounded corners,fill=blue!10,inner sep=2pt},
Harrow/.style={->,>=stealth[round],shorten >=1pt,line width=.4mm,coolblack},
Marrow/.style={->,>=stealth[round],shorten >=1pt,line width=.4mm,deepcarmine}
]
\begin{scope}[yshift=-0cm]
\node at (7.75,-.5) {{\large {\fontfamily{qcs}\selectfont\textsc{$\ell=2$ Series}}}};
\end{scope}
\begin{scope}[yshift=-2cm]
\fill[color=gray!10, rounded corners] (0,1) rectangle (15.9,-5.4);
\fill[color=amaranth!30, rounded corners] (9.4,.6) rectangle (14.3,-5);
\node (SE6r) at (1,0) [,align=center] {$\CS^{(r)}_{E_6,2}$};
\node (TE6r) at (3,0) [,align=center] {$\CT^{(r)}_{E_6,2}$};
\node (SE6r1) at (5,0) [,align=center] {$\CS^{(r-1)}_{E_6,2}$};
\node (TE6r1) at (7,0) [,align=center] {$\CT^{(r-1)}_{E_6,2}$};
\node (DE6) at (8.5,0) [,align=center] {\boldsymbol{\cblack{$\cdots$}}};
\node (SE61) at (10.5,0) [,align=center] {\small{$[II^*,C_5]$}};
\node (TE61) at (13,0) [,align=center] {\small{$[IV^*,E_6]+\bH$}};
\node (SE60) at (15,0) [,align=center] {\small{$\H^4\vphantom{[E_6^6]}$}};
\end{scope}
\begin{scope}[yshift=-3.5cm]
\node (SD4r) at (1,0) [,align=center] {$\CS^{(r)}_{D_4,2}$};
\node (TD4r) at (3,0) [,align=center] {$\CT^{(r)}_{D_4,2}$};
\node (SD4r1) at (5,0) [,align=center] {$\CS^{(r-1)}_{D_4,2}$};
\node (TD4r1) at (7,0) [,align=center] {$\CT^{(r-1)}_{D_4,2}$};
\node (DD4) at (8.5,0) [,align=center] {\boldsymbol{\cblack{$\cdots$}}};
\node (SD41) at (10.5,0) [,align=center] {\small{$[III^*,C_3A_1]$}};
\node (TD41) at (13,0) [,align=center] {\small{$[I_0^*,D_4]+\bH$}};
\node (SD40) at (15,0) [,align=center] {\small{$\H^2\vphantom{[E_6^6]}$}};
\end{scope}
\begin{scope}[yshift=-5cm]
\node (SA2r) at (1,0) [,align=center] {$\CS^{(r)}_{A_2,2}$};
\node (TA2r) at (3,0) [,align=center] {$\CT^{(r)}_{A_2,2}$};
\node (SA2r1) at (5,0) [,align=center] {$\CS^{(r-1)}_{A_2,2}$};
\node (TA2r1) at (7,0) [,align=center] {$\CT^{(r-1)}_{A_2,2}$};
\node (DA2) at (8.5,0) [,align=center] {\boldsymbol{\cblack{$\cdots$}}};
\node (SA21) at (10.5,0) [,align=center] {\small{$[IV^*,C_2U_1]$}};
\node (TA21) at (13,0) [,align=center] {\small{$[IV,A_2]+\bH$}};
\node (SA20) at (15,0) [,align=center] {\small{$\H\vphantom{[E_6^6]}$}};
\end{scope}
\begin{scope}[yshift=-6.5cm]
\node (Sr) at (1,0) [align=center] {{$\blue{\CS^{(r)}_{\varnothing,2}}$}};
\node (Tr) at (3,0) [align=center] {{$\blue{\CT^{(r)}_{\varnothing,2}}$}};
\node (Sr1) at (5,0) [align=center] {{$\blue{\CS^{(r-1)}_{\varnothing,2}}$}};
\node (Tr1) at (7,0) [align=center] {{$\blue{\CT^{(r-1)}_{\varnothing,2}}$}};
\node (D) at (8.5,0) [align=center] {\boldsymbol{\cblack{$\cdots$}}};
\node (S1) at (10.5,0) [,align=center] {{$\blue{[I_0^*,A_1]}$}};
\node (T1) at (13,0) [,align=center] {\small\blue{{$[I_0,\varnothing]$}}};
\node (S0) at (15,0) [,align=center] {\small{$\varnothing$}};
\end{scope}
\draw[Harrow] (SE6r) to (TE6r);
\draw[Harrow] (TE6r) to (SE6r1);
\draw[Harrow] (SE6r1) to (TE6r1);
\draw[Harrow] (DE6) to (SE61);
\draw[Harrow] (SE61) to (TE61);
\draw[Harrow] (TE61) to (SE60);
\draw[shorten >=1pt, line width=.4mm,coolblack] (DE6) to (TE6r1);
\draw[Marrow] (SE6r) to (SD4r);
\draw[Marrow] (TE6r) to (TD4r);
\draw[Marrow] (SE6r1) to (SD4r1);
\draw[Marrow] (TE6r1) to (TD4r1);
\draw[Marrow] (SE61) to (SD41);
\draw[Marrow] (TE61) to (TD41);
\draw[Marrow] (SE60) to (SD40);
\draw[Harrow] (SD4r) to (TD4r);
\draw[Harrow] (TD4r) to (SD4r1);
\draw[Harrow] (SD4r1) to (TD4r1);
\draw[Harrow] (DD4) to (SD41);
\draw[Harrow] (SD41) to (TD41);
\draw[Harrow] (TD41) to (SD40);
\draw[shorten >=1pt, line width=.4mm,coolblack] (DD4) to (TD4r1);
\draw[Marrow] (SD4r) to (SA2r);
\draw[Marrow] (TD4r) to (TA2r);
\draw[Marrow] (SD4r1) to (SA2r1);
\draw[Marrow] (TD4r1) to (TA2r1);
\draw[Marrow] (SD41) to (SA21);
\draw[Marrow] (TD41) to (TA21);
\draw[Marrow] (SD40) to (SA20);
\draw[Harrow] (SA2r) to (TA2r);
\draw[Harrow] (TA2r) to (SA2r1);
\draw[Harrow] (SA2r1) to (TA2r1);
\draw[shorten >=1pt, line width=.4mm,coolblack] (DA2) to (TA2r1);
\draw[Harrow] (DA2) to (SA21);
\draw[Harrow] (SA21) to (TA21);
\draw[Harrow] (TA21) to (SA20);
\draw[Marrow] (SA2r) to (Sr);
\draw[Marrow] (TA2r) to (Tr);
\draw[Marrow] (SA2r1) to (Sr1);
\draw[Marrow] (TA2r1) to (Tr1);
\draw[Marrow] (SA21) to (S1);
\draw[Marrow] (TA21) to (T1);
\draw[Marrow] (SA20) to (S0);


\begin{scope}[yshift=-7.35cm]
\node at (7.75,-.5) {{\large {\fontfamily{qcs}\selectfont\textsc{$\ell=3$ Series}}}};
\end{scope}
\begin{scope}[yshift=-9.5cm]
\fill[color=gray!10, rounded corners] (0,1) rectangle (15.9,-3.9);
\fill[color=amaranth!30, rounded corners] (9.4,.6) rectangle (14.3,-3.5);
\node (SD4r) at (1,0) [,align=center] {$\CS^{(r)}_{D_4,3}$};
\node (TD4r) at (3,0) [,align=center] {$\CT^{(r)}_{D_4,3}$};
\node (SD4r1) at (5,0) [,align=center] {$\CS^{(r-1)}_{D_4,3}$};
\node (TD4r1) at (7,0) [,align=center] {$\CT^{(r-1)}_{D_4,3}$};
\node (DD4) at (8.5,0) [,align=center] {\boldsymbol{\cblack{$\cdots$}}};
\node (SD41) at (10.5,0) [,align=center] {\small{$[II^*,A_3
]$}};
\node (TD41) at (13,0) [,align=center] {\small{$[I_0^*,D_4]+\bH$}};
\node (SD40) at (15,0) [,align=center] {\small{$\H^3\vphantom{[E_6^6]}$}};
\end{scope}
\begin{scope}[yshift=-11cm]
\node (SA1r) at (1,0) [,align=center] {$\CS^{(r)}_{A_1,3}$};
\node (TA1r) at (3,0) [,align=center] {$\CT^{(r)}_{A_1,3}$};
\node (SA1r1) at (5,0) [,align=center] {$\CS^{(r-1)}_{A_1,3}$};
\node (TA1r1) at (7,0) [,align=center] {$\CT^{(r-1)}_{A_1,3}$};
\node (DA1) at (8.5,0) [,align=center] {\boldsymbol{\cblack{$\cdots$}}};
\node (SA11) at (10.5,0) [,align=center] {\small{$[III^*,A_1U_1
]$}};
\node (TA11) at (13,0) [,align=center] {\small{$[III,A_1]+\bH$}};
\node (SA10) at (15,0) [,align=center] {\small{$\H\vphantom{[E_6^6]}$}};
\end{scope}
\begin{scope}[yshift=-12.5cm]
\node (Sr) at (1,0) [align=center] {{$\green{\CS^{(r)}_{\varnothing,3}}$}};
\node (Tr) at (3,0) [align=center] {{$\green{\CT^{(r)}_{\varnothing,3}}$}};
\node (Sr1) at (5,0) [align=center] {{$\green{\CS^{(r-1)}_{\varnothing,3}}$}};
\node (Tr1) at (7,0) [align=center] {{$\green{\CT^{(r-1)}_{\varnothing,3}}$}};
\node (D) at (8.5,0) [align=center] {\boldsymbol{\cblack{$\cdots$}}};
\node (S1) at (10.5,0) [,align=center] {{$\green{[IV^*,U_1]}$}};
\node (T1) at (13,0) [,align=center] {\small\green{{$[I_0,\varnothing]$}}};
\node (S0) at (15,0) [,align=center] {\small{$\varnothing$}};
\end{scope}
\draw[Harrow] (SD4r) to (TD4r);
\draw[Harrow] (TD4r) to (SD4r1);
\draw[Harrow] (SD4r1) to (TD4r1);
\draw[Harrow] (DD4) to (SD41);
\draw[Harrow] (SD41) to (TD41);
\draw[Harrow] (TD41) to (SD40);
\draw[shorten >=1pt, line width=.4mm,coolblack] (DD4) to (TD4r1);
\draw[Marrow] (SD4r) to (SA1r);
\draw[Marrow] (TD4r) to (TA1r);
\draw[Marrow] (SD4r1) to (SA1r1);
\draw[Marrow] (TD4r1) to (TA1r1);
\draw[Marrow] (SD41) to (SA11);
\draw[Marrow] (TD41) to (TA11);
\draw[Marrow] (SD40) to (SA10);
\draw[Harrow] (SA1r) to (TA1r);
\draw[Harrow] (TA1r) to (SA1r1);
\draw[Harrow] (SA1r1) to (TA1r1);
\draw[shorten >=1pt, line width=.4mm,coolblack] (DA1) to (TA1r1);
\draw[Harrow] (DA1) to (SA11);
\draw[Harrow] (SA11) to (TA11);
\draw[Harrow] (TA11) to (SA10);
\draw[Marrow] (SA1r) to (Sr);
\draw[Marrow] (TA1r) to (Tr);
\draw[Marrow] (SA1r1) to (Sr1);
\draw[Marrow] (TA1r1) to (Tr1);
\draw[Marrow] (SA11) to (S1);
\draw[Marrow] (TA11) to (T1);
\draw[Marrow] (SA10) to (S0);


\begin{scope}[yshift=-13.35cm]
\node at (7.75,-.5) {{\large {\fontfamily{qcs}\selectfont\textsc{$\ell=4$ Series}}}};
\end{scope}
\begin{scope}[yshift=-15.5cm]
\fill[color=gray!10, rounded corners] (0,1) rectangle (15.9,-2.4);
\fill[color=amaranth!30, rounded corners] (9.4,.6) rectangle (14.3,-2);
\node (SA2r) at (1,0) [,align=center] {$\CS^{(r)}_{A_2,4}$};
\node (TA2r) at (3,0) [,align=center] {$\CT^{(r)}_{A_2,4}$};
\node (SA2r1) at (5,0) [,align=center] {$\CS^{(r-1)}_{A_2,4}$};
\node (TA2r1) at (7,0) [,align=center] {$\CT^{(r-1)}_{A_2,4}$};
\node (DA2) at (8.5,0) [,align=center] {\boldsymbol{\cblack{$\cdots$}}};
\node (SA21) at (10.5,0) [,align=center] {\small{$[II^*,A_2
]$}};
\node (TA21) at (13,0) [,align=center] {\small{$[IV,A_2]+\bH$}};
\node (SA20) at (15,0) [,align=center] {\small{$\H^2\vphantom{[E_6^6]}$}};
\end{scope}
\begin{scope}[yshift=-17cm]
\node (Sr) at (1,0) [align=center] {{$\green{\CS^{(r)}_{\varnothing,4}}$}};
\node (Tr) at (3,0) [align=center] {{$\green{\CT^{(r-1)}_{\varnothing,4}}$}};
\node (Sr1) at (5,0) [align=center] {{$\green{\CS^{(r)}_{\varnothing,4}}$}};
\node (Tr1) at (7,0) [align=center] {{$\green{\CT^{(r-1)}_{\varnothing,4}}$}};
\node (D) at (8.5,0) [align=center] {\boldsymbol{\cblack{$\cdots$}}};
\node (S1) at (10.5,0) [,align=center] {{$\green{[III^*,U_1]}$}};
\node (T1) at (13,0) [,align=center] {\small\green{{$[I_0,\varnothing]$}}};
\node (S0) at (15,0) [,align=center] {\small{$\varnothing$}};
\end{scope}
\draw[Harrow] (SA2r) to (TA2r);
\draw[Harrow] (TA2r) to (SA2r1);
\draw[Harrow] (SA2r1) to (TA2r1);
\draw[Harrow] (DA2) to (SA21);
\draw[Harrow] (SA21) to (TA21);
\draw[Harrow] (TA21) to (SA20);
\draw[shorten >=1pt, line width=.4mm,coolblack] (DA2) to (TA2r1);
\draw[Marrow] (SA2r) to (Sr);
\draw[Marrow] (TA2r) to (Tr);
\draw[Marrow] (SA2r1) to (Sr1);
\draw[Marrow] (TA2r1) to (Tr1);
\draw[Marrow] (SA21) to (S1);
\draw[Marrow] (TA21) to (T1);
\draw[Marrow] (SA20) to (S0);
\begin{scope}[yshift=-18cm]
\fill[color=amaranth!30, rounded corners] (5,0) rectangle (6,-.5);
\node (L1) at (5,-.75){};
\node (L2) at (6.1,-.75){};
\node (L3) at (5,-1.25){};
\node (L4) at (6.1,-1.25){};
\node[anchor=west] (Text1) at (6,-.25)  {\small{\textsc{: rank-1 theories,}}};
\node[anchor=west] (Text2) at (6,-.75)  {\small{\textsc{: Higgsing,}}};
\node[anchor=west] (Text3) at (6,-1.25) {\small{\textsc{: mass deformation.}}};
\draw[Harrow] (L1) to (L2);
\draw[Marrow] (L3) to (L4);
\end{scope}
\end{tikzpicture}
\caption{Graphical depiction of the RG-relations among the $\CT$ and $\CS$ theories for $\ell=2$, $3$ and $4$. Entries in green, $\green{\CS^{(r)}_{\varnothing,\ell=3,4}}$ and $\green{\CS^{(r)}_{\varnothing,\ell=3,4}}$,  are $\cN=3$ supersymmetric,
while entries in blue, $\blue{\CS^{(r)}_{\varnothing,2}}$ and  $\blue{\CT^{(r)}_{\varnothing,2}}$, are $\cN=4$ super Yang-Mills with gauge group $SO(2r+1)$ and $SO(2r)$, respectively.
We also spelled out the rank-1 theories $\cS^{(1)}_{G,\ell}$ and $\cT^{(1)}_{G,\ell}$ using the standard notation $[\text{Kodaira type},\text{flavor symmetry}]$.
}
\label{ThrRl2}
\end{figure}

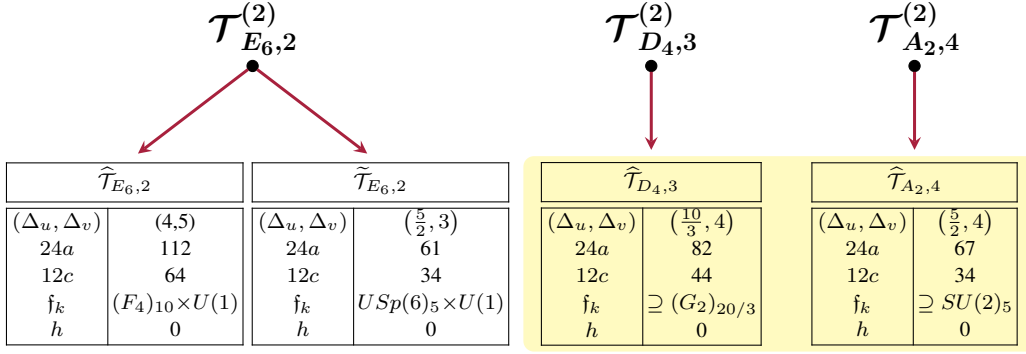
\begin{figure}[h!]
\begin{tikzpicture}[decoration={markings,
mark=at position .5 with {\arrow{>}}},
Marrow/.style={->,>=stealth[round],shorten >=1pt,line width=.4mm,deepcarmine}]
\fill[color=yellow!30, rounded corners] (3.6,-1.2) rectangle (10.45,-3.8);
\begin{scope}
\node[bbc,scale=.5] (p0a) at (0,0) {};
\node[scale=.8] (t0a) at (0,.5) {{\Large$\boldsymbol{\CT^{(2)}_{E_6,2}}$}};
\node (tabp1) at (-1.7,-1.3) {};
\node (tab1) at (-1.7,-2.5) {
  {\scriptsize\begin{tabular}{|@{}c@{}|@{}c@{}|} 
  \hline
  \multicolumn{2}{|c|}{$\hTE$\vphantom{$\Big|$}}\\
  \hline\hline
  $\ (\D_u,\D_v)\ $  &\quad (4,5)\quad{} \\
  $24a$ & 112\\  
  $12c$ & 64 \\
 $\ff_k$ &\,$(F_4)_{10}{\times}U(1)$\,\\ 
$h$&0\\
\hline
  \end{tabular}}};
\node (tabp2) at (1.7,-1.3){};
\node (tab2) at (1.7,-2.5) {
  {\scriptsize\begin{tabular}{|@{}c@{}|@{}c@{}|} 
  \hline
  \multicolumn{2}{|c|}{$\tTE$\vphantom{$\Big|$}}\\
  \hline\hline
  $\ (\D_u,\D_v)\ $  &\quad $\left(\frac52,3\right)$\quad{} \\
  $24a$ & 61\\  
  $12c$ & 34 \\
 $\ff_k$ &\,$USp(6)_{5}{\times}U(1)$\,\\ 
$h$&0\\
\hline
  \end{tabular}}};
\draw[Marrow] (p0a) -- (tabp1);
\draw[Marrow] (p0a) -- (tabp2);
\end{scope}
\begin{scope}[xshift=5.3cm]
\node[bbc,scale=.5] (p0a) at (0,0) {};
\node[scale=.8] (t0a) at (0,.5) {{\Large$\boldsymbol{\CT^{(2)}_{D_4,3}}$}};
\node (tabp1) at (0,-1.3){};
\node (tab1) at (0,-2.5) {
  {\scriptsize\begin{tabular}{|@{}c@{}|@{}c@{}|} 
  \hline
  \multicolumn{2}{|c|}{$\hTG$\vphantom{$\Big|$}}\\
  \hline\hline
  $\ (\D_u,\D_v)\ $  &\quad $\left(\frac{10}3,4\right)$\quad{} \\
  $24a$ & 82\\  
  $12c$ & 44 \\
 $\ff_k$ &\,$\supseteq (G_2)_{20/3}$\,\\ 
$h$&0\\
\hline
  \end{tabular}}};
\draw[Marrow] (p0a) -- (tabp1);
\end{scope}
\begin{scope}[xshift=8.8cm]
\node[bbc,scale=.5] (p0a) at (0,0) {};
\node[scale=.8] (t0a) at (0,.5) {{\Large$\boldsymbol{\CT^{(2)}_{A_2,4}}$}};
\node (tabp1) at (0,-1.3){};
\node (tab1) at (0,-2.5) {
  {\scriptsize\begin{tabular}{|@{}c@{}|@{}c@{}|} 
  \hline
  \multicolumn{2}{|c|}{$\hTA$\vphantom{$\Big|$}}\\
  \hline\hline
  $\ (\D_u,\D_v)\ $  &\quad $\left(\frac52,4\right)$\quad{} \\
  $24a$ & 67\\  
  $12c$ & 34 \\
 $\ff_k$ &\,$\supseteq SU(2)_{5}$\,\\ 
$h$&0\\
\hline
  \end{tabular}}};
\draw[Marrow] (p0a) -- (tabp1);
\end{scope}
\end{tikzpicture}{\caption{\label{newtheories} $\cN=2$ SCFTs which can be obtained by mass deforming rank-2 $\cT$-theories. We displayed in yellow those which to our knowledge have not appeared in the literature before.}}
\end{figure}

We also note that the rank-2 theories $\mathcal{T}^{(2)}_{G,\ell}$ show the symmetry enhancement,
where the isometry $\blue{SU(2)}$ symmetry for $\ell=2$ enhances to $SU(2)^2$ 
while the isometry $\blue{U(1)}$ symmetry for $\ell\neq 2$ enhances to $SU(2)$.
Various mass deformations of these $\CT^{(2)}_{G,\ell}$ theories lead to rank-2  $4d$ SCFTs, some of which to our knowledge have not appeared before.
We summarize their properties in figure \ref{newtheories}. 

The rest of the paper is organized as follows.
In Sec.~\ref{sec:construction}, we first analyze the geometric properties of the \Nequals2 S-folds in F-theory in detail.
We then determine the basic properties of our theories $\cS^{(r)}_{G,\ell}$ and $\cT^{(r)}_{G,\ell}$, the $4d$ theories on the stack of $r$ D3-branes probing these backgrounds.
We also discuss $6d$ constructions of these theories when $\ell\Delta_7=6$.

In the other three sections, we explore more detailed properties of these $4d$ theories.
In Sec.~\ref{sec:stratification}, we discuss the stratifications of the Coulomb branch of the rank-2 cases of our $\cS$ and $\cT$ theories, using the technique recently developed in \cite{Martone:2020nsy,Argyres:2020wmq}.
In Sec.~\ref{sec:magneticquivers}, we determine the magnetic quivers of our theories when $\ell\Delta_7=6$, using their realization as a twisted compactification of $5d$ theories.
Finally, in Sec.~\ref{sec:mass}, we study mass deformations of some of our theories,
both from the $5d$ point of view and from the purely $4d$ point of view. 
Along the way, we encounter a few rank-2 SCFTs which have not been explicitly discussed in the literature, to the authors' knowledge.

\section{\Nequals2 $\CS$-folds and the 4d theories on the probe D3-branes}
\label{sec:construction}

The  $\CN{=}2$ $\CS$-folds constructed in \cite{Apruzzi:2020pmv} combine 7-branes with constant axiodilaton with the $\CN{=}3$ $\CS$-folds studied in \cite{Garcia-Etxebarria:2015wns,Aharony:2016kai}.
The construction involves taking a $\mathbb{Z}_{\ell}$ quotient of a $\mathbb{C}^2$ wrapped by the 7-brane, combined with a $\mathbb{Z}_{\ell}$ quotient of the plane transverse to the 7-brane and the action of a $\mathbb{Z}_{\ell\Delta_7}$ subgroup of the $SL(2,\mathbb{Z})$ duality group of Type IIB string theory to preserve supersymmetry. 
Here, $\mathbb{Z}_{\ell\Delta_7}$ is a symmetry of the theory (therefore making the quotient possible) only if the axiodilaton has a specific value and this must be equal to the value at which the axiodilaton is frozen by the presence of the 7-brane. 

The value of the axiodilaton $\tau$ was already given in Table~\ref{table:7brane}.
The allowed solutions are given by the pairs $(\ell,\Delta_7)$ such that 
\be\label{quotient} \ell\Delta_7=1,2,3,4\;\;\text{or}\;\;6\ee 
with $\Delta_7$ as in Table~\ref{table:7brane} and $\ell$ integer. 
More explicitly, the allowed solutions are
\begin{itemize}
\item $\ell=2$ and $\Delta_7=3/2,2,3$, corresponding to $G=A_2,D_4,E_6$;
\item $\ell=3$ and $\Delta_7=4/3,2$ , corresponding to $G=A_1,D_4$;
\item $\ell=4$ and $\Delta_7=3/2$, corresponding to $G=A_2$;
\item $\ell=5$ and $\Delta_7=6/5$, corresponding to $G=\varnothing$,
\end{itemize}
and the cases with $\Delta_7=1$, which have either \Nequals4 or \Nequals3.
For $\ell=1$ we take the $SL(2,\mathbb{Z})$ quotient to be trivial, so that this case corresponds to having a 7-brane of type $G$ in flat space. 
We do not discuss these well-studied cases of $\ell=1$ further.

In the rest of this section, we study in detail the geometry of these \Nequals2 S-folds and the properties of the 4d theories on $r$ D3-branes probing them.
As this section is somewhat long, here we provide how it is organized.
We start in Sec.~\ref{sec:W} by providing some more detail of the $\bZ_\ell$ quotient on the F-theory geometry, by studying the Weierstrass model.
We then study in Sec.~\ref{sec:4dholonomy} the possible choices of the asymptotic holonomies on the 7-branes on $S^3/\bZ_\ell$,
which we find to correlate well with the choice of the discrete flux of the \Nequals3 S-folds.
In Sec.~\ref{sec:ST} we then explain the computation of the conformal and flavor central charges of our theories $\cS^{(r)}_{(G,\ell)}$ and $\cT^{(r)}_{(G,\ell)}$.
In Sec.~\ref{sec:6d}, we discuss an alternative 6d construction which is available when $\ell\Delta_7=6$, and the duality which relate it to our main F-theory construction.
Now, in either description, the Higgs branch of our theories is to be identified with the instanton moduli space on $\bC^2/\bZ_\ell$. 
We compute its properties in Sec.~\ref{sec:higgs} and confirm their agreements with the results from other analyses.
Finally, we make further comments on the rank-1 cases and their discrete gaugings in Sec.~\ref{sec:rank1},
and on the relation between these theories and SCFTs with more than eight supercharges in Sec.~\ref{section:N3}.

\subsection{Analysis of the F-theory Weierstrass model}
\label{sec:W}
The operation described above can be defined at the level of the F-theory Weierstrass model: We should consider a quotient of the Kodaira singularity describing the given 7-brane which acts as a $\mathbb{Z}_{\ell}$ orbifold of the base of the Weierstrass fibration. 
Here we summarize the discussion in \cite{Apruzzi:2020pmv},
providing the details of the $\ell=5,6$ cases not discussed in that reference.

\begin{table}[!h]
\centering
\begin{tabular}{|c|c|c|l|c|}
\hline
Kodaira type& $G$ &$\Delta_7$ & Weierstrass  & $\tau$ \\
\hline 
$I_0$ & $\varnothing$ & $1$ & $y^2= x^3+fx+g$ & $\tau$ \\
$II$ & $\varnothing$ & $\frac{6}{5}$ & $y^2=x^3+c_{4/5}x+z$ & $e^{\pi i/3}$\\
$III$ & $SU(2)$ &$\frac{4}{3}$ & $y^2=x^3+xz+c_{2/3}z+M_2$ & $e^{\pi i/2}$\\ 
$IV$& $SU(3) $ & $\frac{3}{2}$ & $y^2=x^3+z^2+M_3+x(c_{1/2}z+M_2)$ & $e^{\pi i/3}$\\
$I_0^*$&$SO(8)$& 2 & $y^2=x^3+x(\tau z^2+M_2z+M_4)+ z^3+\widetilde{M}_4z+M_6$ & $\tau$ \\
$IV^*$& $E_6$& 3 & $y^2=x^3+z^4+\sum_{i=2}^4M_{3i}z^{4-i}+x\left(\sum_{i=0}^2M_{2+3i}z^{2-i}\right)$ & $ e^{\pi i/3}$\\
$III^*$& $E_7$&  4 & $y^2=x^3+x(z^3+M_8z+M_{12})+\sum_{i=0}^4M_{2+4i}z^{4-i}$ & $e^{\pi i/2}$\\ 
$II^*$&  $E_8$& 6 & $y^2=x^3+z^5+\sum_{i=2}^5M_{6i}z^{5-i}+x\left(\sum_{i=0}^3M_{2+6i}z^{3-i}\right)$ & $e^{\pi i/3}$\\
\hline
\end{tabular}
\caption{The eight scale-invariant Kodaira singularities with the corresponding Weierstrass forms. We have included explicitly all the deformations including the relevant couplings $c_i$ and mass parameters $M_i$ (where $i$ denotes the scaling dimension) of the corresponding four-dimensional theory living on a probe $D3$ brane.  The coordinate $z$ parametrizes the base of the Weierstrass i.e. the transverse plane to the 7-brane. \label{7braneW}}
\end{table}

If we write the Kodaira singularities in Table \ref{7braneW} in the form $W(x,y,z)=0$, the corresponding holomorphic two-form reads 
\be\label{2holo}\Omega_2=\frac{dzdxdy}{dW}.\ee 
We want our $\mathbb{Z}_{\ell}$ quotient to act on $z$ as $z\rightarrow e^{2\pi i/\ell}z$ and therefore we can introduce the invariant coordinate $U=z^{\ell}$. We then assign a transformation law to $x$ and $y$ in such a way that $y^2$ and $x^3$ transform in the same way and $\Omega_2$ is invariant under the quotient. We also introduce the corresponding invariant coordinates $X$ and $Y$, which are obtained by rescaling $x$ and $y$ by suitable powers of $z$, and require that $\Omega_2$ can be written in terms of $X$, $Y$ and $U$ only. These requirements imply that the invariant coordinates are 
\be\label{coordch} X=xz^{2\ell-2};\quad Y=yz^{3\ell-3}.\ee 
Furthermore, the holomorphic two-form in the new coordinates reads 
\be\label{holonew}\Omega_2=\frac{dUdXdY}{dW(X,Y,U)},\ee 
where we have implicitly assumed that the Kodaira singularity can be rewritten in terms of the invariant coordinates only. This is possible only for solutions of (\ref{quotient}). 

The cases $\ell=2,3,4$ have been discussed in detail in \cite{Apruzzi:2020pmv}. The case $\Delta_7=1$ corresponds to $\CN=3$ $\CS$-folds and the analysis at the Weierstrass level is done starting from the \emph{trivial} Weierstrass  model
\be y^2=x^3+fx+g\ee 
with $f$ and $g$ constant. The procedure described above then leads to 
\be\label{ell6} Y^2=X^3+Xfz^{4\ell-4}+gz^{6\ell-6},\ee 
which can be entirely written in terms of $Y$, $X$ and $U$ for $\ell=2$, $\ell=4$ if $g=0$ and $\ell=3,6$ if $f=0$. This choice $\Delta_7=1$ is the only possibility for $\ell=6$.

 The case $\Delta_7=6/5$ is similar: We start from 
\be y^2=x^3+c_{4/5}x+z\ee 
and introducing a $\mathbb{Z}_{\ell}$ quotient we find 
\be\label{ell5} Y^2=X^3+c_{4/5}Xz^{4\ell-4}+z^{6\ell-5}.\ee 
The term $z^{6\ell-5}$ is a power of $U=z^{\ell}$ only for $\ell=5$, which is indeed the expected solution. It is easy to check that the $\mathbb{Z}_5$ quotient cannot be defined for any other choice of $\Delta_7$. Notice that we have to set $c_{4/5}$ to zero in (\ref{ell5}) so the singularity cannot be deformed. Below we will provide evidence for the existence of the $\mathbb{Z}_5$ $\CN=2$ $\CS$-fold we have just discussed and construct the superconformal field theory living on a stack of D3 branes probing it.

\subsection{Choice of the holonomy of $G$ at infinity}
\label{sec:4dholonomy}
Let us now note that we have a 7-brane carrying a $G$ gauge theory wrapping $\bC^2/\bZ_\ell$.
As the quotient by $\bZ_\ell$ involves a nontrivial $SL(2,\bZ)$ operation,
it is reasonable to assume that it might involve a nontrivial outer-automorphism of $G$.
Then, at spatial infinity $S^3/\bZ_\ell$, one needs to specify an order-$\ell$ automorphism of $G$ which can be  nontrivial as an outer automorphism of order $\ell'$, where $\ell'$ divides $\ell$.

The classification of such holonomies is done via Kac's theorem \cite[Theorem 8.6]{Kac} using the twisted affine Dynkin diagram of type $G^{(\ell')}$;
for a quick summary for string theorists, see \cite[Sec.~3.3]{Tachikawa:2011ch}.
It says that an order-$\ell$ automorphism of $G$
which is an order-$\ell'$ outer automorphism corresponds, up to the diagram automorphism of the twisted Dynkin diagram, to a collection of nodes (where one can choose the same node multiple times) whose Dynkin labels sum to $\ell/\ell'$.
Furthermore, the subgroup of $G$ invariant under the chosen automorphism
has the Dynkin diagram obtained by removing the chosen nodes from the twisted Dynkin diagram.

\begin{table}[h]
\[
\begin{array}{|cc|cc|c|cc|cc|}
\hline
\ell &  G& \ell' & G^{(\ell')} &\text{Dynkin diagram}& \mathcal{T} & H_\cT & \mathcal{S} & H_\cS \\
\hline
\hline
2&E_6&2 &E_6^{(2)}  & \node{1}{\alpha_0}-\node{2}{\alpha_1}-\node{3}{\alpha_2}\Leftarrow\node{2}{\alpha_3}-\node{1}{\alpha_4} & \alpha_0 & (F_4)_1
& \alpha_4 & Sp(4)_1\\
2 & D_4 & 2& D_4^{(2)} & \node{1}{\alpha_0}\Leftarrow \node{1}{\alpha_1}-\node{1}{\alpha_2}\Rightarrow\node{1}{\alpha_3}   & \alpha_0 &  SO(7)_1 & \alpha_2 & Sp(2)^{\alpha_0\alpha_1}_1 SU(2)^{\alpha_3}_2 \\
2 & A_2 & 1 &A_2^{(1)} & \node{1}{\alpha_0} - \node{1}{\alpha_1} -\node{1}{\alpha_2}- & \alpha_0\alpha_0 & SU(3)_1 & \alpha_0\alpha_1  & Sp(1)_1 U(1) \\
\hline
3 & D_4 & 3 & D_4^{(3)} & \node{1}{\alpha_0}-\node{2}{\alpha_1}\Lleftarrow\node{1}{\alpha_2}  & \alpha_0 & (G_2)_1 & \alpha_2 & SU(3)^{\alpha_0\alpha_1}_3 \\
3 & A_1 & 1 & A_1^{(1)} & \node{1}{\alpha_0} - \node{1}{\alpha_1} -  &  \alpha_0\alpha_0\alpha_0 & SU(2)_1 &  \alpha_0\alpha_0\alpha_1 & U(1) \\
\hline
4 & A_2 & 2 & A_2^{(2)} & \node{2}{\alpha_0}\Llleftarrow\node{1}{\alpha_1} & \alpha_0   &SU(2)^{\alpha_1}_1& \alpha_1\alpha_1 & SU(2)^{\alpha_0}_4\\
\hline
\end{array}
\]
\caption{The possible choices of asymptotic holonomies on the S-folds.
Here, the extended Dynkin diagrams of type $A_r^{(1)}$ form a loop.
In the columns $\mathcal{T}$ and $\mathcal{S}$,
we displayed the chosen nodes which specify the asymptotic holonomy
which reproduce the flavor symmetry of these theories.
In the columns $H_\cT$ and $H_\cS$ the subgroup of $G$ commuting with the asymptotic holonomy is given.
The superscripts specify the nodes forming the particular groups,
and the subscripts are embedding indices.
\label{table:holonomies}}
\end{table}

We find that the choices given in Table~\ref{table:holonomies} correctly reproduce the non-isometry part of the generic flavor symmetry $H_\cS$ and $H_\cT$ of the $\mathcal{S}$ and $\mathcal{T}$ theories given in the Introduction.
We note that in each case, there are precisely two choices (up to diagram automorphism) of the collection of nodes such that the sum of Dynkin labels equals $\ell/\ell'$.
We uniformly assign those solely given by $\alpha_0$ to $\cT$
and the other cases to $\cS$.
We also tabulated the embedding indices of $H_{\cS,\cT}\subset G$ in the Table as subscripts.
They can be easily determined from the twisted Dynkin diagram.
Namely, the embedding index is $1$ except when the subdiagram for a particular simple component of $H_{\cS,\cT}$ is contained within the part of the diagram \emph{pointed to by a directed arrow} in the diagram,
in which case the embedding index equals the number of edges in the arrow.
For example,  $SU(3)^{\alpha_0\alpha_1}$ of $(\ell,G)=(3,D_4)$ is pointed to by a triple-edged arrow, and therefore has the embedding index 3,
whereas $SU(2)^{\alpha_0}$ of $(\ell,G)=(4,A_2)$ is pointed to by a four-edged arrow, and has the embedding index 4.

The geometric description implies that the global symmetry of the probe theory should also include the isometry of the $\bC^2/\bZ_\ell$ background, namely $SU(2)$ for $\ell=2$ and $U(1)$ for $\ell>2$.
This is precisely compatible with the flavor symmetry appearing in Table~\ref{table:table}.

In \cite{Aharony:2016kai}, the variants of \Nequals3 S-folds were studied, whose properties are already summarized in Table~\ref{table:N3charges} in the Introduction. 
In particular, for $\ell=2,3,4$, there are two variants, fluxless and fluxful, whose D3-charges are given by \begin{equation}
\epsilon_\text{fluxless}=-\frac{\ell-1}{2\ell},\qquad
\epsilon_\text{fluxful}=+\frac{\ell-1}{2\ell}.
\label{charges}
\end{equation}
Here, in the presence of 7-branes, we also see two choices, albeit from different reasons of having two non-conjugate holonomies of the 7-brane gauge fields at infinity.
We will identify the holonomy giving the $\mathcal{S}$ theories and the $\mathcal{T}$ theories with the fluxful cases and the fluxless cases, respectively.
Various justifications will be provided below.

\subsection{$\CS^{(r)}_{G,\ell}$ theories and $\CT^{(r)}_{G,\ell}$ theories}
\label{sec:ST}

Let us now probe these \Nequals2 S-folds with $r$ D3-branes.
From the geometric setup one can compute holographically the  conformal central charges $a$ and $c$ and the flavor central charge $k$
as was done in \cite{Aharony:2016kai,Apruzzi:2020pmv,Giacomelli:2020jel},
finding the result  
\begin{align}
\label{aformula} a&=\frac{\ell\Delta_7}{4}r^2+\left(\frac{\ell\Delta_7}{2}\epsilon+\frac{2\Delta_7-2}{4}\right)r+O(r^0),\\
\label{cformula} c&=\frac{\ell\Delta_7}{4}r^2+\left(\frac{\ell\Delta_7}{2}\epsilon+\frac{3\Delta_7-3}{4}\right)r+O(r^0),\\
\label{kformula} k_H&= 2 I_{H\hookrightarrow G} \Delta_7 r +O(r^0),
\end{align}
where we only considered the part of the flavor symmetry coming from the subgroup $H$ of $G$.
Here, the embedding index $I_{H\hookrightarrow G}$ was listed in Table~\ref{table:holonomies} as subscripts.

We find the central charges of the $\cS^{(r)}_{G,\ell}$ theories
first computed in \cite{Apruzzi:2020pmv} and also summarized in Table~\ref{table:table}
can be reproduced if we plug $\epsilon_\text{fluxful}$ given in \eqref{charges} in to the formulas above
and set $O(r^0)$ to be the contribution of $\ell(\Delta_7-1)$ hypermultiplets transforming appropriately under $H$.
Note that the $O(r^0)$ terms were fixed in \cite{Apruzzi:2020pmv} 
by demanding that the central charges for the rank-1 cases should be equal to 
$a$ and $c$ of the new rank-1 theories determined field theoretically in \cite{Argyres:2015ffa,Argyres:2015gha,Argyres:2016xmc,Argyres:2016xua}.
We will give a consistency check of this $O(r^0)$ term in Sec.~\ref{sec:higgs}
from the point of view of the instanton moduli spaces on $\bC^2/\bZ_\ell$.

Similarly, by plugging $\epsilon_\text{fluxless}$ given in \eqref{charges} into the equations above and setting $O(r^0)=0$ for both $a$, $c$ and $k$,
we reproduce the central charges of the $\cT^{(r)}_{G,\ell}$ theories
first computed in \cite{Giacomelli:2020jel} and summarized in Table~\ref{table:table}.

Let us now discuss the Coulomb branch spectrum.
For this purpose it is useful to first recall the Coulomb branch spectrum of \Nequals3 theories, determined in \cite{Aharony:2016kai}.
The result is that the Coulomb branch is of the form $\mathbb{C}^{r}/G(\ell,p,r)$,
where  $G(\ell,p,r)$ is a type of complex reflection groups.
It is defined by its action on $\mathbb{C}^r$, spanned by the coordinates $z_i$, and 
is generated by the permutations of the $z_i$ coordinates, together with the transformations:
\be\label{CRG}
(z_1 , z_2 , ... , z_r) \rightarrow (e^{\frac{2\pi a_1 i}{\ell}} z_1 , e^{\frac{2\pi a_2 i}{\ell}} z_2 , ... , e^{\frac{2\pi a_n i}{\ell}} z_r) ,
\ee  
for all $a_i$'s obeying $a_1 + a_2 + ... + a_n = m p$, for some integer $m$. 
It was found in \cite{Aharony:2016kai} that only a subset of the possible groups $G(\ell,p,r)$ actually arises in this construction, specifically, the cases of $\ell=2,3$ and $4$ for $p=1$, and $\ell=2,3,4$ and $6$ for $p=\ell$. The cases of $\ell=2$ correspond to $\CN=4$ SYM theories, as well as the cases of $r=2$, $p=\ell$.  
The Coulomb branch dimensions are then \begin{equation}
\ell, 2\ell, \cdots, (r-1)\ell, r
\end{equation} when $p=\ell$ and \begin{equation}
\ell, 2\ell, \cdots, (r-1)\ell, r\ell
\end{equation} when $p=1$. 
This Coulomb branch spectrum reproduces the central charge combination $2a-c$, using the standard formula.

This allows us to guess the Coulomb branch spectrum of our $\cS^{(r)}_{G,\ell}$ and $\cT^{(r)}_{G,\ell}$ theories easily.
We simply take the coordinates $z_i$ above to have scaling dimension $\Delta_7$.
We then have the Coulomb branch dimensions \begin{equation}
\ell\Delta_7, 2\ell\Delta_7, \cdots, (r-1)\ell\Delta_7, r\Delta_7
\label{TCB}
\end{equation} for the $\cT^{(r)}_{G,\ell}$ theories and \begin{equation}
\ell\Delta_7, 2\ell\Delta_7, \cdots, (r-1)\ell\Delta_7, r\ell\Delta_7
\label{SCB}
\end{equation} for the $\cS^{(r)}_{G,\ell}$ theories.
Again, this Coulomb branch spectrum reproduces $2a-c$.

We note that this analysis applies even to the case $(G,\ell)=(\varnothing,5)$ which was not considered before.

Let us now discuss the special features when the rank is low enough.
For $r=1$, $\CS^{(1)}_{G,\ell}$ exhaust the entire list of rank-1 theories in table 1 of \cite{Argyres:2016xmc}. 
Similarly, for $r=1$, the $\cT^{(1)}_{G,\ell}$ theories simply become equivalent to the old rank-1 theory with $G$ symmetry together with a free hypermultiplet.
This follows once we accept the rank-1 classifications in \cite{Argyres:2015ffa,Argyres:2015gha,Argyres:2016xmc,Argyres:2016xua}
and compare the central charges $a$ and $c$.
We can also consider $\CS^{(0)}_{G,\ell}$ theories, which are simply $\ell(\Delta_7-1)$ free hypermultiplets.

\subsection{6d constructions for $\ell\Delta_7=6$ }
\label{sec:6d} 

\paragraph{The $\CT^{(r)}_{G,\ell}$ theories for $\ell=2,3,4,5,6$:}
As was pointed out in \cite{Giacomelli:2020jel}, the $\CT^{(r)}_{G,\ell}$ theories for $\ell\Delta_7=6$ also have an alternative definition given by the compactification of certain six-dimensional $\CN=(1,0)$ theories on $T^2$ with almost commuting holonomies,
originally considered in \cite{Ohmori:2018ona}.
 The relevant six-dimensional theories can be characterized in terms of the low-energy effective action on their tensor branch (see \cite{Dierigl:2020myk} for a recent discussion about these theories from the F-theory standpoint): 
\be\label{sulquiver}
\begin{tikzpicture}[thick, scale=0.4,baseline=(L2.base)]
\node[rectangle, draw, minimum width=.6cm,minimum height=.6cm](L1) at (-1.5,0){E-string};
\node[](L2) at (3,0){$SU(\ell)$};
\node[](L3) at (7,0){$SU(\ell)$};
\node[](L4) at (10.5,0){$\dots$};
\node[](L5) at (14,0){$SU(\ell)$};
\node[rectangle, draw, minimum width=.6cm,minimum height=.6cm](L8) at (3,-2.5){$\ell$};
\node[rectangle, draw, minimum width=.6cm,minimum height=.6cm](L6) at (17,0){$\ell$};
\node[](L7) at (8.5,2){$r-1$};

\draw[-] (L1) -- (L2);
\draw[-] (L8) -- (L2);
\draw[-] (L2) -- (L3);
\draw[-] (L3) -- (L4);
\draw[-] (L4) -- (L5);
\draw[-] (L6) -- (L5);
\draw[snake=brace]  (2,1) -- (15,1);
\end{tikzpicture}
\ee
Notice that here we are considering a codimension-1 locus of the tensor branch obtained by shrinking (in the F-theory description of the 6d theory) the $-1$ curve which does not support any gauge algebra. 
Then the 4d $\CT^{(r)}_{G,\ell}$ theories are obtained by compactifying these theories on $T^2$, with an almost commuting holonomies for the $SU(\ell)$ flavor symmetry.

In \cite{Giacomelli:2020jel} only the cases $\ell=2,3,4$ were considered, but there is no obstruction in considering the cases $\ell=5,6$ as well, as we will now see. We claim the resulting 4d theories represent the $\CT^{(r)}_{G,\ell}$ models for $\ell=5,6$ and can be realized by probing with $r$ D3 branes the corresponding $\CN=2$ $\CS$-folds in F-theory. 

The analysis of the resulting 4d theories can be carried out uniformly for all cases, and was in fact already provided in detail in \cite{Ohmori:2018ona}.
We will present some of it, emphasizing the two special cases $\ell=5,6$, as they have not received the attention they deserve.
The embedding of the holonomies inside the $E_8$ symmetry of the E-string was already discussed in \cite{Ohmori:2018ona}.
The holonomy has to be chosen in such a way that all fields should be invariant under it, therefore the presence of the bifundamental fields forces us to embed the holonomy in all the $SU(\ell)$ gauge groups and also in the $SU(\ell)$ global symmetries at the two ends of the quiver. 
 As a result at a generic point on the CB of the 4d theory the $SU(\ell)^{r-1}$ gauge group is broken completely and the low-energy degrees of freedom include $r$ vector multiplets and $r$ massless hypermultiplets, as expected for $\CT^{(r)}_{G,\ell}$ theories. 
We find that the CB operators then have dimension $6,12,\dots,6r-6,6r/\ell$,
 using the methods given in \cite[Appendix B]{Ohmori:2018ona}.
This reproduces the spectrum \eqref{TCB} we already saw above.

The rank-1 case is a bit special since there is no gauge group in (\ref{sulquiver}). For $\ell=5$ the resulting theory is the $A_1$ Argyres-Douglas model (CB operator of dimension $6/5$) and for $\ell=6$ it is a free vector multiplet. Notice that by adding a free hypermultiplet we find the worldvolume theory of a single D3 brane probing a flat 7-brane of type $H_0$ for $\ell=5$ and the worldvolume theory of a D3 brane in flat space for $\ell=6$. The rank-2 case is also special because the global symmetry of the 6d theory includes an $SU(2\ell)$ factor instead of the $SU(\ell)^2\times U(1)$ symmetry we see for generic rank. As a result, after the compactification a $SU(2)$ subgroup survives. We instead expect just a $U(1)$ global symmetry for rank $r\geq3$, which fits with the isometry of the $\CS$-fold background for $\ell=5,6$. The global symmetry therefore looks consistent with our claim. 

From the 6d setup we can also compute the central charges $a$ and $c$. Using the formulas derived in \cite{Ohmori:2018ona} we find the recursion relation (valid for $r>2$):
\be\label{acc} (2a-c)_r=(2a-c)_{r-1}+\frac{3d}{\ell}-\frac{1}{4},\ee 
\be\label{ccc} c_r=c_{r-1}-\frac{3}{4}+\frac{3d+3}{\ell},\ee 
where the parameter $d$ is related to a coefficient of the anomaly polynomial of the 6d theory. In the case at hand (\ref{sulquiver}) we have $d=\ell(r-1)+1$. For $r=2$ we can still use (\ref{acc}) and (\ref{ccc}), but we should add to (\ref{ccc}) the contribution of a free hypermultiplet and $(2a-c)_1$, $c_1$ are the central charges of the rank-1 theories we have already discussed. We can easily solve the recursion, finding the result
\be\label{l5}a_r=\frac{30r^2-22r}{20}; \quad c_r=\frac{30r^2-21r}{20}\quad (\text{for}\;\;\ell=5),\ee
\be\label{l6}a_r=c_r=\frac{30r^2-25r}{20}\quad (\text{for}\;\;\ell=6).\ee 
These equations hold for $r>1$ and can be applied to rank-1 theories as well, if we include a free hypermultiplet. Notice that this result is consistent with (\ref{aformula}) and (\ref{cformula}) if we set $\epsilon=-\frac{\ell-1}{2\ell}$ and $O(r^0)=0$ for $\ell=5,6$ as well. For $\ell=5$ we therefore predict that the D3 charge of the corresponding $\CN=2$ $\CS$-fold is $\epsilon=-2/5$ and for $\ell=6$ we find that all the quantities we can compute from the six-dimensional setup are compatible with the $\CN=3$ theories associated with the $\ell=6$ $\CS$-fold. Furthermore, as we mentioned for $r=2$ the generally present $U(1)$ global symmetry enhances to $SU(2)$. This has a straightforward interpretation as for $r=2$ the corresponding $\CN=3$ theory is in fact $\CN=4$ $G_2$ SYM \cite{Aharony:2016kai}. Indeed,  the central charges of the rank-2 theory are those of $\CN=4$ $G_2$ SYM and the CB operators have dimension 2 and 6. To further support this we can also use the formulas derived in \cite{Ohmori:2018ona} to compute the central charge of this $SU(2)$, $k_{SU(2)}=14$, which is indeed equal to the dimension of $G_2$. This fact is also supported by the analysis of the CB stratification performed below. We therefore conclude, somewhat surprisingly, that although the 6d parent theories only have 8 supercharges the resulting 4d models have 12 supercharges for $r>2$ and 16 for $r=2$.

\paragraph{The $\CS^{(r)}_{G,\ell}$ theories for $\ell=2,3,4$:}
We also note that in \cite{Giacomelli:2020jel} the 6d realizations of the $\cS^{(r)}_{G,\ell}$ theories were also found for the cases $\ell\Delta_7=6$.
These are given by taking the 6d theories with the structure \begin{equation}
\begin{aligned}
\ell=2: \quad& 
\begin{tikzpicture}[thick, scale=0.4,baseline=(L2.base)]
\node[rectangle, draw, minimum width=.6cm,minimum height=.6cm](L1) at (0,0){8};
\node[](L2) at (3,0){${SU(2)}$};
\node[](L3) at (7,0){$SU(2)$};
\node[](L4) at (10.5,0){$\dots$};
\node[](L5) at (14,0){$SU(2)$};
\node[rectangle, draw, minimum width=.6cm,minimum height=.6cm](L6) at (17,0){2};
\node[](L7) at (8.5,2){$r$};

\draw[-] (L1) -- (L2);
\draw[-] (L2) -- (L3);
\draw[-] (L3) -- (L4);
\draw[-] (L4) -- (L5);
\draw[-] (L6) -- (L5);
 \draw[snake=brace]  (2,1) -- (15,1);
\end{tikzpicture} \\
\ell=3: \quad &
\begin{tikzpicture}[thick, scale=0.4,baseline=(L2.base)]
\node[rectangle, draw, minimum width=.6cm,minimum height=.6cm](L1) at (0,0){9};
\node[](L2) at (3,0){${ SU(3)}$};
\node[](L3) at (7,0){$SU(3)$};
\node[](L4) at (10.5,0){$\dots$};
\node[](L5) at (14,0){$SU(3)$};
\node[rectangle, draw, minimum width=.6cm,minimum height=.6cm](L6) at (17,0){3};
\node[](L7) at (8.5,2){$r$};

\draw[-] (L1) -- (L2);
\draw[-] (L2) -- (L3);
\draw[-] (L3) -- (L4);
\draw[-] (L4) -- (L5);
\draw[-] (L6) -- (L5);
 \draw[snake=brace]  (2,1) -- (15,1);
\end{tikzpicture} \\
\ell=4: \quad &
\begin{tikzpicture}[thick, scale=0.4,baseline=(L2.base)]
\node[rectangle, draw, minimum width=.6cm,minimum height=.6cm](L1) at (0,0){8};
\node[](L2) at (3,0){${SU(4)}$};
\node[](L3) at (7,0){$SU(4)$};
\node[](L4) at (10.5,0){$\dots$};
\node[](L5) at (14,0){$SU(4)$};
\node[rectangle, draw, minimum width=.6cm,minimum height=.6cm](L6) at (17,0){4};
\node[](L7) at (8.5,2){$r$};
\node[rectangle, draw, minimum width=.6cm,minimum height=.6cm](L8) at (3,-3){1};

\draw[-] (L1) -- (L2);
\draw[-] (L2) -- (L3);
\draw[-] (L3) -- (L4);
\draw[-] (L4) -- (L5);
\draw[-] (L6) -- (L5);
\draw[snake=brace]  (2,1) -- (15,1);
\draw[snake=zigzag,segment aspect=0]  (3,-.5)  -- (3,-2.3);
\end{tikzpicture}
\end{aligned}
\label{Squivers}
\end{equation}
where the zigzag line connecting $SU(4)$ and a square box with 1 stands for a hyper in $\mathbf{6}$.
Then the 4d $\CS^{(r)}_{G,\ell}$ theories are obtained by compactifying these theories on $T^2$, with an almost commuting holonomies for the $SU(\ell)$ flavor symmetry.
Again, we can also determine the Coulomb branch spectrum, using the results in \cite[Appendix B]{Ohmori:2018ona},
which  reproduces the spectrum \eqref{SCB} we just saw.

\paragraph{The $\cS^{(r)}_{G,\ell}$ and $\cT^{(r)}_{G,\ell}$ theories for $\ell\Delta_7=6$ and the nodes of $E_8$ Dynkin diagram:}
We note that the 6d theories given in \eqref{sulquiver} and \eqref{Squivers}
correspond to $r$ M5-branes probing a $\bC^2/\bZ_\ell$ singularity on the $E_8$ wall.
As in the 4d F-theory situation discussed in Sec.~\ref{sec:4dholonomy},
we need to specify the asymptotic $E_8$ holonomy at $S^3/\bZ_\ell$.
Such 6d systems were studied in \cite{Mekareeya:2017jgc}.
The holonomy is again specified using Kac's theorem:
\begin{equation}
E_8^{(1)} : \qquad \node{}{1}-\node{}{2}-\node{}{3}-\node{}{4}-\node{}{5}-\node{\ver{}{3'}}{6}-\node{}{4'}-\node{}{2'} .
\label{e8dynkin}
\end{equation}
The nodes $1, 2, 3, 4, 5, 6$  on the long leg correspond to the $\CT$ theories, 
while the nodes $2', 3', 4'$ on the shorter  legs give the $\CS$ theories. 
Incidentally, this gives another explanation why $\CS$ theories exist only for $\ell=2,3,4$.

The 4d theories are obtained by compactification on $T^2$ with a nontrivial Stiefel-Whitney class in $SU(\ell)/\bZ_\ell$.
These holonomies are embedded into $E_8$ as two commuting order-$\ell$ elements.
For our considerations to be consistent, the holonomies tabulated in Table~\ref{table:holonomies}
when $\ell\Delta_7=6$ should be the $E_8$ holonomies given in \eqref{e8dynkin}
in disguise.

This can be checked using the results in \cite{Borel:1999bx}.
The procedure is as follows.
We  select the nodes in the extended $E_8$ Dynkin diagram above, whose labels are divisible by $\ell$.
We arrange those selected nodes, and label them by the original Dynkin labels divided by $\ell$.
We then place arrows between nodes, realizing a (twisted) affine Dynkin diagram $G^{(\ell')}$.
A general method to determine the placement of arrows is explained in \cite{Borel:1999bx},
but it usually follows just by requiring that the nodes form a (twisted) affine Dynkin diagram.
This then means that the commutant of the two commuting holonomies is $\bZ_{\ell'} \ltimes G$.

Take the case of $\ell=2$ for example.
This gives the diagram \begin{equation}
E_6^{(2)} :\qquad \node{1}{\alpha_0}-\node{2}{\alpha_1}-\node{3}{\alpha_2}\Leftarrow\node{2}{\alpha_3}-\node{1}{\alpha_4} ,
\end{equation}
This means that the subgroup of $E_8$ which commute with two commuting order-$2$ holonomies is $\bZ_2 \ltimes E_6$.
We now pick the third holonomy from this twisted $E_6^{(2)}$ Dynkin diagram to be used as the holonomy on $S^3/\bZ_\ell$ at the asymptotic infinity.
This means that the outer-automorphism holonomies $\rho_\cT$, $\rho_\cS$ of the $(G,\ell)=(E_6,\bZ_2)$ case are in fact two order-2 holonomies of $E_8$ specified by nodes $2$ and $2'$ in \eqref{e8dynkin}.
We can repeat this analysis for $(G,\ell)=(D_4,3)$ and $=(A_2,4)$.

\paragraph{Duality between the F-theoretic and M-theoretic constructions:}

We would now like to very briefly explain why the worldvolume theory of $D3$ branes probing $\mathcal{N}=2$ $\mathcal{S}$-folds is equivalent to the 4d model obtained via twisted compactification of the six-dimensional theories we are discussing here. This connection between $\mathcal{S}$-folds and the torus compactification of the six-dimensional theories can be thought of as a generalization of the basic duality between $r$ $D3$ branes probing a 7-brane of type $E_8$ in  Type IIB string theory 
and 
$r$ M5 branes probing the $E_8$ wall and also wrapping a trivially-fibered $T^2$. 
Which is also equivalent to the well-known fact that the double dimensional reduction of the six-dimensional rank-$r$ E-string theory gives the  rank-$r$ $E_8$ Minahan-Nemeschansky theory in 4d, as originally found in \cite{Ganor:1996pc}

In order to see that the standard F-theory/M-theory duality generalizes to the current set-up, we notice that in both duality frames there is a $\mathbb{C}^2$ transverse to the worldvolume of the $r$ probe branes (along the 7-brane in Type IIB and along the $E_8$ wall in M-theory). The $SO(4)$ acting on this space is identified in both descriptions with the $SU(2)$ R-symmetry times the $SU(2)$ global symmetry of the theory. Since in the $\mathcal{N}=2$ $\mathcal{S}$-fold construction the 7-brane wraps a $\mathbb{C}^2/\mathbb{Z}_{\ell}$, in order to get a dual description it is natural to orbifold the transverse $\mathbb{C}^2$ in M-theory as well. In this way we make contact with orbi-instanton theories. 
Furthermore, as studied in \cite{Ohmori:2018ona} and in particular in its Appendix A, the almost commuting holonomies in $SU(\ell)$ associated to the $\bZ_\ell$ orbifold singularity of M-theory generate the $\bZ_\ell$ orbifolding of the Coulomb branch direction $u$.
Therefore, we naturally have the $\bZ_\ell$ action on the $\bC^2$ along the 7-brane and $\bC$ transverse to the 7-brane.
This is exactly what we have in \Nequals2 S-folds we have been discussing.

In order to make the connection more precise, we should incorporate in the M-theory description the information specifying the $\mathcal{N}=2$ $\mathcal{S}$-fold, specifically the type of 7-brane and the chosen holonomy on $S^3/\mathbb{Z}_{\ell}$. Notice that the duality we are discussing holds only for $\mathcal{S}$-folds satisfying the constraint $\ell\Delta_7=6$, and therefore the choice of 7-brane is equivalent to specifying the value of $\ell$.  These data are mapped in the M-theory description to the $E_8$ holonomy which specifies the 6d theory and the almost commuting holonomies on the torus as described in Section \ref{sec:6d}.

\subsection{Higgs branch as the instanton moduli}
\label{sec:higgs}

Our proposal is that the theories $\cS^{(r)}_{G,\ell}$ and $\cT^{(r)}_{G,\ell}$ are the worldvolume theories on  
$r$ D3-branes probing the \Nequals2 S-fold obtained by the 7-brane of type $G$
superimposed on top of the \Nequals3 S-fold of order $\ell$.

Here we would like to make a further check of this identification by studying their Higgs branches and identifying them with the moduli space of $G$ instantons on $\bH/\bZ_\ell$.
Before proceeding we summarize the results from the field theoretical analysis, namely:
\begin{itemize}
\item 
The Higgs branch of $\cS^{(r)}_{G,\ell}$ has dimension $h^\vee(G)r+\ell(\Delta_7-1)$.
\item
The Higgs branch of $\cT^{(r)}_{G,\ell}$ has dimension $h^\vee(G)r$.
\end{itemize}

As D3-branes can be absorbed into the 7-branes as instantons, 
it is natural to identify these Higgs branches with the moduli spaces of $G$ instantons on $\bR^4/\bZ_\ell$.
Furthermore, at its asymptotic infinity $S^3/\bZ_\ell$, 
we have an order-$\ell$ holonomy around the $\bZ_\ell$ 1-cycle.
We already proposed the choices of this holonomy 
for the theories $\cS^{(r)}_{G,\ell}$ and $\cT^{(r)}_{G,\ell}$ above.
We denote these holonomies by $\rho_\cS$ and $\rho_\cT$ below.

Let us now recall the formula of the dimension of the moduli space of $G$ instantons on $\bC^2/\bZ_\ell$, with two holonomies $\rho_0$ and $\rho_\infty$ on  $S^3/\bZ_\ell$ at the origin and at the asymptotic infinity.
This is given by \cite{KronheimerNakajima} for general $\Gamma\subset SU(2)$: \begin{equation}
\begin{aligned}
\dim_\bH \mathcal{M}_{\rho_\infty,\rho_0}& = 
h^\vee(G) (\int\Tr FF )+  [\eta (\rho_\infty) -\eta(\rho_0)],\\
\int\Tr FF &=  n + CS(\rho_\infty) - CS(\rho_0) 
\end{aligned}
\label{moduli-dim}
\end{equation}
where
$h^\vee(G)$ is the dual Coxeter number of $G$,
$\Tr FF$ is the instanton density normalized to integrate to one on the standard one-instanton configuration,
$n$ is an integer,
and 
$CS(\rho)$ and $\eta(\rho)$ is the classical Chern-Simons invariant and the eta invariant 
of the $G$ bundle on $S^3/\Gamma$ specified by the holonomy $\rho$.
Luckily, there is an explicit formula for the eta invariant:
\begin{equation}
\eta(\rho):=\frac{1}{2|\Gamma|}\sum_{\gamma \neq e} \frac{\chi_{\rho}(\gamma) }{2-\chi_Q(\gamma)}
\end{equation}
where the holonomy is regarded as a homomorphism $\rho:\Gamma\to \mathfrak{g}$,
 $Q$ is the standard two-dimensional representation of $\Gamma$ from the defining embedding $\Gamma\subset \SU(2)$,
and $\chi_V$ is  the character in the representation $V$.
The value $CS(\rho)$ should in principle be computable directly
from the Kac label when $\Gamma=\bZ_\ell$, but we use tricks instead.

We immediately notice that the formula \eqref{moduli-dim} reduces to \begin{equation}
\dim_\bH \mathcal{M}= h^\vee(G) r
\end{equation} when $\rho_\infty=\rho_0$.
This matches with the dimension of the Higgs branch of the $\cT$ theories.
We already determined that $\rho_\infty=\rho_\cT$ in this case.
We are then led to identify $\rho_0=\rho_\cT$ too.
Therefore we identify the Higgs branch of (a discretely gauged version of) the $\cT^{(r)}_{G,\ell}$ theory with $\mathcal{M}_{(\int\Tr FF,\rho_\infty,\rho_0)=(r, \rho_\cT, \rho_\cT)}$.

We now consider the Higgsing from $\cT^{(r+1)}$ to $\cS^{(r)}$.
This should correspond to activating the gauge field on the 7-brane so that 
$\rho_\infty=\rho_\cT$ and $\rho_0=\rho_\cS$.
We would like to determine $CS(\rho_\cT)-CS(\rho_\cS)$.
One trick is the following.
Let $H$ be the subgroup invariant under $\rho_\cT$, which is $F_4$ for $E_6$, for example.
One gauge configuration we can activate is the one-instanton configuration of $H$ on $\bR^4$ centered at the origin, identified by $\bZ_\ell$.
It has instanton number $1/\ell$ on $\bR^4$.
Since it is not an integer, $\rho_0$ should be different from $\rho_\infty$.
Since we only have two choices of holonomies, this fixes the holonomy at the origin to be  $\rho_\cS$.
Therefore we conclude\footnote{%
When $\ell\Delta_7=6$, we can also use the discussion in Sec.~\ref{sec:6d} to determine the  Chern-Simons invariant $CS(\rho_{\cS,\cT})$ by embedding $\rho_{\cS,\cT}$ into $E_8$. 
A general holonomy with the Kac label $\mathbf{w}$ of order $\ell$ in $E_8$ has the Chern-Simons invariant $
CS(\mathbf{w})=-{\langle\mathbf{w},\mathbf{w}\rangle}/({2\ell}),
$
see \cite[Sec.~2]{Mekareeya:2017jgc}.
Now, $\langle\mathbf{w},\mathbf{w}\rangle$ for fundamental weights are given by \[
\node{}{2}-\node{}{6}-\node{}{12}-\node{}{20}-\node{\ver{}{8}}{30}-\node{}{14}-\node{}{4} .
\]
From this, we can easily compute, say 
$CS(\rho_\cT)=-{12}/{8}$,
$CS(\rho_\cS)=-{14}/{8}$ 
modulo 1 for $\ell=4$.
In a similar manner, we can check that  $
CS(\rho_\cT^{(\ell)})-CS(\rho_\cS^{(\ell)})={1}/{\ell}
$ modulo 1 for all cases.
}
\begin{equation}
\int \Tr FF = n+CS(\rho_\cT) - CS(\rho_\cS) = \frac 1{\ell}.
\end{equation}

We note that this instanton number equals $\epsilon_\text{fluxless}-\epsilon_\text{fluxful}$ modulo 1 given in  \eqref{charges},
i.e.~the difference of the D3-brane charges of the flux-less and the fluxfull \Nequals3 S-folds.
This is as it should be, since the instanton configuration carries the D3-brane charge.

The dimension of the instanton moduli space with this instanton number  and $\rho_\infty=\rho_\cT$ and $\rho_0=\rho_\cS$ is then \begin{equation}
\dim_\bH \cM_{(\int\Tr FF,\rho_\infty,\rho_0)=(\frac1\ell, \rho_\cT, \rho_\cS)}= \frac{h^\vee(G)}{\ell} + \eta(\rho_\cT)-\eta(\rho_\cS)
\end{equation} 
which happens to be \begin{equation}
=h^\vee(H)-1,
\end{equation} which can be checked case by case.\footnote{%
For example, $\eta(\rho_\cT)=13/8$  and $\eta(\rho_\cS)=-3/8$ for $(G,\ell)=(E_6,2)$,
leading to $\dim_\bH\cM= 12/2 + 13/8+3/8 = 8 = h^\vee(F_4)-1$.
}
We now note that $h^\vee(H)-1$ is exactly the dimension of the centered 1-instanton moduli of $H$ on $\bR^4$ we started with.
This means that the known configurations saturate the dimension calculated from the index theorem,
allowing us to identify \begin{equation}
\cM_{(\int\Tr FF,\rho_\infty,\rho_0)=(\frac1\ell, \rho_\cT, \rho_\cS)}
= \cM_\text{centered 1-inst}(H).
\label{H}
\end{equation}
This means that the Higgs branch of $\cT^{(r+1)}_{G,\ell}$ theory
contains a stratum given by $\cM_\text{centered 1-inst}(H)$ 
on which the low-energy theory is given by $\cS^{(r)}_{G,\ell}$.
We then identify the Higgs branch of (the discrete gauged version of) $\cS^{(r)}_{G,\ell}$
with $\cM_{(\int\Tr FF,\rho_\infty,\rho_0)=(r+1-\frac1\ell, \rho_\cS, \rho_\cT)}$,
which has the dimension \begin{align}
\dim_\bH \cM_{(\int\Tr FF,\rho_\infty,\rho_0)=(r+1-\frac1\ell, \rho_\cS, \rho_\cT)}
&=h^\vee(G) r + h^\vee(G)-h^\vee(H)+1\\
&=h^\vee(G) r+\ell(\Delta_7-1),
\end{align}
where the equality \begin{equation}
\ell(\Delta_7-1)=h^\vee(G)-h^\vee(H)+1.
\end{equation}
can again be checked by a case-by-case analysis\footnote{%
For example, we have $2(3-1)=12-9+1$ for $(G,\ell)=(E_6,2)$.
}.

Let us study the extreme cases when $r=1$.
It appears to us that the instanton moduli on $\bH/\bZ_\ell$ is equal to the Higgs branch of the \emph{discretely gauged} version of the $\cS$ and $\cT$ theories.
We further assume that the $\cT^{(r=1)}_{G,\ell}$ theories are equal to the old rank-1 theory with $G$ symmetry together with a free hypermultiplet, whose
Higgs branch is given by \begin{equation}
\cM_\text{1-inst}(G)=\cM_\text{centered 1-inst}(G) \times \bH.
\end{equation}
Then it should be that \begin{equation}
\cM_{(\int\Tr FF,\rho_\infty,\rho_0)=(1,\rho_\cT,\rho_\cT)}=\cM_\text{1-inst}(G)/\bZ_{\ell}.
\label{G}
\end{equation}
Before the quotient,  the generic point of $\cM_\text{1-inst}(G)$ is smooth, 
and the stratum $\cM_\text{centered 1-inst}(H)$ is embedded within the fixed locus of $\bZ_{\ell}$.
Then the transverse slice of \eqref{H} within \eqref{G}
is simply the $\bZ_{\ell}$ quotient of a flat space, \begin{equation}
\cM_{(\int\Tr FF,\rho_\infty,\rho_0)=(1-\frac1\ell, \rho_\cS, \rho_\cT)} =
\bH^{h^\vee(G)-h^\vee(H)+1} /\bZ_{\ell}
=\bH^{\ell (\Delta_7-1)} /\bZ_{\ell}
\label{HZ}
\end{equation}
which is indeed compatible with  the identification of 
$\cS^{(0)}_{G,\ell}$ with (a discrete quotient of) $\ell(\Delta_7-1)$ free hypermultiplets.

Let us  identify this space \eqref{HZ} geometrically in a direct manner.
We distinguish two cases, namely $\ell=2$ and $\ell=3,4$.

First, we treat the case $\ell=2$.
In this case the invariant subgroup of $G$ under $\rho_\cS$ is $Sp(k)$
for some $k$.
Then, we can consider one-instanton  configurations of $Sp(k)$ on $\bR^4$ centered at the origin,
which can be considered to be on $\bR^4/\bZ_2$ of instanton number $1/2$.
Since the instanton number is fractional, so $\rho_0\neq \rho_\infty$, forcing $\rho_0=\rho_\cT$.
The formula \eqref{moduli-dim} after a short computation then tells that the dimension of the moduli space is $k$.
Therefore the dimension of the one-instanton moduli of $Sp(k)$ saturates the dimension from the index theorem, leading us to identify \begin{equation}
\cM_{(\int\Tr FF,\rho_\infty,\rho_0)=(1-\frac1\ell, \rho_\cS, \rho_\cT)}
= \cM_\text{centered 1-inst}(Sp(k))=\bH^k/\bZ_2.
\end{equation}
Again, a case-by-case analysis shows $k=\ell(\Delta_7-1)$. 

Next, we consider the case $\ell=3,4$. 
In this case $G$ is a classical group, and therefore the moduli space of instantons on $\bC^2/\bZ_\ell$ should admit an explicit description as the Higgs branch of a quiver gauge theory.
Note that the analysis of $U(N)$ instantons on $\bC^2/\Gamma$ with no nontrivial outer-automorphism is a classic result of \cite{KronheimerNakajima,Douglas:1996sw},
and that it was extended to other classical groups more recently in \cite[Appendix A.4]{Nakajima:2015txa} and \cite{Nakajima:2018bpn} (see also \cite{Dey:2013fea}).
It should not be too difficult to extend their analysis to the case with nontrivial outer-automorphisms,
which would then allow us to determine not only \eqref{HZ} but instanton moduli spaces with larger instanton number.

Here we only discuss the case $(\ell,G)=(3,A_1)$, for which no nontrivial outer-automorphism is involved, so we can simply quote a result in the existing literature.
Then the moduli space is exactly the one studied in \cite{KronheimerNakajima,Douglas:1996sw},
and is the Higgs branch of the quiver gauge theory of the form \begin{equation}
\begin{tikzpicture}[baseline=(A.base)]
\node[rectangle,draw] (A) at  (0,0) {$1$};
\node[circle,draw] (B) at  (1,0) {$1$};
\node[circle,draw] (C) at  (2,0) {$1$};
\node[rectangle,draw] (D) at  (3,0) {$1$};
\draw (A)--(B)--(C)--(D);
\end{tikzpicture}
\end{equation}
which clearly gives $\bC^2/\bZ_3$.
More generally, the $SU(2)$ instanton moduli space on $\bC^2/\Gamma$ with $\Gamma\subset SU(2)$ and $\int\Tr FF=1-1/|\Gamma|$ 
such that the holonomy at infinity is given by $\Gamma$ itself 
was known to be $\bC^2/\Gamma$ itself \cite[Theorem (0.3)]{Nakajima}, 
which was a precursor to \cite{KronheimerNakajima}.

\subsection{Rank-1 theories and discrete gaugings}
\label{sec:rank1}

We have seen that $\CT^{(1)}_{G,\ell}$ theories coincide with the 1 $G$-instanton theories together with a free hypermultiplet (i.e. the center of mass mode), hence their Higgs branch is $\mathcal{M}_\text{1-inst}(G)$. As was pointed out in \cite{Argyres:2016yzz}, the 1 $G$-instanton theories have a $\mathbb{Z}_{\ell}$ symmetry acting on the Coulomb branch which is gaugeable. The presence of such a symmetry is also supported by the six-dimensional realization of $\CT^{(1)}_{G,\ell}$ theories discussed in Section \ref{sec:6d}: This is due to the fact that for $r=1$ we are compactifying on $T^2$ the rank-1 E-string theory and the Coulomb branch of the resulting model in 4d is a $\ell$-fold cover of the Coulomb branch of the $E_8$ Minahan-Nemeschansky theory, see  \cite{Ohmori:2018ona}.

Upon gauging this $\mathbb{Z}_{\ell}$ symmetry the Higgs branch becomes $\mathcal{M}_\text{1-inst}(G)/\mathbb{Z}_{\ell}$ and we recover the 1-instanton moduli space on $\mathbb{H}/\mathbb{Z}_{\ell}$ as was discussed in Section \ref{sec:higgs}. Notice that the $\mathbb{Z}_{\ell}$ symmetry acts on the free hypermultiplet as well, and this fits perfectly with the Type IIB realization we are proposing: If we break the $G$ symmetry completely with a mass deformation, on the one hand $\CT^{(1)}_{G,\ell}$ flows to an $\mathcal{N}=4$ vector multiplet and the $\mathbb{Z}_{\ell}$ gauging we are discussing reduces to the $\mathcal{N}=3$-preserving discrete gauging discussed in \cite{Argyres:2016yzz}. On the other hand, the mass deformation in Type IIB is implemented by removing the 7-brane completely and the geometric background becomes the ordinary $\mathcal{N}=3$ $\mathcal{S}$-fold without flux. We therefore find perfect agreement between the geometric and field-theoretic analysis.

We would also like to point out that, upon turning on a mass for the hypermultiplet (or equivalently for the flavor symmetry factor associated with the isometry of $\mathbb{C}^2/\mathbb{Z}_{\ell}$), the action of $\mathbb{Z}_{\ell}$ on the resulting theory is equivalent to the $\CN=2$-preserving discrete gauging of the 1 $G$-instanton theories described in \cite{Argyres:2016yzz}. This construction therefore provides a stringy realization of discretely gauged 1 $G$-instanton theories.


\subsection{Relations with \Nequals3 SCFTs}
\label{section:N3}
We mentioned that the $\CS$ and $\CT$ theories can be mass deformed to $\CN=3$ SCFTs. Specifically, we claim that $\CS^{(r)}_{G,\ell}$ theories can be mass deformed to the $\CN=3$ SCFTs with moduli space $\mathbb{C}^{3r}/G(\ell,1,r)$, 
while the $\CT^{(r)}_{G,\ell}$ theories, for $\ell \neq 5$, can be mass deformed to the $\CN=3$ SCFTs with moduli space $\mathbb{C}^{3r}/G(\ell,\ell,r)$.

There are several indications that this is the case. First, we note that the spectrum of $\CS^{(r)}_{G,\ell}$ and $\CT^{(r)}_{G,\ell}$ theories, and the spectrum of possible $\CN=3$ SCFTs of this type precisely agree. Second, in the rank-1 case, we indeed have that the $\CS^{(r)}_{G,\ell}$ are known to have mass deformations ending with $\CN=3$ SCFTs with moduli space $\mathbb{C}^{3}/G(\ell,1,1)$.

Finally, we can show that this is true for the case of $\ell=2$. For this we use the representation of $\CS^{(r)}_{G,\ell}$ and $\CT^{(r)}_{G,\ell}$ theories, for $\ell\Delta_7=6$, as the $T^2$ compactifications with almost commuting holonomies of certain six-dimensional $\CN=(1,0)$ theories, as pointed out in \cite{Giacomelli:2020jel}. We can analyze the compactification by first reducing to $5d$ and then further reducing to $4d$, as done in \cite{Ohmori:2018ona}. Next, we shall concentrate on the $\CS^{(r)}_{E_6,2}$ and $\CT^{(r)}_{E_6,2}$ theories.

 In these cases the theories are associated with the $T^2$ compactifications with almost commuting holonomies of $6d$ $\CN=(1,0)$ SCFTs that UV complete the $5d$ gauge theories $SU(2r+1)_0+2AS+8F$ for $\CS^{(r)}_{E_6,2}$ and $SU(2r)_0+2AS+8F$ for $\CT^{(r)}_{E_6,2}$ \cite{Zafrir:2015rga} (see also \cite{Zafrir:2018hkr}). Reducing first to $5d$, we can argue that the resulting $4d$ theories are given by a twisted compactification of the $5d$ SCFTs UV completing the $5d$ gauge theories $SU(2r+1)_0+2AS+6F$ for $\CS^{(r)}_{E_6,2}$ and $SU(2r)_0+2AS+6F$ for $\CT^{(r)}_{E_6,2}$. Here the twist is done by a $\mathbb{Z}_2$ symmetry, acting on the $5d$ gauge theories through charge conjugation (see \cite{Zafrir:2016wkk} for a study of this type of twisted reductions). 

Mass deformations of these theories can then be studied by considering mass deformations of the $5d$ SCFTs. By using these we can eventually get to the $4d$ theories associated with the twisted compactification of the $5d$ gauge theories $SU(2r+1)_0+2AS$ for $\CS^{(r)}_{E_6,2}$ and $SU(2r)_0+2AS$ for $\CT^{(r)}_{E_6,2}$, where here we have used mass deformations to send the $5d$ SCFTs to these $5d$ gauge theories and integrate away the six fundamental hypers. As the $5d$ gauge theories are IR free, it is straightforward to analyze the reduction and determine that we get the $4d$ gauge theories of $SO(n)$ with an antisymmetric hyper, for $n=2r+1$ in the $\CS^{(r)}_{E_6,2}$ case and $n=2r$ in the $\CT^{(r)}_{E_6,2}$ case. These are non-other then the $\CN=4$ SYM theories associated with $G(2,1,r)$ for the $\CS^{(r)}_{E_6,2}$ case and $G(2,2,r)$ for the $\CT^{(r)}_{E_6,2}$ case\footnote{This is also consistent with the results in \cite{Beratto:2020wmn}, where it was found that the $3d$ mirror of the $3d$ reduction of the $\CT^{(2)}_{A_2,2}$ theory can be deformed via an FI deformation to the $3d$ mirror of $\CN=4$ $SO(4)$ SYM.}. This fits the structure we proposed. Further evidence for this claim will be provided in section \ref{sec:4dMass} where we will study these mass deformations directly from the $4d$ perspective.

This leaves the $\CT^{(r)}_{G,5}$ case, which are not associated with $\CN=3$ SCFTs. 
As $G(5,5,r)$ is non-crystallographic, it is in fact impossible to have an $\CN=3$ SCFT with the moduli space $\mathbb{C}^{3r}/G(5,5,r)$. However, there is no such restriction for $\CN=6$ SCFTs in $3d$, and indeed there are $\CN=6$ SCFTs with moduli space $\mathbb{C}^{4r}/G(5,5,r)$, see for instance the discussion in \cite{Tachikawa:2019dvq}. It is then possible that if we further reduce to $3d$, our statement will also hold for the the $\ell=5$ case, but now with $3d$ $\CN=6$ SCFTs. This can be motivated as follows. The S-fold construction, when reduced to $3d$, is expected to be related to the $\mathbb{R}^8/\mathbb{Z}_k$ orbifolds used to engineer the ABJM and ABJ theories\cite{Aharony:2008ug,Aharony:2008gk}. Therefore, while there is no $\mathbb{Z}_5$ S-fold, the orbifold $\mathbb{R}^8/\mathbb{Z}_5$ exists and there is a corresponding ABJM theory. Thus, it seems reasonable that the $3d$ versions of the $\CS$ and $\CT$-theories can be mass deformed to $\CN=6$ SCFTs, specifically, those of ABJM type with $k=\ell$. For all $\ell$ but 5, this is implied by our suggested relation between the $\CS$ and $\CT$-theories and $\CN=3$ SCFTs. 

It is interesting in this regard to consider the Higgs branch for the $\ell=5$ case. This is as the Higgs branch is invariant under the $3d$ reduction, and therefore, must have a specific form for this to work. We will not preform here a detailed study of the Higgs branch, but we do note that from \eqref{l5} we have that $24(c_r-a_r) = r(1+\frac{1}{5})$. This suggests that the Higgs branch has dimension $r$, and on a generic point of which the theory reduces to $r$ decoupled $H_0$ AD theories. This comes about as the number $24(c_r-a_r)$ is related to the anomalies of the $U(1)$ $\CN = 2$ superconformal R-symmetry, which is not broken on the Higgs branch. As such its anomaly must be matched by the resulting low-energy theory, where we also note that for a free hyper we have that $24(c-a) = 1$, while for $H_0$ AD theory we have that $24(c-a) = \frac{1}{5}$. As the $H_0$ AD theory reduces to a free twisted hyper\cite{Nanopoulos:2010bv}, in $3d$ we expect an $r$ dimensional Higgs branch on a generic point of which we get $r$ copies of $\mathbb{C}^4$. This is consistent with the $\CN=6$ theory.

\section{Stratification of the Coulomb Branch of rank-2 theories}
\label{sec:stratification}

In this subsection we will study in more detail the structure of the Coulomb Branch (CB) of both the $\CS$ and $\CT$ theories for the special case of rank-2. The analysis here will be then leveraged below to understand the mass deformations of these theories and, in particular, those which give rise to new SCFTs. This way we will be able to conjecture the existence of new rank-2 $\cN=2$ SCFTs in four dimensions. 

The low-energy theory on a generic point of the CB $\cC$ is almost as boring as it gets; a free $\cN=2$ supersymmetric $\U(1)^r$ gauge theory with no massless charged states. $r$ is called the \emph{rank} of the theory and coincides with the complex dimensionality of $\cC$, ${\rm dim}_\C\cC=r$; we will indicate the global collective coordinates of $\cC$ as $\bu$. $\cC$ is a singular space and its singular locus, which is a closed subset of $\cC$ and will be denoted as $\bar{\XS}$, coincides with the locus of the CB where the low-energy theory is more interesting and potentially not-free. The smooth part of the CB is $\cCrg := \cC \setminus \XSb$ and thus $\cCrg$ is an open subset of $\cC$. When the $\cN=2$ theory is superconformal the symmetry group includes an $\R^+ \times \U(1)_R$ (we are neglecting the $\SU(2)_R$ factor as it acts trivially on $\cC$) which can be spontaneously broken, and so acts non-trivially on $\cC$, and combines to give a $\C^*$ action on the CB.  The entire structure of $\cC$ has to be compatible with the $\C^*$ action and in particular $\XSb$ and $\cCrg$ have to be closed under it. In the rank-2 case we use the following convention $\bu:=(u,v)$, where $u$ has the lowest scaling dimension of the two CB coordinates.


$\XSb$ has a very rich structure which can be leveraged to great extent to learn new features about $\cN=2$ SCFTs, see for a brief summary of this philosophy \cite{Argyres:2020nrr}. Considerable progress in the understanding of the CB of theories of rank higher than 1, has been achieved in a series of recent papers \cite{Martone:2020nsy,Argyres:2020wmq}. These techniques will be applied here to the $\CT$ and $\CS$ theories but before we start with this analysis, let's remind the reader about the key ideas:

\begin{itemize}
    \item[1.] The central charge $|Z_Q|$ of the four dimensional $\cN=2$ Supersymmetry algebra, is a lower bound on the mass of a state with charge $Q$,  therefore $Z_Q(\bu)$ vanishes for any $\bu\in\XSb$. Assuming away some pathological behavior and carefully keeping track of the structure of the CB geometry, it is possible to prove that $\XSb$ is an $r-1$ complex dimensional algebraic subvariety of $\cC$, which is the union of connected, irreducible, components $\XSb_i$:
    \beq\label{stra1}
    \XSb := \bigcup_{i\in I}\XSb_i^{(1)},\qquad 
    \XSb_i^{(1)} := \Bigl\{\bu\in\cC \,\Big|\, Z_Q\left(\s(\bu)\right)=0, \quad \forall \, Q\in\L_i \Bigr\}.
    \eeq
    Each $\XSb_i^{(1)}$ is defined by the vanishing of the central charge for charges in the lattice $\Lambda_i$ corresponding to the set of BPS states in the theory which become massless there. The superscript $(1)$ indicates the complex co-dimension of the components.

    \item[2.] Since $\XSb$ is a complex co-dimension one algebraic subvariety of $\cC$, it can be cut out by a single polynomial on the CB, which is a product of polynomials whose zero locus corresponds to distinct connected components \eqref{stra1}. If this polynomial is reduced, then it is unique up to an overall constant factor. We then define the \emph{discriminant locus} to be the following quantity:
    \beq\label{DisLoc}
    D_x:=\prod_{i\in I} P_i(\bu),\qquad \XSb_i=:\Big\{\bu\in\cC\Big| P_i(\bu)=0\Big\}
    \eeq
    where the $P_i(\bu)$ are distinct and irreducible for all $i\in I$. Because our initial theory is superconformal, the $P_i(\bu)$ are weighted homogeneous polynomials in the $\bu$ and their scaling dimension $\D^{\rm sing}_i$ plays a special role in what follows:
    \beq\label{Dsing}
    \D_i^{\rm sing}:=\D\Big(P_i(\bu)\Big).
    \eeq
    \item[3.] The special K\"ahler structure of $\cC$ naturally induces a stratification on $\XSb$. First, higher complex codimension components, $\XSb^{(\ell)}_j$, with $\ell>1$, arise by either intersections or as a singular locus of the $\XSb^{(1)}_i$s. The complex codimension of each component has a precise correspondence with the rank of the low energy theory supported there or, more precisely, supported on an open subset which we call the \emph{strata} associated to the component $\XSb^{(\ell)}_j$ and we will indicate without the bar $\XS^{(\ell)}_j$. Let's spell this point out a bit more clearly.
    
    Call the rank-$r$ theory at the superconformal vacuum $\XT$ and call $\XT_\bu$ the low-energy effective description of $\XT$  at the generic point $\bu$. For example we have:
    \beq
    \XT_\bu\equiv \text{free $\cN=2$ $\U(1)^r$},\qquad \bu\in \cCrg.
    \eeq
    In general $\XT_\bu$, if $\bu\in\XSb^{(\ell)}_j$, is a theory of rank-$\ell$ which could be either IR-free or an SCFT. The rank-1 theories supported on the complex co-dimension one strata play a special role in our analysis and we will indicate them as:
    \beq\label{pairT}
    \XT_i\equiv \XT_\bu,\qquad \bu\in\XS^{(1)}_i, \quad i\in I.
    \eeq
    and the quantities indexed by $i\in I$, $(c_i,k_i,h_i)$, label the central charges of these rank-1 theories $\XT_i$ and will be used to compute the central charges of the SCFT at the superconformal vacuum $\XT$ (see below). We also use $u_i$ to label the coordinate parametrizing the one complex dimensional CB of $\XT_i$ and define:
    \beq\label{Di}
    \D_i:=\D(u_i)
    \eeq
    which defines the last quantity entering the central charge formulae which we will shortly define.

    \item[4.] The stratification of the CB singular locus is richer than initially thought; the transverse slice to each component $\XS^{(\ell)}_j$ inherits naturally a Special K\"ahler structure from its interpretation as the CB of $\XT_\bu$, for $\bu\in\XS^{(\ell)}_j$. It is less trivial to show that the Special K\"ahler structure of the ambient space $\cC$ consistently restricts on $\XS^{(\ell)}_j$ which is itself a Special K\"ahler space \cite{Argyres:2020wmq}. To understand how to compute the Special K\"ahler structure induced on the strata, we refer the interested reader to the original literature. The combination of these results gives rise to a \emph{Special K\"ahler stratification} that resembles in many ways the stratification of Symplectic singularities \cite{beauville1999symplectic,kaledin2006symplectic}, with the remarkable difference that in the Special K\"ahler case the minimal transition are always complex dimension one.

    \item[5.] Generalizing \cite{Shapere:2008zf}, it is possible to derive explicit formulae expressing the central charges of an arbitrary $\cN=2$ SCFT in terms of corresponding quantities of the rank-1 theories $\XT_i$'s \cite{Martone:2020nsy}:
    \begin{subequations}
    \begin{align}
    \label{actotaint}
    24 a &= 5r + h + 6 \left(\sum_{\ell=1}^r\D_{\bu_\ell}-1\right) +\sum_{i\in I}\D^{\rm sing}_i \frac{12 c_i-2-h_i}{\D_i} ,
    \\\label{actotbint}
    12 c &= 2r + h + \sum_{i\in I}\D^{\rm sing}_{i} \frac{12 c_i-2-h_i}{\D_i},\\\label{actotcint}
    k_\ff&=\sum_{i\in I_{\ff}}\frac{\D_i^{\rm sing}}{d_i\D_i} \left(k^i-T({\bf2}\bh_i)\right)+T({\bf2}\bh).
    \end{align}
    \end{subequations}
    Here, $r$ is the rank of the SCFT, $h$ is the quaternionic dimension of the theory's extended Coulomb branch and $\D_{\bu_\ell}$ is the scaling dimension of the theory's $\ell$-th component of the CB coordinate vector $\bu$. The sums indexed by $i$ are performed over all complex codimension one components $\XS_i^{(1)}$: $\D^{\rm sing}_i$ and $\D_i$ are defined in \eqref{Dsing} and \eqref{Di},  all the remaining quantities  indexed by $i$ (except $d_i$) refer to corresponding quantities of $\XT_i$ defined in \eqref{pairT}. 
Finally $d_i$ is the embedding index of the flavor symmetry.    
 We call these formulae \emph{central charge formulae} and their great service is that they allow to re-write the SCFT data of a rank-$r$ SCFT in terms of easily accessible geometric data (e.g. the scaling dimension of their CB parameter or dimension of its Extended Coulomb Branch) and the SCFT data of rank-1 theories which have been fully classified.

    \item[6.] Finally, a key role in identifying the correct stratification is played by the \emph{UV-IR simple flavor condition} \cite{Martone:2020nsy} which states that all mass deformations of a rank-$r$ $\cN=2$ SCFT deform the CB asymptotically and are realized, in the low-energy limit, as mass deformations of the rank-1 theories $\XT_i$. Below, we will use this property extensively when analyzing the mass deformations of $\CT$ and $\CS$ theories.
\end{itemize}

\subsection{$\CS^{(2)}_{G,\ell}$ stratification}

\begin{figure}[h!]
\begin{tikzpicture}[decoration={markings,
mark=at position .5 with {\arrow{>}}}]
\begin{scope}[scale=1.7]

\node[bbc,scale=.5] (p0a) at (0,0) {};
\node[bbc,scale=.5] (p0b) at (0,-2) {};
\node[scale=.8] (t0a) at (0,.3) {$\cC$};
\node[scale=.8] (t0b) at (0,-2.3) {0};
\node[scale=.8] (p1) at (-.7,-1) {$\boldsymbol{[I^*_0,C_1]}{\times}\H^{\ell(\D_7-1)}$\ \ };
\node[scale=.8] (p2) at (.7,-1) {\ \ $\boldsymbol{\cS^{(1)}_{(G,\ell)}}{\times}\H$};
\draw[red] (p0a) -- (p1);
\draw[red] (p0a) -- (p2);
\draw[red] (p1) -- (p0b);
\draw[red] (p2) -- (p0b);
\node[scale=.8] (t1a) at (-.6,-.4) {$I_0^*$};
\node[scale=.8] (t1bb) at (-.7,-1.8) {$\big[u^2=v\big]$};
\node[scale=.8] (t1ba) at (-.7,-1.55) {$K_{\ell\D_7}$};
\node[scale=.8] (t2a) at (.6,-.4) {$K_{\ell\D_7}$};
\node[scale=.8] (t2ba) at (.7,-1.55) {$K_{\ell\D_7}$};
\node[scale=.8] (t2bb) at (.7,-1.8) {$\big[v=0\big]$};
\end{scope}
\end{tikzpicture}
{\caption{\label{stra:Sth}Special K\"ahler stratification for $\CS^{(2)}_{G,\ell}$, with $\ell=2,\ 3$ and $4$.}}
\end{figure}
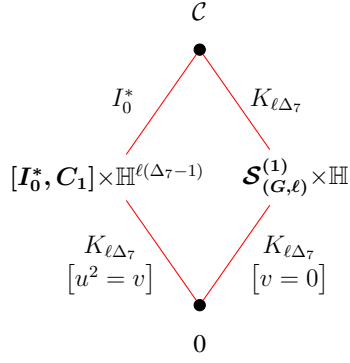

As a warm-up, let us discuss the CB stratification of the $\cS$-theories. This analysis was already performed in  \cite{Argyres:2020wmq} for the $\ell=2$ case. Given the F-theory realization of these theories, and the fact that the CB moduli are realized in the string theory picture as the transverse positions of the two D3 with respect to the 7-brane, we can straightforwardly derive the structure of the singular loci and accurately identify the corresponding low-energy description. The result, that applies uniformly to all $\CS$ theories, is depicted in figure \ref{stra:Sth}.
There, we use $K_\D$ to indicate the rank-1 Kodaira geometry with uniformizing parameter of scaling dimension $\D$ . Explicitly $K_i$ corresponds to a $I_0$, $II$, $III$, $IV$, $I^*_0$, $IV^*$, $III^*$ and $II^*$ for $i=1$, $\frac 65$, $\frac43$, $\frac32$, 2, 3, 4 and 6 respectively.
We also specify the extra quaternionic factor $\H^n$ to keep track of the ECB of the intermediate strata.

The special positions of the D3 branes giving rise to extra charged massless states are easy to identify. When a single D3 brane probes the 7-brane + fluxful S-fold singularity, the states becoming massless correspond to strings stretching from the 7-brane to one D3, we therefore conclude that the low-energy theory there will be a rank-1 $\CS$ theory. Alternatively, when the positions of two D3 branes coincide, the massless string states correspond to those stretching among the two D3s giving rise to a low energy $\cN=4$ $SU(2)$ theory. Summarizing, we expect that the singular locus has two disconnected component $\{\XSb_1,\XSb_2\}$ and the theories supported on their corresponding strata are:
\be\label{thStrS}
\begin{array}{l}
    \XT_1\quad\to \quad \CS^{(1)}_{G,\ell}\times \H,\\
    \XT_2\quad\to \quad (\cN=4\ SU(2))\times \H^{\ell(\D_7-1)}.
\end{array}
\ee

To determine the Special K\"ahler structure induced on $\XS_{1,2}$ we need to write their closures as an algebraic subvariety of $\cC$. This can also be done by leveraging the intuition coming from the F-theory picture. Call $(z_1,z_2)$ the coordinates of the two D3 branes transverse to the 7-brane, which, due to the presence of the 7-brane + S-fold, carry scaling dimension $\ell\D_7$. The fluxful S-fold induces a $G(\ell,1,2)$ action on the $z_i$ \cite{Aharony:2016kai}, see \eqref{CRG}. The result is that the CB of the S-theories is described by:
\be
u=\frac{z_1+z_2}2\qquad{\rm and}\qquad v=z_1 z_2
\ee
from which it immediately follows that the closure of the two strata can be written algebraically as:
\beq\label{straSth}
\XSb_1:=\{(u,v)\in\cC\mid v=0\}\quad ;\quad \XSb_2:=\{(u,v)\in\cC\mid u^2=v\}.
\eeq
To be able to identify the Special K\"ahler structure of the strata, we need to compute the scaling dimension of the uniformizing parameter describing this one complex dimensional variety. Given the algebraic expression \eqref{stra:Sth}, a straightforward calculation shows that the strata are of Kodaira type $K_{\ell\D_7}$ \cite{Argyres:2020wmq}. 

It is instructive to perform a check of this analysis by matching the central charges of the $\CS$ theories using \eqref{actotaint}-\eqref{actotcint}, for example we can calculate explicitly the $c$ central charge in the case of $\CS_{E_6,2}^{(2)}$. The sum in \eqref{actotbint} will be over the two strata in \eqref{straSth} and, using \eqref{thStrS} and \eqref{straSth}, we have the following:
\be
\XS_1:\left\{
\begin{array}{l}
\D^{\rm sing}_1=12\\
12 c_1=49\\
h_1=5\\
\D_1=6
\end{array}\right.
\qquad ;\qquad
\XS_2:\left\{
\begin{array}{l}
\D^{\rm sing}_2=12\\
12 c_2=9\\
h_2=1\\
\D_2=2
\end{array}
\right.
\ee
from which, using $r=2$ and $h=6$, \eqref{actotcint} gives $12c=130$ matching the result in table \ref{table:table}. We leave it up to the reader to compute $a$, $k_{C_4}$ and $k_{SU(2)}$\footnote{The $SU(2)$ flavor symmetry is realized in the low-energy as diagonal subgroup of the $SU(2)$ factors carried on each strata. This observation is key to reproduce the appropriate level, see \cite{Argyres:2020wmq} for more details. }.

\subsection{$\CT^{(2)}_{G,\ell}$ stratification}

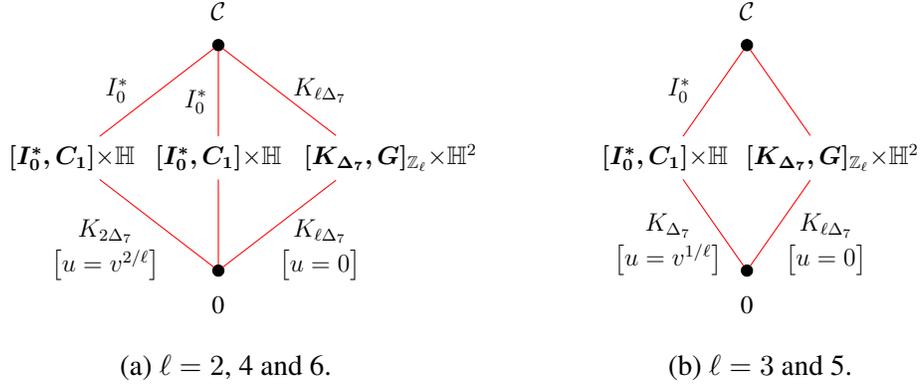
\begin{figure}[t!]
\ffigbox{
\begin{subfloatrow}
\ffigbox[6cm][]{
\begin{tikzpicture}[decoration={markings,
mark=at position .5 with {\arrow{>}}}]
\begin{scope}[scale=1.5]
\node[bbc,scale=.5] (p0a) at (0,0) {};
\node[bbc,scale=.5] (p0b) at (0,-2) {};
\node[scale=.8] (t0a) at (0,.3) {$\cC$};
\node[scale=.8] (t0b) at (0,-2.3) {0};
\node[scale=.8] (p1) at (-1.3,-1) {$\boldsymbol{[I^*_0,C_1]}{\times}\H$};
\node[scale=.8] (p2) at (0,-1) {$\boldsymbol{[I^*_0,C_1]}{\times}\H$};
\node[scale=.8] (p3) at (1.3,-1) {\qquad $\boldsymbol{[K_{\D_7},G]}_{\mathbb{Z}_\ell}{\times}\H^2$};
\node[scale=.8] (t1a) at (-.9,-.4) {$I_0^*$};
\node[scale=.8] (t1b) at (-.2,-.5) {$I_0^*$};
\node[scale=.8] (t1c) at (-1,-1.65) {$K_{2\D_7}$};
\node[scale=.8] (t1c) at (-1,-1.95) {$\big[u=v^{2/\ell}\big]$};
\node[scale=.8] (t2a) at (.9,-.4) {$K_{\ell\D_7}$};
\node[scale=.8] (t2b) at (-.3,-1.5) {};
\node[scale=.8] (t2c) at (.9,-1.65) {$K_{\ell\D_7}$};
\node[scale=.8] (t1c) at (.9,-1.95) {$\big[u=0\big]$};
\draw[red] (p0a) -- (p1);
\draw[red] (p0a) -- (p2);
\draw[red] (p0a) -- (p3);
\draw[red] (p1) -- (p0b);
\draw[red] (p2) -- (p0b);
\draw[red] (p3) -- (p0b);
\end{scope}
\end{tikzpicture}}
{\caption{$\ell=2$, 4 and 6.}}
\end{subfloatrow}\hspace{1cm}
\begin{subfloatrow}
\ffigbox[6cm][]{
\begin{tikzpicture}[decoration={markings,
mark=at position .5 with {\arrow{>}}}]
\begin{scope}[scale=1.5]
\node[bbc,scale=.5] (p0a) at (0,0) {};
\node[bbc,scale=.5] (p0b) at (0,-2) {};
\node[scale=.8] (t0a) at (0,.3) {$\cC$};
\node[scale=.8] (t0b) at (0,-2.3) {0};
\node[scale=.8] (p1) at (-.7,-1) {$\boldsymbol{[I^*_0,C_1]}{\times}\H$\ \ };
\node[scale=.8] (p2) at (.7,-1) {\ \ $\boldsymbol{[K_{\D_7},G]}_{\mathbb{Z}_\ell}{\times}\H^2$};
\node[scale=.8] (t1a) at (-.6,-.4) {$I_0^*$};
\node[scale=.8] (t2c) at (-.7,-1.6) {$K_{\D_7}$};
\node[scale=.8] (t1c) at (-.7,-1.9) {$\big[u=v^{1/\ell}\big]$};
\node[scale=.8] (t2c) at (.7,-1.6) {$K_{\ell\D_7}$};
\node[scale=.8] (t1c) at (.7,-1.9) {$\big[u=0\big]$};
\draw[red] (p0a) -- (p1);
\draw[red] (p0a) -- (p2);
\draw[red] (p1) -- (p0b);
\draw[red] (p2) -- (p0b);
\end{scope}
\end{tikzpicture}}
{\caption{$\ell=3$ and 5.}}
\end{subfloatrow}}{\caption{\label{fig:2str}Special K\"ahler stratification of $\CT^{(2)}_{G,\ell}$. (a) is for $\ell=2$, 4 and 6 while (b) is for $\ell=3$ and 5.}}
\end{figure}

The analysis of the $\cT$-theories is in many ways analogous to the case just analyzed but in this case we will find a richer structure which is reflected in figure \ref{fig:2str} (a) and (b) respectively.

As before, we expect that the two brane configurations which give rise to extra charged massless states are either two coincident D3 branes or a single D3 probing the 7-brane + flux-less S-fold set up. We therefore conclude that:
\be\label{thStrT}
\begin{array}{l}
    \XT_1\quad\to \quad [\CT^{(1)}_{G,\ell}]_{\bZ_\ell}\times \H^2\\
    \XT_2\quad\to \quad (\cN=4\ SU(2))\times \H
\end{array}
\ee
where again the factor of $\H^n$ are added to account for the ECB,
and the subscript $_{\bZ_{\ell}}$ means that we need to perform the discrete $\bZ_\ell$ gauging, as we discussed in Sec.~\ref{sec:rank1}.
The equation \eqref{thStrT} would suggest that the T-theories have two disconnected complex co-dimension one singular components but the careful reader might object that the diagrams in figure \ref{fig:2str} show instead three disconnected components, at least for the $\ell=2$, $4$ and $6$. Let us see how this comes about. 

As we did in the previous subsection, in order to write down the closures of the strata as algebraic subvariety of $\cC$, we need to analyze how the CB coordinates are written in terms of $(z_1,z_2)$, the coordinates of the D3 branes transverse to the 7-brane. Again the $z_i$ have scaling dimension $\ell\D_7$ but now the flux-less S-fold induces a $G(\ell,\ell,2)$ action on them. It is a straightforward calculation to compute the appropriate invariant for this action and we find:
\be
u=\sqrt[\ell]{z_1z_2}\quad,\quad v=\frac{z_1+z_2}2
\ee
from which we find the equivalent of \eqref{straSth} in this case:
\beq\label{straTth}
\XSb_1:=\{(u,v)\in\cC\mid u=0\}\quad \&\quad \XSb_2:=\{(u,v)\in\cC\mid u^\ell= v^2\}.
\eeq
The third disconnected component arises because the algebraic variety $u^\ell=v^2$ is irreducible only if ${\rm gcd}(2,\ell)=1$. If $\ell$ is even the polynomial can be written instead as a product of two irreducible polynomials in which case we instead have:
\be
\XSb_2\quad\to\quad\left\{
\def\arraystretch{1.3}
\begin{array}{l}
\XSb_{2a}:=\{(u,v)\in\cC\mid u^{\ell/2}= v\}\\
\XSb_{2b}:=\{(u,v)\in\cC\mid u^{\ell/2}= -v\}
\end{array}
\right.
\ee
and therefore $\XSb_2$ splits in two disconnected components as in figure \ref{fig:2str} (a).

\begin{figure}[h!]
\begin{tikzpicture}[decoration={markings,
mark=at position .5 with {\arrow{>}}}]
\begin{scope}[scale=1.5]
\node[bbc,scale=.5] (p0a) at (0,0) {};
\node[bbc,scale=.5] (p0b) at (0,-3) {};
\node[scale=.8] (n0) at (0,.3) {$\cH$};
\node[scale=.8] (n1) at (0,-.7) {$\CT^{(1)}_{G,1}$};
\node[scale=.8] (n2a) at (.7,-1.4) {{$\CT^{(1)}_{G,1}\times\CT^{(1)}_{G,1}$}};
\node[scale=.8] (n3) at (-.7,-1.8) {$\CS^{(1)}_{G,\ell}$};
\node[scale=.8] (n4) at (.7,-2.5) {$\CT^{(2)}_{G,1}$};
\node[scale=.8] (n5) at (0,-3.3) {0};
\node[scale=.8] (t1) at (-.4,-2.6) {$\mathfrak{f}$};
\node[scale=.8] (t2) at (.4,-2.9) {$\mathfrak{a}_1$};
\draw[blue] (p0a) -- (n1);
\draw[blue] (n1) -- (n2a);
\draw[blue] (n1) -- (n3);
\draw[blue] (n2a) -- (n4);
\draw[blue] (n3) -- (p0b);
\draw[blue] (n4) -- (p0b);
\end{scope}
\end{tikzpicture}}{\caption{\label{fig:HasseHB}Higgs Branch Hasse diagram of the $\CT^{(2)}_{G,\ell}$ theories from the Coulomb Branch stratification. In the graphic depiction above we used the notation for which $\mathfrak{g}$ indicates the minimal nilpotent orbit of the Lie algebra $G$ and $\mathfrak{f}$ is non-geometric non-abelian symmetry of the theory.}
\end{figure}
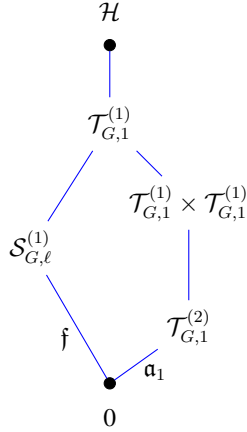

Applying the central charge formulae to the stratification in figure \ref{fig:2str}, it is also possible to straightforwardly derive the values for the central charges of the $\cT$-theories reported in table \ref{table:table}, directly from the rank-1 data. Since this calculation is completely analogous to the one performed above, we will not reproduce it here. But instead we will elaborate on how to use the CB stratification to reproduce the Higgs Branch structure of these theories, which was already discussed in \cite{Giacomelli:2020jel} and, for the reader's convenience, is reproduced in figure \ref{fig:HasseHB}.

Turning on Higgs moduli of the theories supported on various strata of the CB, it is possible to explore the entire moduli space of the theory \cite{Argyres:2020wmq}, although at the moment there is no systematic way to do so. In fact the Higgs branches of the theories visible from the CB analysis give direct information on the lowest HB transitions but only partial indications on the subsequent ones, which instead depend on the details of the Higgsing pattern on the HB. This latter cannot be inferred systematically from the CB. For simple enough cases, it is possible to quickly converge to an educated guess which can be then checked a posteriori using the chiral algebra techniques developed in \cite{Beem:2019tfp,CCLMW2020}. In the case of $\CT^{(2)}_{G,\ell}$, the transition corresponding to $\mathfrak{f}$ corresponds to moving on the HB of the $[K_{\Delta_7},G]_{\mathbb{Z}_\ell}$ while the $\mathfrak{a}_1$ corresponds instead to moving along the HB of the $[I_0^*,C_1]$. The chiral algebra for these theories was constructed in \cite{Giacomelli:2020jel} precisely using the aforementioned techniques, so we will not perform this extra check here. We will come back to these techniques in the sections below to compute the SCFT data of new $\mathcal{N}=2$ SCFTs which can be obtained by mass deforming the $\CT^{(2)}_{G,\ell}$ theories.


\section{Magnetic quivers}
\label{sec:magneticquivers}

Another method by which we can study the Higgs branch is using magnetic quivers. The latter are $3d$ $\CN = 4$ quivers whose Coulomb branch gives the Higgs branch of the studied theory, here the $4d$ theories discussed so far. There are various methods by which these can be determined. For the case at hand, we can use the realizations of these theories as twisted compactifications of $5d$ SCFTs. Specifically, we mentioned that the $\CT$-theories, for $\ell \Delta_7 = 6$, can be realized by the compactifications of certain $6d$ $(1,0)$ SCFTs on a torus with almost commuting holonomies, and the same is also true for the $\CS$-theories (see \cite{Giacomelli:2020jel} for the details). As we mentioned previously, by reducing along one of the circles, these can be mapped to twisted compactifications of $5d$ gauge theories (see \cite{Ohmori:2018ona} for the details). The magnetic quivers can then be derived from these, as done in \cite{Bourget:2020asf}, which studied the magnetic quivers for the rank-1 $\CS$-theories. More specifically, as explained in \cite{Ohmori:2018ona}, we can use the $6d$ picture to get a description of the $5d$ SCFT, whose twisted compactification leads to the $4d$ theories, in terms of brane webs. We can then use the prescription in \cite{Cabrera:2018jxt} to read the magnetic quivers from the brane webs. Here we need to take into account the effect of the twist, which implies that only directions invariant under the twist can be accessed. See \cite{Bourget:2020asf}, for how this affects the magnetic quivers.

\begin{figure}
\center
\includegraphics[width=0.75\textwidth]{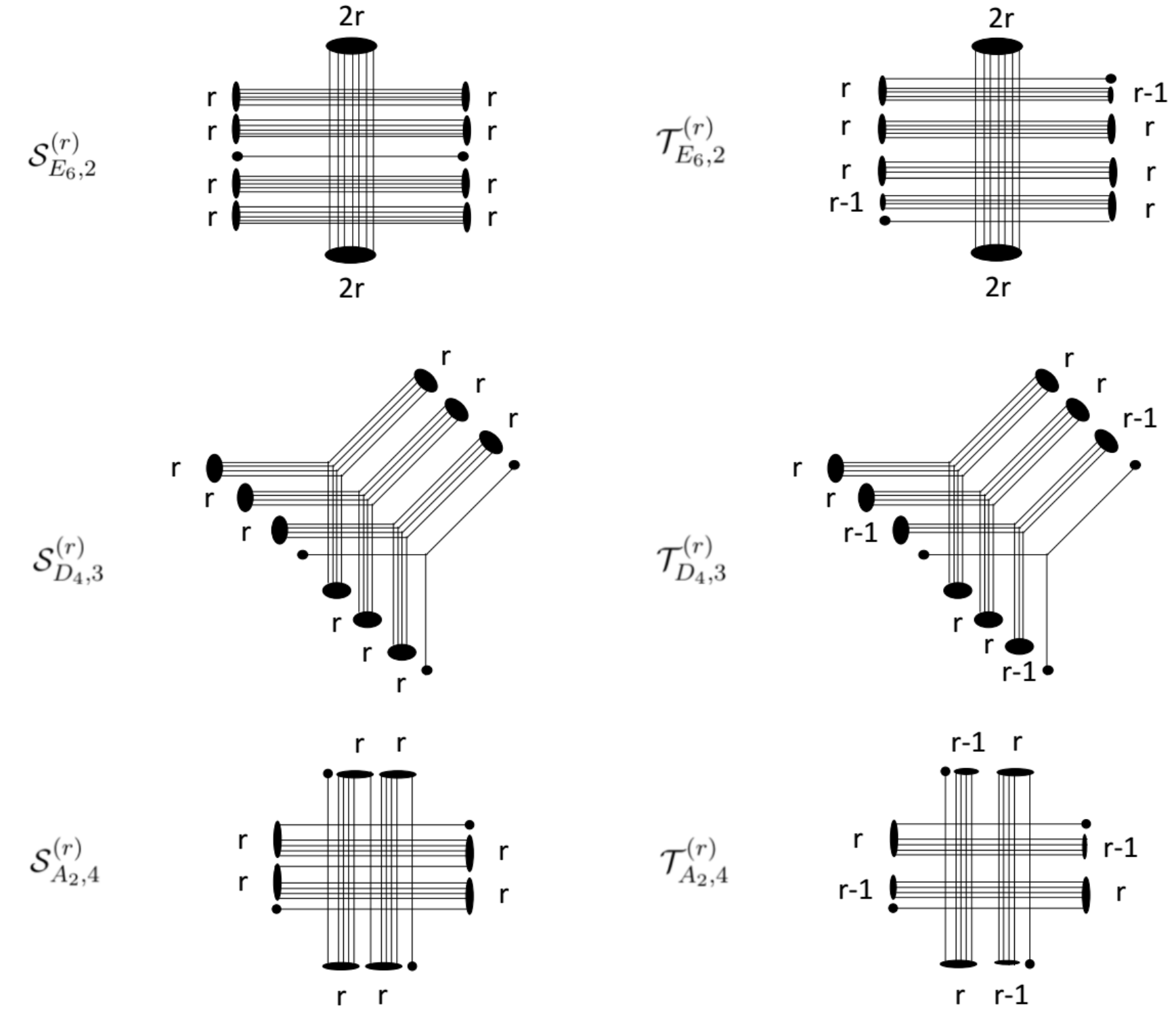} 
\caption{The brane webs describing the $5d$ SCFTs whose twisted compactifications yield the $4d$ theories written to the left of the web. Here the black dots represent 7-branes of the type determined by the 5-branes ending on them. If bigger than 1, the number of 5-branes ending on each 7-brane is written next to the 7-brane. For the top two theories the compactification is done with a $\mathbb{Z}_2$ twist corresponding to a $\pi$ rotation of the web and the $SL(2,\mathbb{Z})$ transformation $-I$. For the two middle theories the compactification is done with a $\mathbb{Z}_3$ twist corresponding to a $\frac{2\pi}{3}$ rotation of the web and the $SL(2,\mathbb{Z})$ transformation $S T$ (which is a symmetry of the web if the axiodilaton is set to the invariant value). For the bottom two theories the compactification is done with a $\mathbb{Z}_4$ twist corresponding to a $\frac{\pi}{2}$ rotation of the web and the $SL(2,\mathbb{Z})$ transformation $S$.}
\label{MQbranewebs}
\end{figure}

\begin{figure}
\center
\includegraphics[width=0.75\textwidth]{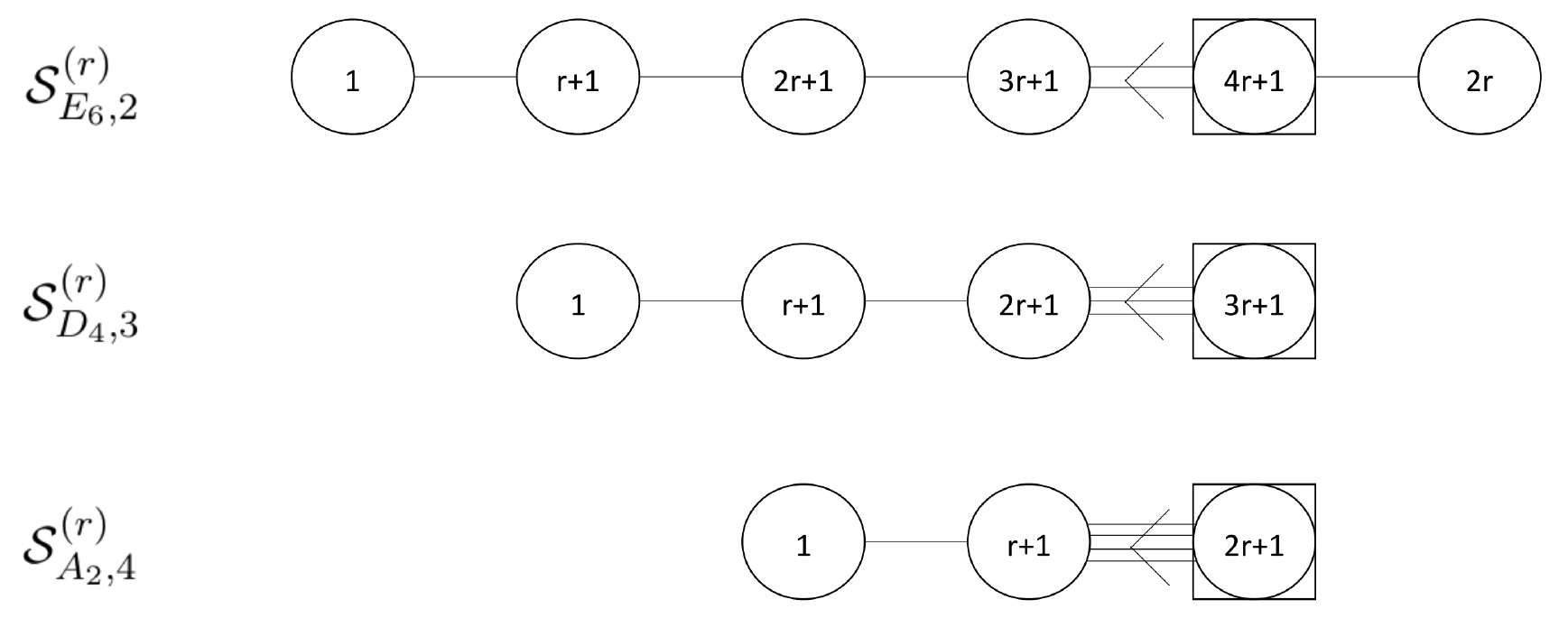} 
\caption{The magnetic quivers for some of the $\CS$-theories. Here the square surrounding one of the nodes represents the node where the ungauging is done.}
\label{MQSth}
\end{figure}

\begin{figure}
\center
\includegraphics[width=0.75\textwidth]{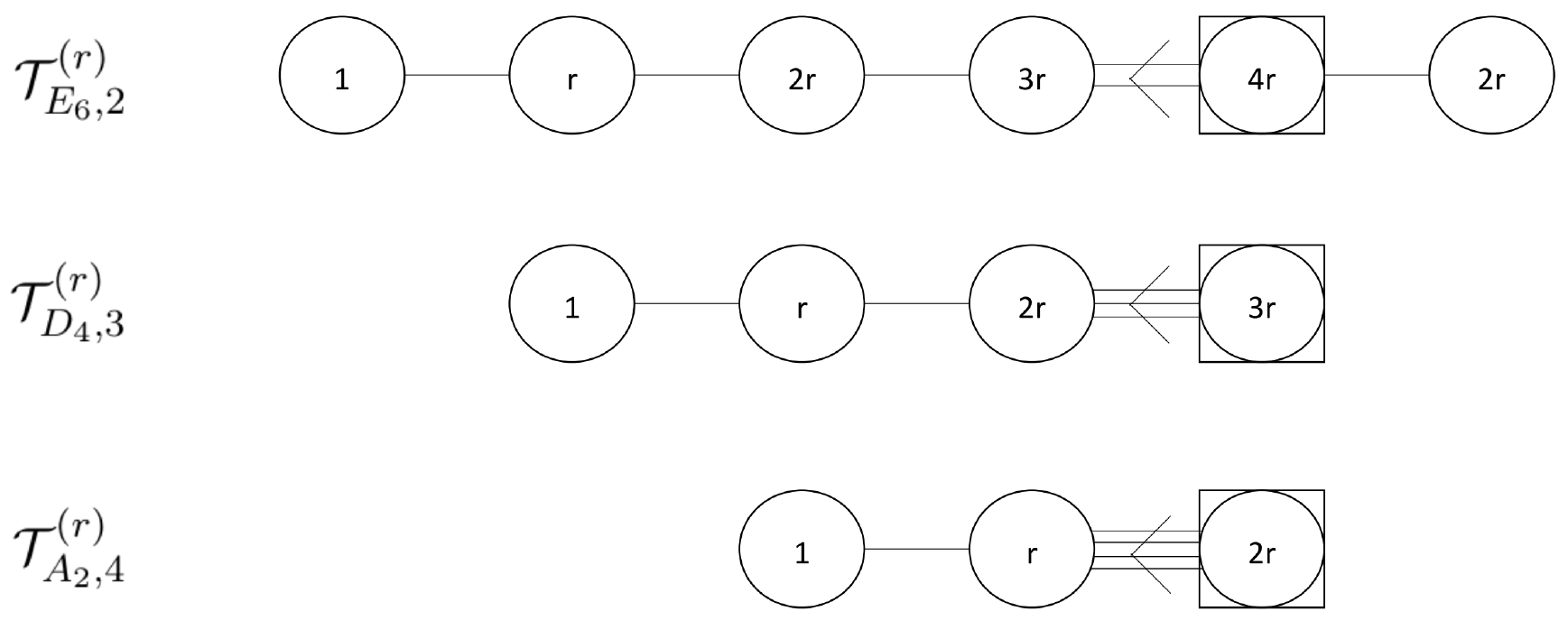} 
\caption{The magnetic quivers for some of the $\CT$-theories. Here the square surrounding one of the nodes represents the node where the ungauging is done.}
\label{MQTth}
\end{figure}

It is possible to employ the methods used there to also produce the magnetic quivers for the higher rank cases. Here for simplicity, we only consider cases with $\ell \Delta_7 = 6$, and for $\ell=2,3$ and $4$. The brane webs describing the $5d$ SCFTs whose twisted compactifications yield the associated $4d$ theories are given in figure \ref{MQbranewebs}. The resulting quivers are presented in figure \ref{MQSth} for the $\CS$-theories, and in figure \ref{MQTth} for the $\CT$-theories.

While we shall not present a detailed study of the magnetic quivers here, which is performed in \cite{Imperial}, we do wish to mention some features that can be immediately uncovered from them. Specifically, there is a basic Higgs branch generator associated with every node in the quiver, which is given by the basic magnetic monopole associated with that node. Being a Higgs branch generator, its ground state is a scalar charged under the $SU(2)$ part of the R-symmetry, but not under the abelian part. For a $U(n_c)$ node, seeing a total number of $n^T_f$ flavors\footnote{When counting the number of flavors, non-simply laced connections of order $k$ count as $k$ bifundamental for the node the arrow exits from, but only as one bifundamental for the node the arrow enters to.}, the specific representation is the one of dimension $(\bold{3+n^T_f-2n_c})$. If $n^T_f = 2n_c$, the node is called balanced, and the corresponding Higgs branch operator contains a conserved current, causing the symmetry on the Coulomb branch to enhance beyond the $U(1)$ per node minus one that is naively expected. In fact when considering basic monopole operators charged under any combination of balanced nodes, one finds that there are sufficient number of conserved currents to enhance the symmetry to the group whose Dynkin diagram is formed by the collection of balanced nodes. 

The non-balanced nodes usually give other Higgs branch generators. One can again show that if you consider basic monopole operators charged under the said node and any combination of balanced nodes, you get additional operators such that they form a representation of the symmetry given by the Dynkin diagram formed by the collection of balanced nodes, that is determined by the balanced nodes that this node is connected to. Using this, the magnetic quivers allow us to quickly infer various properties of the associated Higgs branch.

Consider the $\CS$ theories, whose magnetic quivers are presented in figure \ref{MQSth}. Assuming $r>1$, we see that in all three cases all nodes save for the two edge ones are balanced. This gives an expected global symmetry of at least $Sp(4)\times U(1)$ for $\CS^{(r)}_{E_6,2}$, $SU(3)\times U(1)$ for $\CS^{(r)}_{D_4,3}$ and $SU(2)\times U(1)$ for $\CS^{(r)}_{A_2,4}$. This is consistent with the expectations given in table \ref{table:table}. Note that for $\CS^{(r)}_{E_6,2}$, we expect a further enhancement of $U(1)\rightarrow SU(2)$, though that is not visible from just looking at the balanced nodes.

The unbalanced nodes lead us to expect two basic Higgs branch generators. First, from the left unbalanced node, we expect a Higgs branch generator whose lowest component is a scalar in the $\bold{r+2}$ dimensional representation of $SU(2)_R$. Additionally, as that node is connected to the leftmost balanced node, these are expected to be in the fundamental representation of the associated flavor symmetry group. The rightmost unbalanced node gives an additional Higgs branch generator. For $\CS^{(r)}_{E_6,2}$, it is in the $\bold{4}$ dimensional representation of $SU(2)_R$, and the $\bold{42}$ dimensional representation of $Sp(4)$. For $\CS^{(r)}_{D_4,3}$, it is also in the $\bold{4}$ dimensional representation of $SU(2)_R$, and the $\bold{10}$ dimensional representation of $SU(3)$, where here we have taken the three index symmetric representation of the fundamental as the unbalanced node is connected to the balanced node associated with the fundamental by an arrow of order $3$. Similarly, for $\CS^{(r)}_{A_2,4}$, it is now in the $\bold{5}$ dimensional representation of $SU(2)_R$, and also the $\bold{5}$ dimensional representation of the $SU(2)$ global symmetry.   

We can match this against the expectation from the $6d$ construction, reviewed in section \ref{sec:6d}, see \eqref{Squivers}. Specifically, as pointed out in \cite{Giacomelli:2020jel}, these SCFTs can also be constructed by the compactification of a family of $6d$ SCFTs on a torus with almost commuting holonomies, see \cite{Giacomelli:2020jel} for the details. We can understand the spectrum of Higgs branch operators also from the $6d$ description by considering the Higgs branch operators in the $6d$ SCFT that are invariant under the holonomies. This is expected as the Higgs branch should be invariant under dimensional reduction, and so should be affected only by the holonomies. 

Consider first the case of $\CS^{(r)}_{E_6,2}$. A survey of the basic Higgs branch operators of this theory was done in \cite{Zafrir:2018hkr}, and here we shall use these results. The global symmetry of the $6d$ SCFT is expected to be $SU(2)_E \times SU(2)_F \times SO(16)$, and besides the moment map operators associated with these symmetries, there are two additional Higgs branch operators. One is in the $(\bold{2},\bold{r+1},\bold{16})$ of the global symmetry and in the $\bold{r+2}$ of $SU(2)_R$, while the other is in the $(\bold{1},\bold{2},\bold{128})$ of the global symmetry and in the $\bold{4}$ of $SU(2)_R$. Here the former comes from the gauge invariant made from all the bifundamentals and the flavors at the edges, while the other is of non-perturbative origin. The compactification is done with two almost commuting holonomies in the diagonal $SU(2)$ of $SU(2)_E$ and an $SU(2)$ subgroup of $SO(16)$ such that the commutant is $Sp(4)$. Under the embedding of the $SU(2)$ in $SO(16)$, we have that:

\begin{align}
\bold{16}_{SO(16)} &\rightarrow (\bold{2}_{SU(2)},\bold{8}_{Sp(4)}) \; , \\
 \bold{128}_{SO(16)}& \rightarrow (\bold{5}_{SU(2)},\bold{1}_{Sp(4)})\oplus (\bold{3}_{SU(2)},\bold{27}_{Sp(4)}) \oplus (\bold{1}_{SU(2)},\bold{42}_{Sp(4)}). 
\end{align}

From this we see that we expect to get from the operator with charges $(\bold{2},\bold{r+1},\bold{16})$ an operator in the $(\bold{r+1},\bold{8})$ of $SU(2)_F \times Sp(4)$ and in the $\bold{r+2}$ of $SU(2)_R$ \footnote{Here we use the fact that the holonomies are in the diagonal $SU(2)$, and we have that there is an $SU(2)$ singlet in the product $\bold{2} \otimes \bold{2}$.}, while from the other we expect an operator in the $(\bold{2},\bold{42})$ of the global symmetry and in the $\bold{4}$ of $SU(2)_R$. These indeed match the operators we find from the magnetic quiver, though here we can also infer their expected charge under $SU(2)_F$.

This analysis can be repeated for the $\CS^{(r)}_{D_4,3}$ and $\CS^{(r)}_{A_2,4}$ theories, where again we find consistent results between the magnetic quivers and the $6d$ construction. Briefly, the operator in the $\bold{r+2}$ of $SU(2)_R$, again comes from the gauge invariant made from all the bifundamentals and the flavors at the edges. The second one, though, now comes from the baryons of the $SU$ group on the $-1$ curve.

We can perform a similar analysis on the $\CT$ theories. From the balanced nodes, we see that the global symmetry should be at least $F_4\times U(1)$ for $\CT^{(r)}_{E_6,2}$, $G_2\times U(1)$ for $\CS^{(r)}_{D_4,3}$ and $SU(2)\times U(1)$ for $\CS^{(r)}_{A_2,4}$. This is again consistent with the expectations given in table \ref{table:table}, save that for $\CT^{(r)}_{E_6,2}$, we expect a further enhancement of $U(1)\rightarrow SU(2)$. From studying the unbalanced nodes, we see that there should be two additional Higgs branch generators. One, present in all three cases, is a flavor singlet in the $\bold{r+1}$ representation of $SU(2)_R$. The second, is in the $\bold{4}$ representation of $SU(2)_R$ and in the fundamental of the flavor symmetry\footnote{When $r=2$, the first operator becomes the moment map operator of the extra currents for the additional $SU(2)$, while the second one becomes charged under it in the fundamental.}. 

We can again match this against the expectation from the $6d$ construction, reviewed in section \ref{sec:6d}, see \eqref{sulquiver}. Specifically, the flavor singlet comes from the gauge invariant made from all the bifundamentals and the flavors at the edges, where for $\CT^{(r)}_{E_6,2}$, we can see that it is also in the $\bold{r}$ dimensional representation of the flavor $SU(2)$, see \cite{Zafrir:2018hkr} for a discussion on the Higgs branch operators in this $6d$ SCFT. The additional operator comes from the gauge invariant made from the extra current operators of the E-string theory that are gauge charged in the fundamental of the $SU$ group attached to them, and the fundamental hypermultiplets of that group. 

\section{Mass deformations}
\label{sec:mass}

We can also consider mass deformations of the $\CS$ and $\CT$ theories. These may lead to other $4d$ SCFTs, or to theories containing an IR free part. We shall not perform an exhaustive search here, instead we shall concentrate on specific cases that are adequately approachable by the methods available to us. 
\subsection{Analysis using $5d$ descriptions }
In our analysis we will start by using the $5d$ description of the $\CS$ and $\CT$ theories from which we can more easily infer which mass deformations lead to $4d$ $\mathcal{N}=2$ SCFTs and extract some basic properties of these fixed points. We then use the large set of four dimensional consistency conditions which arise from a careful analysis of Coulomb and Higgs branches to fully characterize these theories and make sharp predictions about the existence of new rank-2 $4d$ SCFTs.

As we previously mentioned, the $4d$ theories can also be described by a twisted compactification of specific $5d$ SCFTs. The $5d$ SCFTs can be described by a brane web, where the symmetry we twist by is given by a combination of a rotation in the plane of the web and an $SL(2,\mathbb{Z})$ transformation. Mass deformation are then given by motions of the external $7$-branes of the web that respect the symmetry that we twist by. These lead to either a new $5d$ SCFT or to a phase containing an IR free gauge theory. When reduced to $4d$ the mass deformations give similar flows in $4d$, leading to new $4d$ theories. If the $5d$ deformations lead to a phase containing an IR free gauge theory, then we expect the corresponding $4d$ flow to also lead to a phase containing an IR free part. 

The $4d$ reductions of $5d$ deformations that lead to $5d$ SCFTs are more varied. These may lead to $4d$ SCFTs, but can also lead to phases containing an IR free gauge theory. For the class of $5d$ SCFTs whose twisted reduction was studied in \cite{Zafrir:2016wkk,Ohmori:2018ona}, and theories related to them by Higgs branch flows, there is evidence that the twisted reductions lead to $4d$ SCFTs. We will be somewhat agnostic about cases not of that form. 

We shall next describe some mass deformations that can be observed with this method for each case, where we shall mostly concentrate on the cases of $\ell=2,3$ and $4$. As our starting theory, we shall take the cases obeying $\ell \Delta_7=6$, as we expect the rest to be reachable via mass deformations from these. The brane diagrams associated with the $5d$ SCFTs whose twisted compactifications give these theories were already given in figure \ref{MQbranewebs}. 

\subsubsection{Mass deformations to other $\cN=2$ SCFTs}

We begin by considering mass deformations of the $5d$ SCFT leading to other $5d$ SCFTs. Here, for most cases, we only observe the mass deformations leading us from $\CS^{(r)}_{G,\ell}$ to $\CS^{(r)}_{G',\ell}$ or $\CT^{(r)}_{G,\ell}$  to $\CT^{(r)}_{G',\ell}$, where the group $G$ flows in the following manner $E_6 \rightarrow D_4 \rightarrow A_2$. Finally, from $A_2$ we can flow to the cases expected to have higher supersymmetry. This recovers the flow pattern shown in figure \ref{ThrRl2}. In the special case of $\CT^{(2)}_{G,\ell}$ there is an additional $SU(2)$, and additional mass deformations related to it. In this case, we indeed find additional mass deformations leading to new $4d$ SCFTs. 

\begin{figure}
\center
\includegraphics[width=0.9\textwidth]{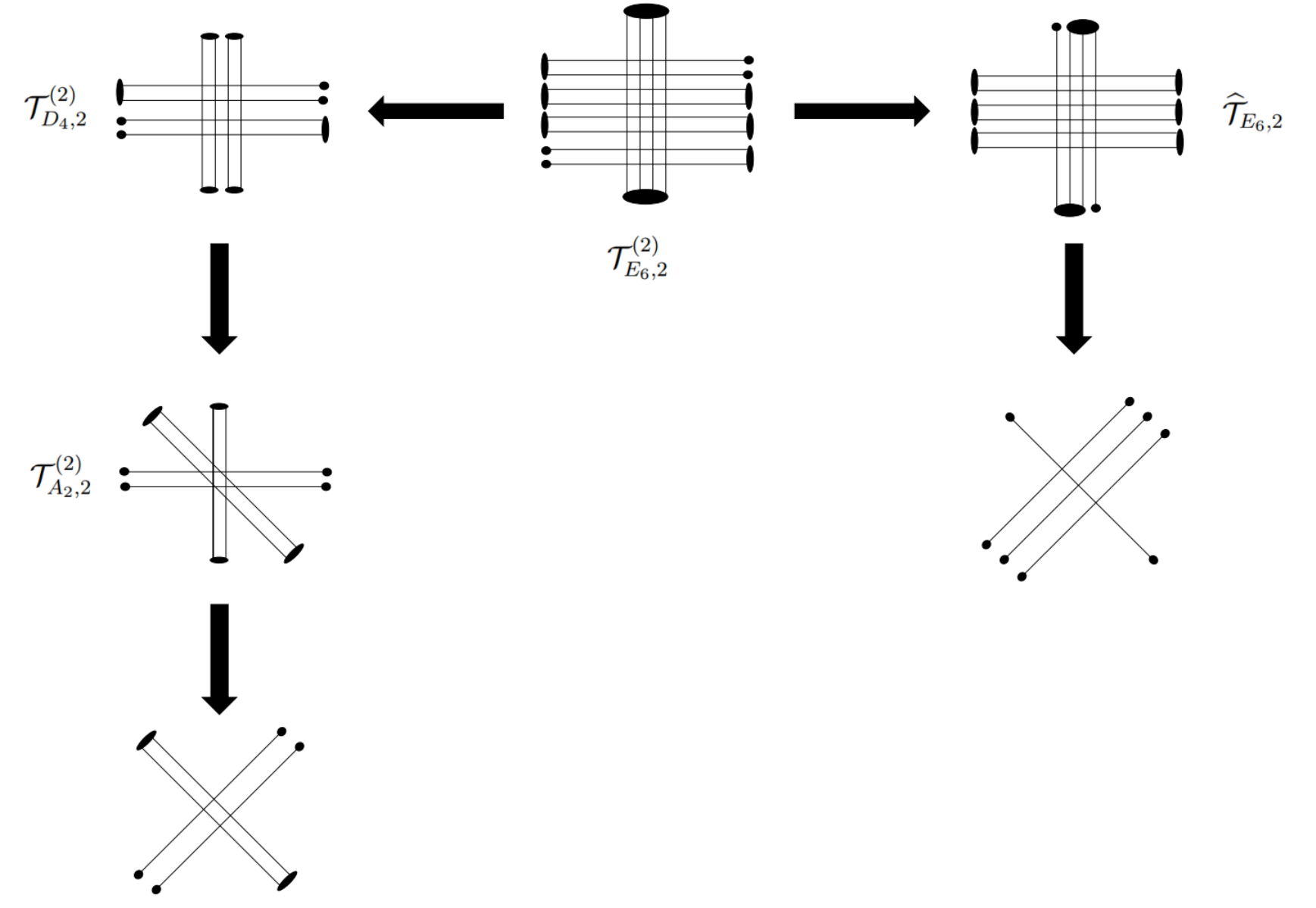} 
\caption{An illustration of the mass deformations for the case of $\CT^{(2)}_{E_6,2}$. Written are $5d$ SCFTs, represented through their brane webs, whose twisted compactification yields the $4d$ $\CT^{(2)}_{E_6,2}$ and some of its mass deformations. Here the compactification is done with a $\mathbb{Z}_2$ twist corresponding to a $\pi$ rotation of the web and the $SL(2,\mathbb{Z})$ transformation $-I$. Next to each web is written the $4d$ SCFT which we expect to result from the twisted reduction of the associated $5d$ SCFT. If nothing is written, then we either do not expect or do not know whether the resulting $4d$ theory is an SCFT.}
\label{T2E62FC}
\end{figure}

To illustrate the method, we shall present the case of $\CT^{(2)}_{E_6,2}$ in detail. The flow pattern in this case is shown in figure \ref{T2E62FC}. The starting point is the $5d$ SCFT shown in the top middle of the figure, whose $\mathbb{Z}_2$ twisted compactification leads to the $\CT^{(2)}_{E_6,2}$ $4d$ SCFT. One set of mass deformations leads to the flow pattern shown on the left. Here the top left theory is of the form studied in \cite{Zafrir:2016wkk}, and we expect the twisted compactification to give a $4d$ SCFT. The $5d$ picture allows us to infer information about the Higgs branch of the quiver, notably the associated magnetic quiver, from which we can see that the global symmetry should be at least $SO(7)\times SU(2)$ and the Higgs branch dimension should be $12$. This motivate us to identify this $4d$ SCFT with $\CT^{(2)}_{D_4,2}$. Similar considerations motivate us to identify the $4d$ theory we get from the twisted compactification of the middle left theory with $\CT^{(2)}_{A_2,2}$. 

Finally, the bottom left theory is not of the form for which there is evidence in favor of a $4d$ SCFT. However, this $5d$ SCFT has a mass deformation leading to the $5d$ gauge theory $SU(4)_0 + 2AS$, and following the reasoning explained in section \ref{section:N3}, we expect the resulting $4d$ theory to be the $\CN = 4$ $SO(4)$ super Yang-Mills theory. As this exhausts the $4d$ SCFTs we expect from F-theory, we are led to conclude that the bottom left theory most likely does not reduce to a $4d$ SCFT.

While here we have shown only the case of $\CT^{(2)}_{E_6,2}$, the left flow pattern generalizes to all $\CT^{(r)}_{E_6,2}$ theories, and there is also an analogous flow pattern for the $\CS^{(r)}_{E_6,2}$ theories. The flow pattern on the right, though, is special for the $\CT^{(2)}_{E_6,2}$ case. Particularly, the top right web is of the form considered in \cite{Zafrir:2016wkk}, and we expect its twisted reduction to give a different $4d$ SCFT, which we dub $\hat{\CT}_{E_6,2}$. We expect this SCFT to have the $F_4$ part of the global symmetry, as the mass deformation does not appear to break it, and a Higgs branch of dimension $16$, which we can read from the web. This theory can be identified with example 14 in \cite{Wang:2018gvb}\footnote{%
This theory had also appeared previously as entry 6 of the table in \cite[Sec.~3.3]{Chacaltana:2015bna}, although only $SO(9)\subset F_4$ was identified. } and will be further characterized below. The web can be further deformed to the one shown in the middle right. This theory is not of the form for which there is evidence in favor of a $4d$ SCFT, so we will refrain from making any concrete claim about the conformality of the expected $4d$ theory at this point.

The flow pattern between the various $5d$ SCFTs appearing in figure \ref{T2E62FC} can also be understood by considering these theories as UV completions of $5d$ gauge theories. Specifically, the $5d$ SCFT shown in the top middle of the figure is the UV completions of several dual $5d$ gauge theories. First, is the $5d$ $SU(4)_0 + 2AS + 6F$ theory, that we have mentioned in previous sections. Additionally, it is also the UV completion of a $5F+USp(4)\times SU(2)+1F$ gauge theory, with a bifundamental hyper between the two groups, and the $1F+SU(2)\times SU(2)\times SU(2)+1F$ gauge theory, with two fundamental hypers for the middle $SU(2)$ group and bifundamental hypers between the middle and two edge $SU(2)$ groups. Of special interest here are the $SU(4)$ and $SU(2)^3$ descriptions as the $\mathbb{Z}_2$ symmetry we twist by is manifest in these, given by charge conjugation on the former and quiver reflection on the latter. 

The mass deformations shown in the figure have a natural interpretation in the gauge theories, such that the resulting $5d$ SCFTs are UV completions of gauge theories that are given by mass deformations of the gauge theories we mentioned for the $\CT^{(2)}_{E_6,2}$ case. Notably, consider the $SU(4)_0 + 2AS + 6F$ theory. One possible set of mass deformations is to integrate away the fundamental hypers. To be consistent with the discrete symmetry we twist by, these must be integrated out in pairs with masses of opposite signs so that the Chern-Simons term remains zero\footnote{The $5d$ Chern-Simons term is not invariant under charge conjugation so, if the mass deformations generate a non-zero Chern-Simons term then they do not respect the discrete symmetry.}. These mass deformations lead to the $5d$ gauge theories $SU(4)_0 + 2AS + (6-2i)F$ for $i=1,2$ and $3$. The $5d$ SCFTs that UV complete these gauge theories are the ones on the left in figure \ref{T2E62FC}, where the case of $i=1$ corresponds to the top web, $i=2$ to the middle one and $i=3$ to the bottom one.

We can also consider deformations from the other gauge theory frames, notably, the $SU(2)^3$ one. One deformation we can consider is integrating out the fundamental flavors for the two edge groups, which must be integrated together to respect the discrete symmetry we twist by. In fact, these can be integrated in two distinct ways, depending on the sign of the masses taken. While both lead to an $SU(2)^3$ gauge theory with two fundamental hypers for the middle $SU(2)$ group and bifundamental hypers between the middle and two edge $SU(2)$ groups, they differ by the $\theta$ angles of the edge $SU(2)$ groups\footnote{We recall here that $5d$ $Sp$ type gauge theories have a $\mathbb{Z}_2$ valued $\theta$ angle arising from the fact that $\pi_4 (Sp) = \mathbb{Z}_2$. When fundamental flavors are present, then the $\theta$ angle can be changed by changing the sign of the mass term for an odd number of such flavors. As a result, in the presence of matter for which this is possible, the $\theta$ angles become physically irrelevant.}. These must be equal, to respect the discrete symmetry, but can be either both $0$ or $\pi$. The latter choice gives a dual description for the $SU(4)_0 + 2AS + 4F$ theory, while the former gives a new $5d$ SCFT as its UV completion, which is the one shown on the top right of figure \ref{T2E62FC}. We can continue on and integrate the two fundamentals for the middle group, leading to the $SU(2)^3_0$ gauge theory, whose SCFT UV completion is given by the web on the middle right of figure \ref{T2E62FC}.

Finally we can consider the deformation given by integrating the two antisymmetric hypers for the $SU(4)_0 + 2AS + 6F$ theory, or the two middle flavors from the $SU(2)^3$ theory. This gives the dual gauge theories $SU(4)_0 + 6F$ and $1F+SU(2)\times SU(2)_0 \times SU(2)+1F$. These are UV completed by a $5d$ SCFT, of the type considered in \cite{Zafrir:2016wkk}. In fact, this theory is one of the theories that were studied in that reference, and we can use the results there for the $4d$ theory. Notably, we expect it to have at least a $USp(6)_{5}\times U(1)$ global symmetry, and have the central charges: $a=\frac{61}{24}$, $c=\frac{17}{6}$. This appears to be a new $4d$ SCFT, on account of not being equal to other SCFTs in the class of theories discussed here and will be further characterized below. 

It is possible to use various properties of this SCFT, observed from the $5d$ construction, to argue that the dimensions of the Coulomb branch operators should be $\D=\{\frac{5}2,3\}$. Specifically, we observe the following properties of this SCFT:

\begin{itemize}
\item The Coulomb branch is expected to be two dimensional.
\item We expect that $n_v = 4(2a-c)=9$, and using  the results of \cite{Shapere:2008zf}, we have that $n_v = 9 = 2\D_1 + 2\D_2 -2$, for $\D_1$ and $\D_2$ the dimensions of the two Coulomb branch operators\footnote{This relation between $n_v$ and the dimensions of Coulomb branch operators is known to fail in cases involving discrete gauge symmetries. It seems reasonable to us that this should not occur for this case.}.
\item This theory can be reached via Higgsing of the rank $2$ theory dubbed $R_{2,4}$ that was introduced in \cite{Chacaltana:2014nya}. The Coulomb branch of the $R_{2,4}$ theory is spanned by operators of dimensions $3$ and $5$. 
\end{itemize}

Consider the process of Higgsing the $R_{2,4}$ theory to the $USp(6)_{5}\times U(1)$ SCFT we mentioned. In this process, the Coulomb branch spanning operators can either be lifted or they can be decomposed to a product of operators of smaller dimensions. For instance, when we Higgs $SO(2N+1)+(N_f+1)F$ to $SO(2N)+N_fF$, the Coulomb branch operator of dimension $2N$ decomposes to the square of a Coulomb branch operator of dimension $N$. As here we expect the rank to remain two, the Coulomb branch spanning operators cannot be lifted, so must decompose to Coulomb branch operators of smaller dimension. It is straightforward to see that the only scenario consistent with the expected value of $n_v$ is that the dimension $5$ Coulomb branch operator decomposes to the square of a dimension $\frac{5}{2}$ Coulomb branch operator, leading to the Coulomb branch being spanned by operators of dimension $\frac{5}{2}$ and $3$. This conclusion is completely consistent with the $4d$ moduli space analysis presented below.

For generic $\CS^{(r)}_{E_6,2}$ and $\CT^{(r)}_{E_6,2}$ theories the $5d$ $SU(4)$ gauge theory generalizes to the $SU(n)_0 + 2AS + 6F$ theory, where $n=2r+1$ for the $\CS$ case and $n=2r$ for the $\CT$ case. The $SU(2)^3$ description, however, does not generalizes to generic cases. We can then consider similar mass deformations. The ones given by integrating out the fundamental flavors are related to $5d$ SCFTs associated with the $\CS^{(r)}_{G,2}$ and $\CT^{(r)}_{G,2}$ theories for other groups $G$. The ones given by integrating out the antisymmetric are generically not of the form for which there is evidence in favor of a $4d$ SCFT, so we will refrain from making any concrete claim about the conformality of the expected $4d$ theory at this point.  

Similarly, for $\CT^{(2)}_{D_4,3}$ and $\CT^{(2)}_{A_2,4}$, we find deformations leading to $5d$ SCFTs, whose twisted compactification is expected to yield $4d$ SCFTs. From the $5d$ description, we see that the global symmetry should contain $G_2$ for $\CT^{(2)}_{D_4,3}$ and $SU(2)$ for $\CT^{(2)}_{A_2,4}$, and their Higgs branch dimension should be $6$ for $\CT^{(2)}_{D_4,3}$ and $2$ for $\CT^{(2)}_{A_2,4}$. We expect these to give new $4d$ SCFTs. For brevity we shall not explicitly present all the webs for these, which can be generated using similar moves as in the $\CT^{(2)}_{E_6,2}$ example but we will instead describe in detail their $4d$ moduli space below.

\subsubsection{Mass deformations to IR free gauge theories}

We can also consider mass deformations leading to IR free gauge theories, which in the $5d$ description are given by mass deformations leading to theories with an IR free part. For example, we noted that the $\CS^{(r)}_{E_6,2}$ and $\CT^{(r)}_{E_6,2}$ theories are given by twisted compactifications of the $5d$ SCFTs that UV complete the $5d$ gauge theory $SU(N)_0+2AS+6F$, where $N=2r+1$ for the $\CS$ case and $N=2r$ for the $\CT$ case. Here, the $\mathbb{Z}_2$ symmetry we twist by acts as charge conjugation on the gauge theory. This implies that the $\CS^{(r)}_{E_6,2}$ theories possess a mass deformation leading to the IR free gauge theory $SO(2r+1)_0+1AS+3V$, while the $\CT^{(r)}_{E_6,2}$ theories possess a mass deformation leading to the IR free gauge theory $SO(2r)_0+1AS+3V$. The flow along the route $E_6 \rightarrow D_4 \rightarrow A_2$ is implemented in the $5d$ gauge theory by giving masses to pairs of flavors. This implies that the $\CS^{(r)}_{D_4,2}$ theories possess a mass deformation leading to the IR free gauge theory $SO(2r+1)_0+1AS+2V$, the $\CT^{(r)}_{D_4,2}$ theories possess a mass deformation leading to the IR free gauge theory $SO(2r)_0+1AS+2V$, and similarly for the other cases.

It should be possible to find additional such mass deformations also for the other cases, though we will not perform an extensive study here.

\subsection{Direct $4d$ analysis}\label{sec:4dMass}

The $5d$ discussion above suggests that, at rank-2, there are mass deformations which lead to $\cN=2$ $4d$ SCFTs and are not visible from the F-theory picture. The $5d$ analysis allows us also to quickly identify the flavor symmetry of these theories and some other basic features which we will use here to fully characterize these SCFTs. 

The mass deformations of the rank-2 theory are all realized as mass deformations of the rank-1 theories supported on the CB singular strata\footnote{In essence this is the content of the UV-IR simple flavor condition \cite{Martone:2020nsy}.}. Since the mass deformations of rank-1 theories are by now completely understood, we can use this knowledge to study the mass deformations of the rank-2 theories as well. Unfortunately it is not yet obvious which of the mass deformations of the rank-1 theories lead to rank-2 SCFTs and it is very likely that we are missing some extra consistency conditions which ought to be imposed. Therefore we don't expect our analysis to be in any ways complete.

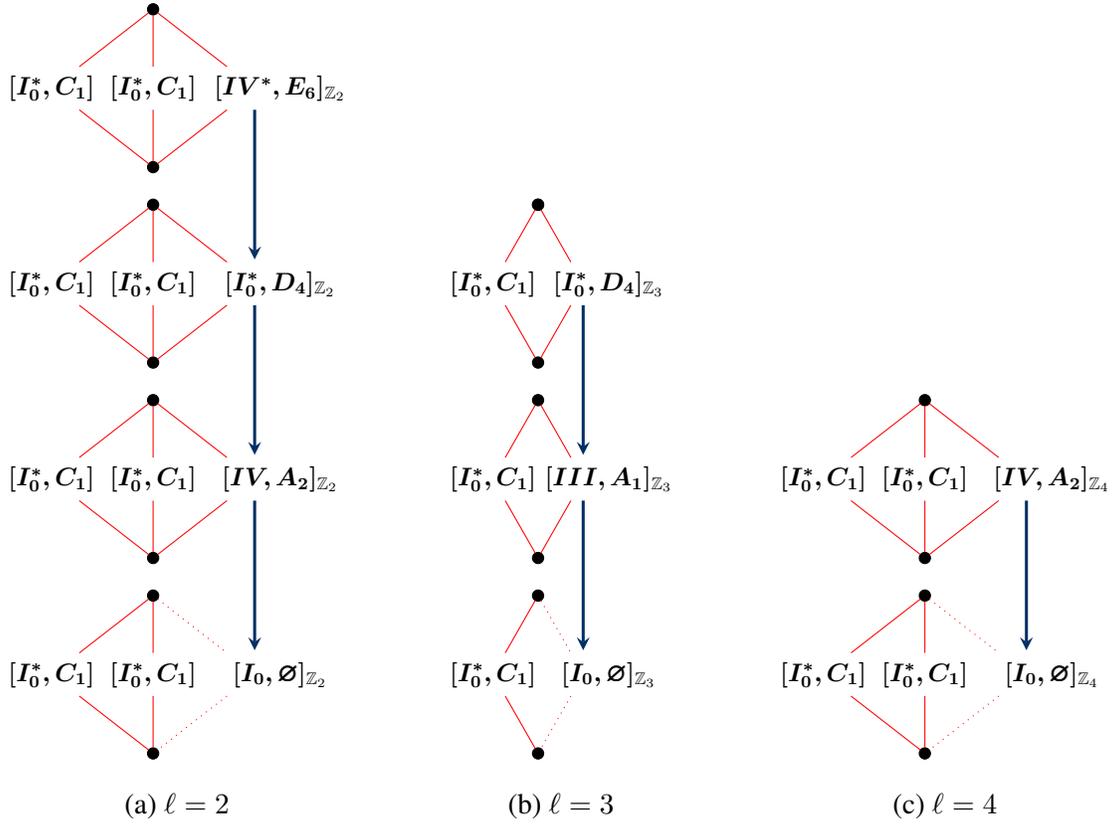
\begin{figure}
\ffigbox{
\begin{subfloatrow}
\ffigbox[6cm][]{\begin{tikzpicture}[decoration={markings,
mark=at position .5 with {\arrow{>}}},
Marrow/.style={->,>=stealth[round],shorten >=1pt,line width=.4mm,coolblack}]
\begin{scope}[scale=1.5]
\node[bbc,scale=.5] (p0a) at (0,0) {};
\node[bbc,scale=.5] (p0b) at (0,-1.4) {};
\node[scale=.8] (p1) at (-.9,-.7) {$\boldsymbol{[I^*_0,C_1]}$};
\node[scale=.8] (p2) at (0,-.7) {$\boldsymbol{[I^*_0,C_1]}$};
\node[scale=.8] (p3) at (.9,-.7) {\qquad $\boldsymbol{[IV^*,E_6]}_{\mathbb{Z}_2}$};
\draw[red] (p0a) -- (p1);
\draw[red] (p0a) -- (p2);
\draw[red] (p0a) -- (p3);
\draw[red] (p1) -- (p0b);
\draw[red] (p2) -- (p0b);
\draw[red] (p3) -- (p0b);
\end{scope}
\begin{scope}[yshift=-2.6cm, scale=1.5]
\node[bbc,scale=.5] (Ap0a) at (0,0) {};
\node[bbc,scale=.5] (Ap0b) at (0,-1.4) {};
\node[scale=.8] (Ap1) at (-.9,-.7) {$\boldsymbol{[I^*_0,C_1]}$};
\node[scale=.8] (Ap2) at (0,-.7) {$\boldsymbol{[I^*_0,C_1]}$};
\node[scale=.8] (Ap3) at (.9,-.7) {\qquad $\boldsymbol{[I_0^*,D_4]}_{\mathbb{Z}_2}$};
\draw[red] (Ap0a) -- (Ap1);
\draw[red] (Ap0a) -- (Ap2);
\draw[red] (Ap0a) -- (Ap3);
\draw[red] (Ap1) -- (Ap0b);
\draw[red] (Ap2) -- (Ap0b);
\draw[red] (Ap3) -- (Ap0b);
\end{scope}
\begin{scope}[yshift=-5.2cm, scale=1.5]
\node[bbc,scale=.5] (Bp0a) at (0,0) {};
\node[bbc,scale=.5] (Bp0b) at (0,-1.4) {};
\node[scale=.8] (Bp1) at (-.9,-.7) {$\boldsymbol{[I^*_0,C_1]}$};
\node[scale=.8] (Bp2) at (0,-.7) {$\boldsymbol{[I^*_0,C_1]}$};
\node[scale=.8] (Bp3) at (.9,-.7) {\qquad $\boldsymbol{[IV,A_2]}_{\mathbb{Z}_2}$};
\draw[red] (Bp0a) -- (Bp1);
\draw[red] (Bp0a) -- (Bp2);
\draw[red] (Bp0a) -- (Bp3);
\draw[red] (Bp1) -- (Bp0b);
\draw[red] (Bp2) -- (Bp0b);
\draw[red] (Bp3) -- (Bp0b);
\end{scope}
\begin{scope}[yshift=-7.8cm, scale=1.5]
\node[bbc,scale=.5] (Cp0a) at (0,0) {};
\node[bbc,scale=.5] (Cp0b) at (0,-1.4) {};
\node[scale=.8] (Cp1) at (-.9,-.7) {$\boldsymbol{[I^*_0,C_1]}$};
\node[scale=.8] (Cp2) at (0,-.7) {$\boldsymbol{[I^*_0,C_1]}$};
\node[scale=.8] (Cp3) at (.9,-.7) {\qquad $\boldsymbol{[I_0,\varnothing]}_{\mathbb{Z}_2}$};
\draw[red] (Cp0a) -- (Cp1);
\draw[red] (Cp0a) -- (Cp2);
\draw[red,dotted] (Cp0a) -- (Cp3);
\draw[red] (Cp1) -- (Cp0b);
\draw[red] (Cp2) -- (Cp0b);
\draw[red,dotted] (Cp3) -- (Cp0b);
\end{scope}
\draw[Marrow] (p3) to (Ap3);
\draw[Marrow] (Ap3) to (Bp3);
\draw[Marrow] (Bp3) to (Cp3);
\end{tikzpicture}}{\caption{$\ell=2$}}
\end{subfloatrow}\hspace{-1cm}
\begin{subfloatrow}
\ffigbox[6cm][]{\begin{tikzpicture}[decoration={markings,
mark=at position .5 with {\arrow{>}}},
Marrow/.style={->,>=stealth[round],shorten >=1pt,line width=.4mm,coolblack}]
\node[bbc,scale=.5] (p0a) at (0,1.3) {};
\begin{scope}[yshift=1.3cm,scale=1.5]
\node[bbc,scale=.5] (p0a) at (0,0) {};
\node[bbc,scale=.5] (p0b) at (0,-1.4) {};
\node[scale=.8] (p1) at (-.4,-.7) {$\boldsymbol{[I^*_0,C_1]}$};
\node[scale=.8] (p3) at (.4,-.7) {\qquad $\boldsymbol{[I_0^*,D_4]}_{\mathbb{Z}_3}$};
\draw[red] (p0a) -- (p1);
\draw[red] (p0a) -- (p3);
\draw[red] (p1) -- (p0b);
\draw[red] (p3) -- (p0b);
\end{scope}
\begin{scope}[yshift=-1.3cm, scale=1.5]
\node[bbc,scale=.5] (Ap0a) at (0,0) {};
\node[bbc,scale=.5] (Ap0b) at (0,-1.4) {};
\node[scale=.8] (Ap1) at (-.4,-.7) {$\boldsymbol{[I^*_0,C_1]}$};
\node[scale=.8] (Ap3) at (.4,-.7) {\qquad $\boldsymbol{[III,A_1]}_{\mathbb{Z}_3}$};
\draw[red] (Ap0a) -- (Ap1);
\draw[red] (Ap0a) -- (Ap3);
\draw[red] (Ap1) -- (Ap0b);
\draw[red] (Ap3) -- (Ap0b);
\end{scope}
\begin{scope}[yshift=-3.9cm, scale=1.5]
\node[bbc,scale=.5] (Bp0a) at (0,0) {};
\node[bbc,scale=.5] (Bp0b) at (0,-1.4) {};
\node[scale=.8] (Bp1) at (-.4,-.7) {$\boldsymbol{[I^*_0,C_1]}$};
\node[scale=.8] (Bp3) at (.4,-.7) {\qquad $\boldsymbol{[I_0,\varnothing]}_{\mathbb{Z}_3}$};
\draw[red] (Bp0a) -- (Bp1);
\draw[red,dotted] (Bp0a) -- (Bp3);
\draw[red] (Bp1) -- (Bp0b);
\draw[red,dotted] (Bp3) -- (Bp0b);
\end{scope}
\draw[Marrow] (p3) to (Ap3);
\draw[Marrow] (Ap3) to (Bp3);
\end{tikzpicture}}{\caption{$\ell=3$}}
\end{subfloatrow}\hspace{-1cm}
 \begin{subfloatrow}
\ffigbox[6cm][]{\begin{tikzpicture}[decoration={markings,
mark=at position .5 with {\arrow{>}}},
Marrow/.style={->,>=stealth[round],shorten >=1pt,line width=.4mm,coolblack}]
\begin{scope}[scale=1.5]
\node[bbc,scale=.5] (p0a) at (0,0) {};
\node[bbc,scale=.5] (p0b) at (0,-1.4) {};
\node[scale=.8] (p1) at (-.9,-.7) {$\boldsymbol{[I^*_0,C_1]}$};
\node[scale=.8] (p2) at (0,-.7) {$\boldsymbol{[I^*_0,C_1]}$};
\node[scale=.8] (p3) at (.9,-.7) {\qquad $\boldsymbol{[IV,A_2]}_{\mathbb{Z}_4}$};
\draw[red] (p0a) -- (p1);
\draw[red] (p0a) -- (p2);
\draw[red] (p0a) -- (p3);
\draw[red] (p1) -- (p0b);
\draw[red] (p2) -- (p0b);
\draw[red] (p3) -- (p0b);
\end{scope}
\begin{scope}[yshift=-2.6cm, scale=1.5]
\node[bbc,scale=.5] (Ap0a) at (0,0) {};
\node[bbc,scale=.5] (Ap0b) at (0,-1.4) {};
\node[scale=.8] (Ap1) at (-.9,-.7) {$\boldsymbol{[I^*_0,C_1]}$};
\node[scale=.8] (Ap2) at (0,-.7) {$\boldsymbol{[I^*_0,C_1]}$};
\node[scale=.8] (Ap3) at (.9,-.7) {\qquad $\boldsymbol{[I_0,\varnothing]}_{\mathbb{Z}_4}$};
\draw[red] (Ap0a) -- (Ap1);
\draw[red] (Ap0a) -- (Ap2);
\draw[red,dotted] (Ap0a) -- (Ap3);
\draw[red] (Ap1) -- (Ap0b);
\draw[red] (Ap2) -- (Ap0b);
\draw[red,dotted] (Ap3) -- (Ap0b);
\end{scope}
\draw[Marrow] (p3) to (Ap3);
\end{tikzpicture}}{\caption{$\ell=4$}}
\end{subfloatrow}
}{\caption{\label{fig:MD}F-theory mass deformations of the $\CT^{(2)}_{G,\ell}$ theories from the point of view of the Special K\"ahler stratification; (a) is for $\ell=2$, (b) is for $\ell=3$ and (c) is for $\ell=4$.}}
\end{figure}

Let us start with a warm-up and establish how the F-theory mass deformations can be seen from the CB stratification point of view, the result are depicted in figure \ref{fig:MD}. As it is apparent there, these deformations correspond to the mass deformations pattern of the rank-1 theories studied in \cite{Argyres:2016yzz}, see in particular table 1 with the caveat that we are using here a slightly different notation where we keep track of the parent theory (for examples $[IV^*,E_6]_{\mathbb{Z}_2}$ is labeled as $[II^*,F_4]$ in \cite{Argyres:2016yzz}). The flow in figure \ref{fig:MD} also perfectly reproduces the relation between these theories and $\cN=4$ theories arising at the end of their mass deformation flows, as explained in detail in Section \ref{section:N3}.

Embolden by this nice result we might reasonably expect that the other mass deformations found from $5d$ and which only involve the $SU(2)$ flavor factors, deform the left side of the stratification leaving the right side invariant. 
Let us see explicit examples.

\subsubsection{$\hat{\CT}_{E_6,2}$}
We will start from the analysis of the mass deformation of the $\CT_{E_6,2}^{(2)}$ leading to the $4d$ limit of the top right theory of figure \ref{T2E62FC} for which we already have a candidate, namely example 14 in \cite{Wang:2018gvb}.
Let us start by recalling what are the properties of this putative theory:
\be\label{hTE}
\hTE:\left\{
\begin{array}{r@{\,}l}
\D&=\{4,5\},\\
12c&=64,\\
24a&=112,\\
\mathfrak{f}&= F_4 \times U(1),\\
k_{F_4}&=10.
\end{array}
\right.
\ee
Given this information it is fairly straightforward to come up with a guess for its CB stratification.

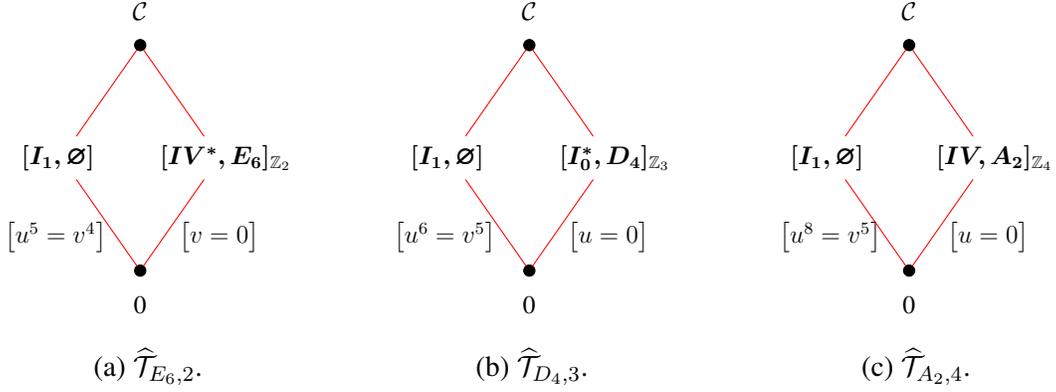
\begin{figure}[t!]
\ffigbox{
\begin{subfloatrow}
\ffigbox[6cm][]{
\begin{tikzpicture}[decoration={markings,
mark=at position .5 with {\arrow{>}}}]
\begin{scope}[scale=1.5]
\node[bbc,scale=.5] (p0a) at (0,0) {};
\node[bbc,scale=.5] (p0b) at (0,-2) {};
\node[scale=.8] (t0a) at (0,.3) {$\cC$};
\node[scale=.8] (t0b) at (0,-2.3) {0};
\node[scale=.8] (p1) at (-.7,-1) {$\boldsymbol{[I_1,\varnothing]}$\ \ };
\node[scale=.8] (p2) at (.7,-1) {\ \ $\boldsymbol{[IV^*,E_6]}_{\mathbb{Z}_2}$};
\node[scale=.8] (t1c) at (-.75,-1.7) {$\big[u^5=v^4\big]$};
\node[scale=.8] (t1c) at (.7,-1.7) {$\big[v=0\big]$};
\draw[red] (p0a) -- (p1);
\draw[red] (p0a) -- (p2);
\draw[red] (p1) -- (p0b);
\draw[red] (p2) -- (p0b);
\end{scope}
\end{tikzpicture}}
{\caption{$\hTE$.}}
\end{subfloatrow}\hspace{-1cm}
\begin{subfloatrow}
\ffigbox[6cm][]{
\begin{tikzpicture}[decoration={markings,
mark=at position .5 with {\arrow{>}}}]
\begin{scope}[scale=1.5]
\node[bbc,scale=.5] (p0a) at (0,0) {};
\node[bbc,scale=.5] (p0b) at (0,-2) {};
\node[scale=.8] (t0a) at (0,.3) {$\cC$};
\node[scale=.8] (t0b) at (0,-2.3) {0};
\node[scale=.8] (p1) at (-.7,-1) {$\boldsymbol{[I_1,\varnothing]}$\ \ };
\node[scale=.8] (p2) at (.7,-1) {\ \ $\boldsymbol{[I_0^*,D_4]}_{\mathbb{Z}_3}$};
\node[scale=.8] (t1c) at (-.75,-1.7) {$\big[u^6=v^5\big]$};
\node[scale=.8] (t1c) at (.7,-1.7) {$\big[u=0\big]$};
\draw[red] (p0a) -- (p1);
\draw[red] (p0a) -- (p2);
\draw[red] (p1) -- (p0b);
\draw[red] (p2) -- (p0b);
\end{scope}
\end{tikzpicture}}
{\caption{$\hTG$.}}
\end{subfloatrow}\hspace{-1cm}
\begin{subfloatrow}
\ffigbox[6cm][]{
\begin{tikzpicture}[decoration={markings,
mark=at position .5 with {\arrow{>}}}]
\begin{scope}[scale=1.5]
\node[bbc,scale=.5] (p0a) at (0,0) {};
\node[bbc,scale=.5] (p0b) at (0,-2) {};
\node[scale=.8] (t0a) at (0,.3) {$\cC$};
\node[scale=.8] (t0b) at (0,-2.3) {0};
\node[scale=.8] (p1) at (-.7,-1) {$\boldsymbol{[I_1,\varnothing]}$\ \ };
\node[scale=.8] (p2) at (.7,-1) {\ \ $\boldsymbol{[IV,A_2]}_{\mathbb{Z}_4}$};
\node[scale=.8] (t1c) at (-.7,-1.7) {$\big[u^8=v^5\big]$};
\node[scale=.8] (t1c) at (.7,-1.7) {$\big[u=0\big]$};
\draw[red] (p0a) -- (p1);
\draw[red] (p0a) -- (p2);
\draw[red] (p1) -- (p0b);
\draw[red] (p2) -- (p0b);
\end{scope}
\end{tikzpicture}}
{\caption{$\hTA$.}}
\end{subfloatrow}}{\caption{\label{fig:MDth}Special K\"ahler stratification of the $\cN=2$ SCFTs obtained by turning on the masses corresponding to the $SU(2)$ flavor factors of (a) $\CT^{(2)}_{E_6,2}$, (b) $\CT^{(2)}_{D_4,3}$ and (c) $\CT^{(2)}_{A_2,4}$.}}
\end{figure}

Because of the UV-IR simple flavor condition, the simple factors of the flavor symmetry of the UV theory need to be realized as flavor symmetry of the rank-1 theories supported on a co-dimension one strata. The $F_4$ factor must be realized by the $[IV^*,E_6]_{\mathbb{Z}_2}$ while both strata on the right need to be lifted. We then speculate that from the $4d$ perspective, the mass deformation which leads to this theory is the mass associated with the diagonal $SU(2)$ inside the $SU(2)\times SU(2)$ of $\CT^{(2)}_{E_6,2}$. 

Using \eqref{actotcint} we can solve for the algebraic form of the $\XSb_{F_4}$, the closure of the strata supporting the $[IV^*,E_6]_{\mathbb{Z}_2}$\footnote{It is well known that in doing computations which are derived from the twisted partition function, we have to ``ignore'' discrete gauging. So while the CB scaling dimension of $[IV^*,E_6]_{\mathbb{Z}_2}$ is 6, the appropriate $\D_i$ to use in \eqref{actotcint} is the one of the parent theory, that is $\D=3$.}:
\be
\XSb_{F_4}:\{(u,v)\in\cC \mid v=0\}.
\ee
Since $\hTE$ is not a product theory, and $\XSb_{F_4}$ is an unknotted stratum, this cannot be the whole story \cite{Argyres:2018urp}. The rest of the stratification can be easily inferred by matching the $c$ and $a$ central charges using \eqref{actotaint} and \eqref{actotbint}. The final result of our analysis is shown in figure \ref{fig:MDth} which neatly confirms our expectations.

\begin{figure}[t!]
\ffigbox{
\begin{subfloatrow}
\ffigbox[6cm][]{
\begin{tikzpicture}[decoration={markings,
mark=at position .5 with {\arrow{>}}}]
\begin{scope}[scale=1.5]
\node[bbc,scale=.5] (p0a) at (0,0) {};
\node[scale=.8] (tp2) at (0.3,-.5) {$\mathfrak{d}_4$};
\node[bbc,scale=.5] (p0b) at (0,-3) {};
\node[scale=.8] (t0a) at (0,.3) {$\cH$};
\node[scale=.8] (t0b) at (0,-3.3) {0};
\node[scale=.8] (p2) at (0,-2) {$\CS^{(1)}_{D_4,2}$\ \ };
\node[scale=.8] (tp2) at (0.3,-1.5) {$\mathfrak{c}_3$};
\node[scale=.8] (p1) at (0,-1) {$\CT^{(1)}_{D_4,1}$};
\node[scale=.8] (tp1) at (0.3,-2.5) {$\mathfrak{f}_4$};
\draw[blue] (p0a) -- (p1);
\draw[blue] (p1) -- (p2);
\draw[blue] (p2) -- (p0b);
\end{scope}
\end{tikzpicture}}
{\caption{$\hTE$.}}
\end{subfloatrow}\hspace{-1cm}
\begin{subfloatrow}
\ffigbox[6cm][]{
\begin{tikzpicture}[decoration={markings,
mark=at position .5 with {\arrow{>}}}]
\begin{scope}[scale=1.5]
\node[bbc,scale=.5] (p0a) at (0,0) {};
\node[bbc,scale=.5] (p0b) at (0,-3) {};
\node[scale=.8] (t0a) at (0,.3) {$\cH$};
\node[scale=.8] (t0b) at (0,-3.3) {0};
\node[scale=.8] (p2) at (0,-2) {$\CS^{(1)}_{A_1,3}$\ \ };
\node[scale=.8] (tp2) at (0.3,-1.5) {$\mathfrak{c}_3$};
\node[scale=.8] (p1) at (0,-1) {$\CT^{(1)}_{A_1,1}$};
\node[scale=.8] (tp1) at (0.3,-2.5) {$\mathfrak{g}_2$};
\draw[blue] (p0a) -- (p1);
\draw[blue] (p1) -- (p2);
\draw[blue] (p2) -- (p0b);
\end{scope}
\end{tikzpicture}}
{\caption{$\hTG$.}}
\end{subfloatrow}\hspace{-1cm}
\begin{subfloatrow}
\ffigbox[6cm][]{
\begin{tikzpicture}[decoration={markings,
mark=at position .5 with {\arrow{>}}}]
\begin{scope}[scale=1.5]
\node[scale=.8] (p0a) at (0,-1) {$\CT^{(1)}_{\varnothing,4}$};
\node[bbc,scale=.5] (p0b) at (0,-3) {};
\node[scale=.8] (t0a) at (0,-0.5) {$\cH$};
\node[scale=.8] (t0b) at (0,-3.3) {0};
\node[scale=.8] (tp2) at (0.3,-1.5) {$\mathfrak{a}_1$};
\node[scale=.8] (p1) at (0,-2) {$\CS^{(1)}_{\varnothing,4}$\ \ };
\node[scale=.8] (tp1) at (0.3,-2.5) {$\mathfrak{a}_1$};
 \draw[blue] (p0a) -- (p1);
\draw[blue] (p1) -- (p0b);
\end{scope}
\end{tikzpicture}}
{\caption{$\hTA$.}}
\end{subfloatrow}}{\caption{\label{fig:MDHBs}Higgs branches of the $\cN=2$ SCFTs obtained by mass deforming (a) $\CT^{(2)}_{E_6,2}$, (b) $\CT^{(2)}_{D_4,3}$ and (c) $\CT^{(2)}_{A_2,4}$.}}
\end{figure}
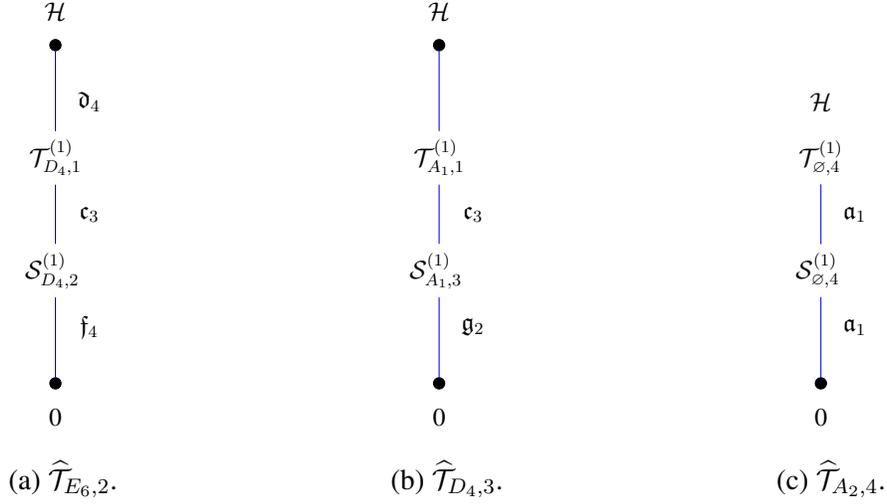

Before concluding this analysis, it is useful to also understand the HB of this theory. Since the mass deformations which we turned on completely break the $SU(2)$ factors, we expect to no longer have the branches which start with an $\mathfrak{a}_1$ transition. Therefore the strata of the HB of $\hTE$ should form a totally ordered set. From the CB stratification we can read off the first transitions of the HB. We then conclude that the HB of $\hTE$ should start with the next to minimal nilpotent orbit of $F_4$, which is indeed the HB of $[IV^*,E_6]_{\mathbb{Z}_2}$. To complete our analysis we need to identify the rank-1 theory\footnote{Since this theory has no ECB, the theory supported on the second stratum of the HB cannot be a rank-2 theory. Under the assumption that no interacting rank-0 SCFT exists, we are left with rank-1 as our only option.} supported on the second stratum of the HB, which is isomorphic to $\mathfrak{f}_4$. There are multiple ways to do that, but we can use a trick since from our CB analysis we also know the second minimal transition in the HB, which is $\mathfrak{c}_3$. This transition is naturally interpreted as first HB transition of the theory supported on $\mathfrak{f}_4$. This observation is enough to single out $\CS^{(1)}_{D_4,2}$ as the appropriate choice and therefore the stratification depicted in figure \ref{fig:MDHBs} (a) follows.

There are multiple checks that can be performed that we obtained the correct HB stratification:
\begin{itemize}
    \item[1.] The quaternionic dimension of the HB is 16 as expected both from the $5d$ analysis and the $U(1)_r^3$ anomaly matching.
    \item[2.] The symplectic stratification in figure \ref{fig:MDHBs} provides enough information to employ the techniques in \cite{Beem:2019tfp,CCLMW2020} and construct the chiral algebras associated to these theories. This calculation depends on the geometry of the stratum (in this case $\mathfrak{f}_4$) and the theory supported there. While the details of the chiral algebra construction are complicated and we have not worked them out in detail, it is possible to very quickly reconstruct the $c$ and $k$ central charge using the by now established relations \cite{Beem:2013sza}:
    \be
    12 c^{4d}=-c^{2d}\qquad ; \qquad k^{4d}_{\mathfrak{f}}=-\frac{k^{2d}_{\mathfrak{f}}}{2}.
    \ee
    Performing this quick calculation again perfectly matches the expected values in \eqref{hTE}
\end{itemize}

\subsubsection{$\hat{\CT}_{D_4,3}$}

Let us now analyze the mass deformation leading to $\hTG$. From the $5d$ analysis we could infer that this theory has the full $G_2$ flavor symmetry of the $\CT_{D_4,3}^{(2)}$ and the HB dimension, which is 8. In this case, we do not know of any candidate $\cN=2$ SCFT which could match these properties and we will need to use all the intuition which we learned from the previous section to fully characterize this theory. We will make the assumption that the CB and HB stratification of this putative theory are those depicted in figure \ref{fig:MDth} (b) and \ref{fig:MDHBs} (b) respectively.

To fully characterize the theory we will impose the following constraints:
\begin{itemize}
    \item Realizing the $G_2$ currents, $\mathcal{J}_{G_2}$, of the $\hTG$ chiral algebra in terms of the generalized free field constructions \cite{Beem:2019tfp,Giacomelli:2020jel}, immediately provides the level:
    \be\label{CAk}
    3k_{G_2}[\hTG]=10+k_{A_1}[\CS^{(1)}_{A_1,3}]\quad\Rightarrow\quad k_{G_2}[\hTG]=\frac{20}3
    \ee
    where we introduced the notation $k_{\mathfrak{f}}[\XT]$ to refer to the level of the simple $\mathfrak{f}$ factor of the theory $\XT$.
    \item The generalized free-field construction of the chiral algebra also immediately allows us to compute the $c$ central charge for this theory \cite{Beem:2019tfp}:
    \be
    12 c^{\hTG}= 12 c^{\CS^{(1)}_{A_1,3}}+2\left(3\frac{k_{G_2}[\hTG]}2-1\right)+2\quad\Rightarrow\quad 12 c^{\hTG}=44
    \ee
    where the 2 at the end arises as the quaternionic dimension of the strata supporting the $\CS^{(1)}_{A_1,3}$ minus one.
    \item Using \eqref{actotbint}, and assuming that $\XSb_{G_2}$ is an unknotted component, we can immediately compute the value of one of the two CB coordinates:
    \be
    \frac{20}3=\frac{4\D_u}2\quad\Rightarrow\quad \D_u=\frac{10}3.
    \ee
    \item Finally matching the $U(1)_r^3$ with the free hypermultiplets at the generic point of the HB we can compute $a$ and therefore $\D_v$: 
    \be
    24(c^{\hTG}-a^{\hTG})=8\quad\Rightarrow\quad\begin{array}{r@{\,}l}
    24a&=82\\
    \D_v&=4 
    \end{array}
    \ee
    which is a compatible result given that $\{10/3,4\}$ is indeed an admitted pair of scaling dimensions \cite{Caorsi:2018zsq}.
\end{itemize}

Summarizing we find the following data:
\be\label{hTG}
\hTG:\left\{
\begin{array}{r@{\,}l}
\D&=\{\frac{10}3,4\}\\
12c&=44\\
24a&=82\\
\mathfrak{f}&\supseteq G_2\\
k_{G_2}&=\frac{20}3
\end{array}
\right.
\ee

\subsubsection{$\hat{\CT}_{A_2,4}$}

From the $5d$ analysis we concluded that there is a mass deformation deforming the $\CT^{(2)}_{A_2,4}$ theory to yet another new rank-2 $\cN=2$ SCFT which we will label $\hTA$ and which has at least an $SU(2)$ flavor symmetry. Assuming that the correct Coulomb and Higgs stratification are those depicted in figure \ref{fig:MDth} (c) and \ref{fig:MDHBs} (c), we can again fully characterize this theory. As in the previous section, we will leverage the tight constraints which directly follows from the structure of the moduli space. In this case the analysis will be slightly more involved because we cannot use a formula analogous to \eqref{CAk} to compute the level of the $SU(2)$ as the non-abelian factor is fully broken by the nilpotent vev initiating the Higgsing to $\CS_{\varnothing,4}^{(1)}$.

Imposing that the $c$ central charge is compatible both with \eqref{actotbint} and the generalized free field chiral algebra construction, that the level of the $SU(2)$ flavor symmetry is compatible with \eqref{actotcint} and that the value of $a$ is such that the $U(1)_r^3$ anomaly is matched by the low energy theory on a generic point of the Higgs branch (which in this case is the combination of a single $\cN=2$ vector multiplet and two hypermultiplets) we obtain the following data:
\be
\hTA:\left\{
\begin{array}{r@{\,}l}
\D&=\{\frac52,4\}\\
12c&=34\\
24a&=67\\
\mathfrak{f}&\supseteq SU(2)\\
k_{SU(2)}&=5
\end{array}
\right.
\ee
To obtain this result we have used as input the allowed scaling dimensions at rank-2 \cite{Argyres:2018zay} and found that $\D_u=\frac 52$ was the only consistent solution. While $\{\frac52,4\}$ does not appear in the list of allowed pairs in \cite{Caorsi:2018zsq}, it is indeed allowed if we consider the by now well-known extension of this list using different branches of the logarithm, see \cite[footnote 16]{Ohmori:2018ona} and \cite[footnote 17]{Giacomelli:2020jel}.

\subsubsection{Others}

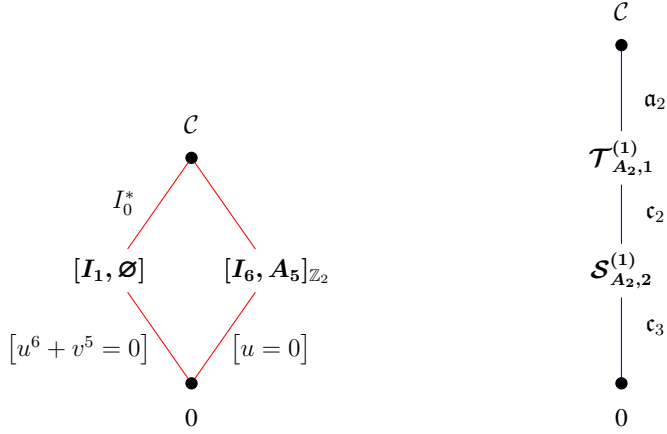
\begin{figure}[t!]
\ffigbox{
\begin{subfloatrow}
\ffigbox[6cm][]{
\begin{tikzpicture}[decoration={markings,
mark=at position .5 with {\arrow{>}}}]
\begin{scope}[scale=1.5]
\node[bbc,scale=.5] (p0a) at (0,0) {};
\node[bbc,scale=.5] (p0b) at (0,-2) {};
\node[scale=.8] (t0a) at (0,.3) {$\cC$};
\node[scale=.8] (t0b) at (0,-2.3) {0};
\node[scale=.8] (p1) at (-.7,-1) {$\boldsymbol{[I_1,\varnothing]}$\ \ };
\node[scale=.8] (p2) at (.7,-1) {\ \ $\boldsymbol{[I_6,A_5]}_{\mathbb{Z}_2}$};
\node[scale=.8] (t1a) at (-.6,-.4) {$I_0^*$};
\node[scale=.8] (t1c) at (-1,-1.7) {$\big[u^6+v^5=0\big]$};
\node[scale=.8] (t1c) at (.7,-1.7) {$\big[u=0\big]$};
\draw[red] (p0a) -- (p1);
\draw[red] (p0a) -- (p2);
\draw[red] (p1) -- (p0b);
\draw[red] (p2) -- (p0b);
\end{scope}
\end{tikzpicture}}
{\caption{Coulomb branch of the $\tTE$.}}
\end{subfloatrow}
\begin{subfloatrow}
\ffigbox[6cm][]{
\begin{tikzpicture}[decoration={markings,
mark=at position .5 with {\arrow{>}}}]
\begin{scope}[scale=1.5]
\node[bbc,scale=.5] (p0a) at (0,0) {};
\node[bbc,scale=.5] (p0b) at (0,-3) {};
\node[scale=.8] (t0a) at (0,.3) {$\cC$};
\node[scale=.8] (t0b) at (0,-3.3) {0};
\node[scale=.8] (p1) at (0,-1) {$\boldsymbol{\CT^{(1)}_{A_2,1}}$};
\node[scale=.8] (p2) at (0,-2) {$\boldsymbol{\CS^{(1)}_{A_2,2}}$};
\node[scale=.8] (t1c) at (.3,-.5) {$\mathfrak{a}_2$};
\node[scale=.8] (t2c) at (.3,-1.5) {$\mathfrak{c}_2$};
\node[scale=.8] (t1c) at (.3,-2.5) {$\mathfrak{c}_3$};
\draw[blue] (p0a) -- (p1);
\draw[blue] (p1) -- (p2);
\draw[blue] (p2) -- (p0b);
\end{scope}
\end{tikzpicture}}
{\caption{Higgs branch of the $\tTE$.}}
\end{subfloatrow}}{\caption{\label{fig:tTE}The Coulomb (a) and Higgs (b) stratification of $\tTE$.}}
\end{figure}

Let us conclude our discussion with determining the $4d$ consistency of yet another $\cN=2$ SCFT whose existence is suggested by the $5d$ analysis and which can be reached by mass deforming the $\CT^{(2)}_{E_6,2}$. We call this theory $\tTE$. Let us recall the properties of this theory:
\be
\tTE:\left\{
\begin{array}{r@{\,}l}
\D&=\{\frac52,3\}\\
12c&=34,\\
24a&=61,\\
\mathfrak{f}&= USp(6)\times U(1),\\
k_{USp(6)}&=5.
\end{array}
\right.
\ee

We want to now check that the above properties, derived from the $5d$ analysis, are indeed consistent with the $4d$ moduli space constraints. Because the UV flavor symmetry needs to be realized on the CB and we are assuming that $\tTE$ can be obtained by turning on a mass deformation of the original $\CT^{(2)}_{E_6,2}$, a natural guess of the rank-1 theory realizing the $USp(6)$ factor is a $\mathbb{Z}_2$ discretely gauged $\cN=2$ $U(1)$ gauge theory with six massless hypermultiplets, which we will also denote as $[I_6,A_5]_{\mathbb{Z}_2}$. This theory has originally a $U(6)_2$ flavor symmetry which is broken to $USp(6)_2$ by the gauging. Then from \eqref{actotcint} immediately follows that the CB coordinate with the lowest scaling dimension should have $\D_u=\frac52$, as argued above, and that
\be
\XSb_{USp(6)}:=\{(u,v)\in\cC \mid u=0\}. 
\ee
Matching the $a$ and $c$ central charges it also follows that $\D_v=3$ and that the CB stratification should be as in figure \ref{fig:tTE} (a). The Higgs branch of the theory is shown instead in figure \ref{fig:tTE} (b) and it again passes all the non-trivial chiral algebra checks. 

It remains less clear how to exactly identify the $\CT^{(2)}_{E_6,2}$ mass deformation from a purely $4d$ perspective. We speculate that this theory is obtained by turning on the mass deformation associated to the diagonal subgroup of the two $SU(2)$s of $\CT_{E_6,2}^{(2)}$ and the $SU(2)\subset F_4$ whose commutant inside $F_4$ is $USp(6)$.

\section*{Acknowledgments}
The authors are thankful to Jacques Distler  for illuminating discussions on possible class S realizations of the 4d theories we discussed.
The authors also thank Craig Lawrie for helpful comments in an earlier draft of the paper.

The work of S.G. is supported by the ERC Consolidator Grant 682608 Higgs bundles: Supersymmetric Gauge Theories and Geometry (HIGGSBNDL). M.M. gratefully acknowledges the Simons Foundation (Simons Collaboration on the Non-perturbative Bootstrap) grants 488647 and 397411, for the support of his work. YT is in part supported  by WPI Initiative, MEXT, Japan at IPMU, the University of Tokyo,
and in part by JSPS KAKENHI Grant-in-Aid (Wakate-A), No.17H04837 
and JSPS KAKENHI Grant-in-Aid (Kiban-S), No.16H06335.
GZ is supported in part by the ERC-STG grant 637844-HBQFTNCER and by the INFN.


\def\arxivfont{\rm}
\bibliographystyle{ytphys}
\baselineskip=.95\baselineskip
\bibliography{ref}

\end{document}